\documentclass[
        prb,
        amsfonts,
        amssymb,
        amsmath,
        notitlepage,
        superscriptaddress, 
        longbibliography,
        reprint]{revtex4-1}

\usepackage[T1]{fontenc}
\usepackage{times}
\usepackage[colorlinks,urlcolor=blue,linkcolor=blue,citecolor=blue]{hyperref}

\usepackage{xcolor}
\usepackage{graphicx}
\usepackage{cancel,soul,ulem}

\begin{document}

\title{A review of modelling in ferrimagnetic spintronics}

\author{Joseph Barker}
\email{j.barker@leeds.ac.uk}
\affiliation{School of Physics and Astronomy, University of Leeds, Leeds LS2 9JT, United Kingdom}
\affiliation{Bragg Centre for Materials Research, University of Leeds, Leeds LS2 9JT, United Kingdom}

\author{Unai Atxitia}
\email{unaiatxitia@zedat.fu-berlin.de}
\affiliation{Fachbereich Physik, Freie Universit\"{a}t Berlin, Arnimallee 14, 14195 Berlin, Germany}
\affiliation{Dahlem Center for Complex Quantum System, Berlin, Arnimallee 14,  14195 Berlin, Germany}

\begin{abstract}
	In this review we introduce computer modelling and simulation techniques which are used for ferrimagnetic materials. We focus on models where thermal effects are accounted for, atomistic spin dynamics and finite temperature macrospin approaches. We survey the literature of two of the most commonly modelled ferrimagnets in the field of spintronics--the amorphous alloy GdFeCo and the magnetic insulator yttrium iron garnet. We look at how generic models and material specific models have been applied to predict and understand spintronic experiments, focusing on the fields of ultrafast magnetisation dynamics, spincaloritronics and magnetic textures dynamics and give an outlook for modelling in ferrimagnetic spintronics
\end{abstract}

\maketitle

\section{Introduction}

In the field of spintronics \cite{Bader_2010_AnnRevCondensMatterPhys_1_1} the goal is to use the spin degree of freedom to create new devices which are superior to electronic devices in some aspect such as energy consumption. It also provides insights into the fundamental physics of spin conversion with electrons and phonons\cite{Otani_2017_NatPhys_13_9}. For example, creating memory devices which can be controlled with electric or spin currents but with better durability and lower power consumption than electronic memories. There is also a push to develop new types of devices for specialised roles such as machine learning where the additional degrees of freedom in magnetic materials can be exploited \cite{Finocchio_2021_JMagnMagnMater_521}.

\begin{figure}
    \centering
    \includegraphics[width=0.48\textwidth]{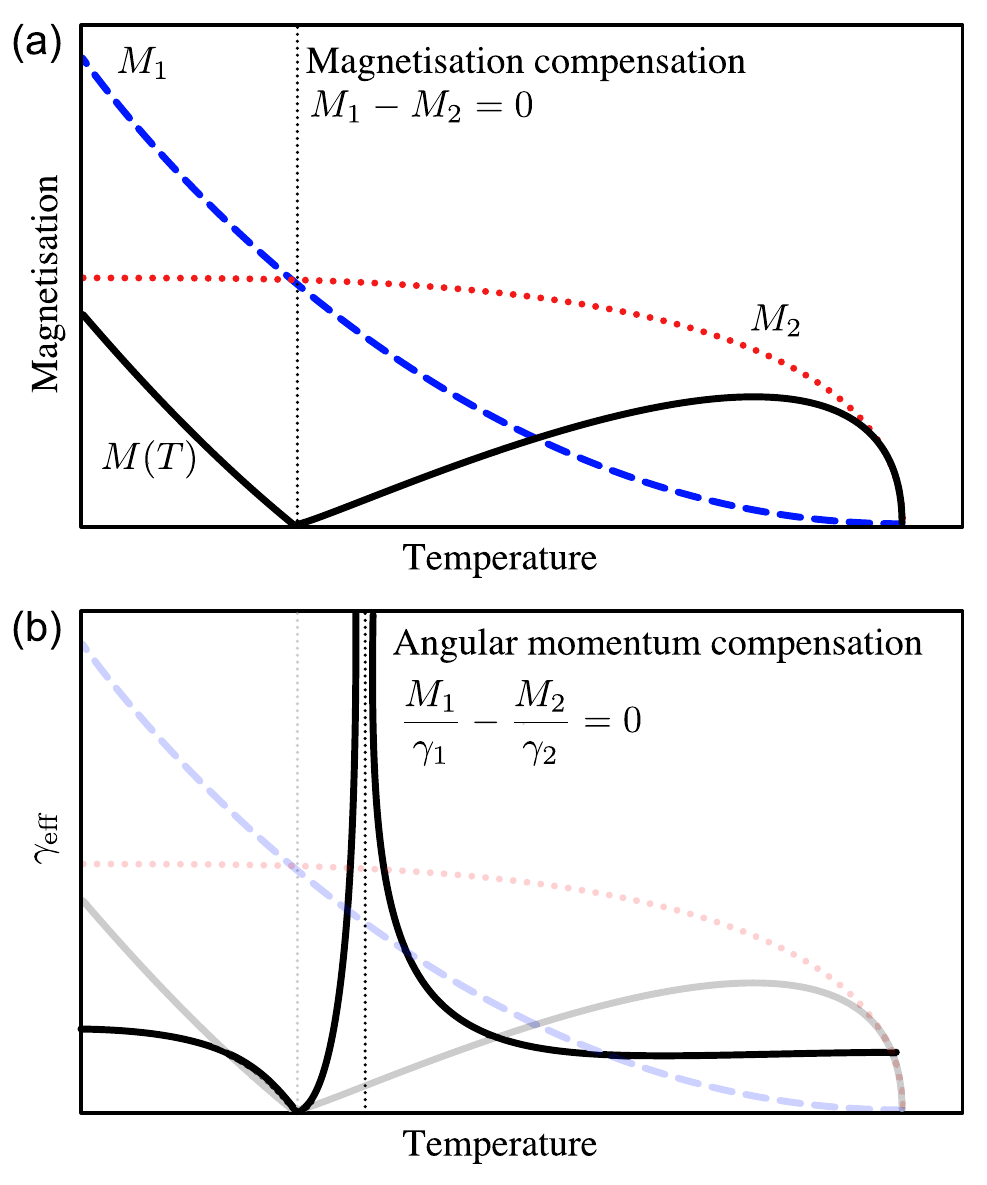}
    \caption{Diagram of the (a) magnetisation compensation and (b) angular momentum compensation points in a two sublattice ferrimagnet. $M_1$ and $M_2$ are the magnetisations of the antiparallel sublattices, which have different gyromagnetic ratios $\gamma_1$ and $\gamma_2$ respectively. $\gamma_{\mathrm{eff}}$ is the effective gyromagnetic ratio defined in Eq.~\eqref{eq:effective_gamma}. If $\gamma_1 = \gamma_2$ then the magnetisation and angular momentum compensation temperatures coincide.}
    \label{fig:compensation_points}
\end{figure}

Ferromagnets provide us with a single magnetisation vector which can be manipulated using external stimuli such as magnetic fields and spin transfer from electrical currents. The interaction of the magnetisation vector with the spin quantisation axis of electrons gives rise to a range of physical effects such as giant magnetoresistance\cite{Parkin_1995_AnnRevMaterSci_25_1, Baibich_1988_PhysRevLett_61_21, Binasch_1989_PhysRevB_39_7}, spin Hall\cite{Dyakonov_1971_SovPhysJETP_13,Dyakonov_1971_PhysLettA_35_6,Hirsch_1999_PhysRevLett_83_9,Kato_2004_Science_306_5703,Wunderlich_2005_PhysRevLett_94_4} and Rashba-Edelstein effects\cite{Edelstein_1990_SolidStateCommun_73_3, Sanchez_2013_NatCommun_4_1}, which can be used to create devices or as fundamental probes in experiments. For quite some time ferromagnetic behaviour has been the focus of studies in spintronics. Whilst ferrimagnetic materials have been used it is generally because of a desirable material property, such as a low magnetic damping or high magneto-optical coupling. The ferrimagnetic nature was incidental and often ignored in theory and simulations due to their minor role in the macroscopic magnetic properties of interest by the time. As research progressed it became clear that some results could only be explained due to the unique characteristics of ferrimagnets which occur with two or more coupled but anti-aligned magnetisation vectors. It has long been known that ferrimagnets have special points in their phase diagram such as the angular momentum and magnetisation compensation points. For a two sublattice ferrimagnet with antiparallel sublattice magnetisations $M_1(T)$ and $M_2(T)$, the magnetisation compensation point is when $M_1(T) = M_2(T)$ (see Fig.~\ref{fig:compensation_points}(a)). There is zero net magnetisation because of the cancellation of the two antiparallel sublattices. If the two sublattices also have different gyromagnetic ratios $\gamma_1$ and $\gamma_2$, then the angular momentum compensation temperature is different from the magnetisation compensation point, being at $M_1(T)/\gamma_1 = M_2(T)/\gamma_2$ as shown in Fig.~\ref{fig:compensation_points}(b). 

Recently, researchers have become interested in using the compensation points to tune the macroscopic properties of ferrimagnets to impart some of the desirable qualities of both ferromagnets--a measurable net magnetisation--and antiferromagnets--high frequency dynamics. The hope is that the versatile spin couplings present in ferrimagnets can be technologically exploited in devices for ultrafast spintronics\cite{Kirilyuk_2010_RevModPhys_82_3}, with the potential to overcome the gigahertz frequency limitations of ferromagnetic based technologies and realise terahertz spintronics\cite{Walowski_2016_JApplPhys_120_14}. Ferrimagnetic insulators may also enable low power devices which work by pure spin currents\cite{Cornelissen_2015_NatPhys_11_12}. Consequently, a large and ever increasing range of ferrimagnetic materials are now used to investigate and push forward the field of spintronics.

Computer modelling of ferrimagnets aims to describe the macroscopic magnetic behaviour, which can be observed in experiments, by considering the microscopic or multiple sublattice variables which are difficult to isolate experimentally. Modelling ferrimagnets involves many of the similar hurdles as modelling antiferromagnets. There are multiple connected sublattices and antiferromagnetic coupling between them can lead to complex phase diagrams and magnetisation dynamics. Especially important in modelling ferrimagnets is the correct accounting of the thermodynamic properties as temperature tunes the system across the compensation points. Simulations are also frequently used in non-equilibrium physics where experimental results can be noisy or of limited resolution and analytic theory can be difficult. In this Review we focus on the two major numerical modelling approaches; atomistic spin dynamics (ASD) \cite{Nowak_Book_ASD, Nowak_2005_PhysRevB_72_17, Skubic_2008_JPhysCondensMatter_20_31, Ma_2011_PhysRevB_83_13, Evans_2014_JPhysCondensMatter_26_10} and the Landau-Lifshitz-Bloch (LLB) equation \cite{Garanin_1997_PhysRevB_55_5, Garanin_2004_PhysRevB_70_21,Chubykalo-Fesenko_2006_PhysRevB_74_9,Evans_2012_PhysRevB_85_1, Atxitia_2012_PhysRevB_86_10, Atxitia_2016_JPhysDApplPhys_50_3} which are able to produce accurate descriptions of both the dynamics and thermodynamics of ferrimagnets. These computational methods are used for different length and timescale limits. Atomistic spin-dynamics simulations are restricted by computational costs to lateral dimensions of 10-100 nm. Timescales are limited to tens of nanoseconds. The simulation of mesoscopic spin textures such as magnetic domains is extremely computationally expensive. For comparison with experiments much larger sample sizes are often necessary. This is usually tackled with micromagnetic approaches. Conventional micromagnetics, rests on a generalisation of the Landau-Lifshitz equation, which is a zero temperature approach and cannot be used when the temperature of the spin system varies in space and time. The primary finite temperature micromagnetic approach is the LLB equation of motion.

We proceed in Section~\ref{sec:modelling} by introducing the basic theory and principles behind the common methods and how they apply to ferrimagnets. In Section~\ref{sec:materials} we explain how two materials--GdFeCo and yttrium iron garnet--have been modelled and why these two materials are so focused on in spintronics. In Section~\ref{sec:survey} we survey the literature where these methods and material models have played an important role either in elucidating the understanding of experiments or predictions of ferrimagnetic behaviour. We conclude in Section~\ref{sec:outlook} with an outlook of the future directions for modelling in ferrimagnetic spintronics.

\section{Modelling Techniques}
\label{sec:modelling}

At a constant temperature and away from the magnetisation or angular momentum compensation points ferrimagnets behave like ferromagnets. Ferrimagnets show a net magnetisation and the low frequency dynamics can be described with equations analogous to a ferromagnet. A considerable amount of modelling of ferrimagnetic materials has used the conventional micromagnetism approach which assumes smooth spatial variation of the net magnetisation. Either by considering the material simply as a ferromagnet, an effective medium approach which adjusts the parameters to mimic a given temperature, or by coupling multiple magnetisation vectors in a cell or close proximity. These approaches describe many experimental situations well\cite{Woo_2018_NatCommun_9_1,Caretta_2018_NatNanotechnol_13_12}, but when temperature becomes a dynamic variable or higher energy modes play a role in dynamics or thermodynamics, more sophisticated approaches are needed. 

\subsection{Atomistic Spin Dynamics (ASD)} 
\label{sec:asd}

Atomic spin models are a natural formalism for ferrimagnetism because they can explicitly account for the different magnetic moments at the atomic scale within the material. They are often complemented with statistical physics methods because with large ensembles of spins, thermodynamic properties can be calculated. A Heisenberg Hamiltonian describes the 
energetics of the system owning to the interactions between the classical `spin' vectors. The word spin is used colloquially in this field and means the classical local magnetic moment ascribed to an atomic site. Additional terms of relativistic origin or due to coupling to external stimulus can be added to the Hamiltonian according to the material being studied. A typical example Hamiltonian is
\begin{equation}   
       \mathcal{H}= - \tfrac{1}{2}\sum_{i \neq j} J_{ij} \mathbf{S}_i \cdot \mathbf{S}_j - \sum_{i} d_z S_{z,i}^2-\sum_{i} \mu_{i}\mathbf{B} \cdot \mathbf{S}_i.
\label{eq:HamiltonianASD}
\end{equation}
where the first term is the Heisenberg exchange with $J_{ij}$ is the electronic exchange energy (the $1/2$ is for the double counting in the sum) and $\mathbf{S}_i$ is a classical spin vector of unit length. The second term is a uniaxial anisotropy in the $z$-direction of energy $d_z$ representing magneto-crystalline effects and the third term is an external magnetic field $\mathbf{B}$ measured in Tesla and $\mu_{i}$ is the size of the magnetic moment of the $i$-th spin in Bohr magneton.

The exchange parameters quantify the interaction energy between the spins upon small rotation from the ground state, as demonstrated by Liechestein et al. \cite{Liechtenstein_1984_JPhysFMetalPhys_14_7}. The exchange interaction here can arise from direct exchange of the overlap of electronic orbitals, typical in transition metals, or indirect such as super-exchange mediated by a third (non-magnetic) ion as is common in oxides. Ferrimagnets can be described by considering this Hamiltonian with two or more distinct types of spin. Two possibilities are i) when there the spin moments of different atoms have different magnitudes, for example a combination of transition metal and rare earth elements as in the alloy GdFeCo ii) in a crystal where the same element exists in different environments which are antiferromagnetically coupled. The magnetic garnets are a typical example of this, where Fe$^{3+}$ ions exist in both octahedral and tetrahedral environments in a 2:3 ratio with antiferromagnetic coupling between the two sublattices. In material specific models the exchange parameters $J_{ij}$ are usually parametrised either from neutron scattering measurements of the magnon spectrum or direct calculation of the electronic structure from first principles methods. Often the exchange interaction contains contributions beyond the nearest neighbours which leads to complex magnon band structures.

Classical atomistic spin models are often solved using Monte Carlo techniques~\cite{Nowak_Book} or atomistic spin dynamics--which provides dynamical quantities of interest. The equation of motion for the spin dynamics is the Landau-Lifshitz-Gilbert (LLG) equation: 
\begin{equation}
\frac{\partial \mathbf{S}_i}{\partial t} = - \frac{|\gamma_i|}{(1+\lambda_i^2)}\left[\left( \mathbf{S}_i \times \mathbf{H}_i \right) - \lambda_i  \left( \mathbf{S}_i \times \left(\mathbf{S}_i \times \mathbf{H}_i \right) \right)\right].
\label{eq:llg}
\end{equation}
$\lambda_i$ is the (dimensionless) local intrinsic damping assigned to the $i$-th site, $\gamma_i = g$ is the gyromagnetic ratio on the $i$-th site in radians per second per Tesla. $\mathbf{H}_i= - (1/\mu_{i})(\partial \mathcal{H}/\partial \mathbf{S}_i)$ is the effective field local to a site $i$ derived from the Hamiltonian. The LLG equation is usually solved numerically for large collections of coupled spins in a manner similar to molecular dynamics for atomic motion. Besides the antiferromagnetic interactions between different sublattices, ferrimagnets are modelled by using distinct parameters on different atomic lattice sites. Elements can have magnetic moments of differing values, different gyromagnetic ratios or exchange interactions. This atomic scale detail is something not possible in micromagnetic approaches based on the continuum approximation.

Temperature is included in this dynamical model following Brown \cite{Brown_1963_PhysRev_130_5} by adding a stochastic field $\boldsymbol{\zeta}_i$ to the effective field $\mathbf{H}_i= \boldsymbol{\zeta}_i(t) - (1/\mu_{i})(\partial \mathcal{H}/\partial \mathbf{S}_i)$. Equation~(\ref{eq:llg}) now becomes a Langevin equation. Conventionally the classical fluctuation dissipation theorem is used (as this is a classical spin model) and the statistical properties of the stochastic field are therefore a white noise with correlations~\cite{Atxitia_2009_PhysRevLett_102_5}:
\begin{equation}
\langle \zeta_{i,\alpha}(t) \rangle = 0 \quad \text{and} \quad \langle \zeta_{i,\alpha}(0) \zeta_{j,\beta}(t) \rangle = \frac{2 \lambda_i k_\text{B} T  \delta_{ij}\delta_{\alpha\beta}\delta(t)}{\mu_{i}\gamma_{i}}.
\label{eq:noise-correlator}
\end{equation}
where $k_B T$ is the thermal energy provided by the heat bath and $\alpha$, $\beta$ are Cartesian components. Within ASD spin-spin interactions occur naturally due to the coupled system of LLG equations and hence their effect is present without explicit inclusion--something which often limits analytic thermodynamic models. The magnon-magnon interactions become significant for thermodynamics at temperatures above approximately half the Curie temperature ($T_{\mathrm{C}}$). Recently advances have been made in the thermostating to use the quantum fluctuation dissipation theorem instead~\cite{Barker_2019_PhysRevB_100_14, Anders_2020_arXiv_2009_00600}. This is essentially the semi-classical limit often used in analytic theory where spin moments are considered in the classical limit but the quantum (Planck) distribution is used as the thermal distribution for the magnons. The use of quantum statistics enables quantitative calculations of thermodynamics even at low temperatures\cite{Ito_2019_PhysRevB_100_6,Barker_2019_PhysRevB_100_14}.

The thermostated spin dynamics gives a canonical ensemble from which thermodynamics can be calculated. The total magnetisation at a given time is $\mathbf{M}(t) = \sum_{i} \mu_{s,i}\mathbf{S}_{i}(t)$ and a thermodynamic average can be calculated from a long time series in equilibrium. For ferrimagnets it is often useful to calculate the sublattice magnetisation, something which only a few experimental methods can probe.

\begin{equation}
    \mathbf{M}_{\nu}(t) = \sum_j \mu_{s,j}\mathbf{S}_{j}(t) 
\end{equation}
where $j$ are the indices belonging to the $\nu$-th sublattice. 

Resolving quantities by sublattice links into the Landau-Lifshitz-Bloch method (described below in Section~\ref{sec:llb}) which requires these thermodynamic quantities. The thermal spin fluctuations cause the system to sample broad regions of the free energy and so although the anisotropy, $d_z$, and exchange energy, $J_{ij}$, in the Hamiltonian are independent of temperature, the macroscopic anisotropy (often denoted $K$) and the macroscopic exchange stiffness (often $A$) are temperature dependent and can be calculated with ASD\cite{Bergman_2010_PhysRevB_81_14}. These are also key input parameters for macroscopic formalisms such as the micromagnetic and LLB methods. The statistical thermodynamic aspects of ASD means it can be used to calculate phase diagrams. The accuracy exceeds what is possible with mean field approaches which are known to significantly over estimate critical points due to the absence of magnon-magnon interactions. For a set of material specific parameters, these microscopic details produce the thermodynamic angular momentum and magnetisation compensation points which distinguish ferrimagnets from ferro and antiferromagnets.  Non-collinear phases such as spin flop which occur in high magnetic fields can be modelled and even non-equilibrium and dynamical states\cite{Ostler_2012_NatCommun_3_1}. 

The damping parameter in ASD describes the rate of energy and angular momentum dissipation to the heat bath. It represents the electron and phonon systems. Detailed descriptions of the heat-bath dynamics are rarely\cite{Ma_2012_PhysRevB_85_18, Ruckriegel_2014_PhysRevB_89_18} introduced or modelled in detail. The coupling to electron and phonon systems are  ``direct'' damping pathways\cite{Gurevich_Book} involving charge carriers or the lattice, with spin-orbit coupling being paramount for the transfer of angular momentum from the spin space to the lattice. The atomic damping parameter appearing in Eq. \eqref{eq:llg} is different from the macroscopic damping parameter--the relaxation rate of the total magnetisation--as measured in an experiment. The macroscopic damping includes further contributions from ``indirect'' mechanisms--magnon-magnon interactions--which are internal to the spin system. This results in an increase in the damping of the macroscopic magnetisation with increasing temperature.

ASD is used for both generic modelling as well as material specific modelling. Generic models include many of the complex effects which are heavily approximated in analytic theory, such as magnon-magnon interactions and non-linear dynamics, but without being tied to a single material. This is well exampled in the field of spincaloritronics (Section~\ref{sec:spincaloritronics}) which studies the coupling between spin and heat--an ideal field for the application of ASD.

\subsection{Macrospin models}

Much of the basic theory of ferrimagnets comes from a simple macrospin model of two coupled magnetisation vectors. This provides simple equations for effective bulk properties within the limit that the two macrospins are rigidly coupled, assuming a strong inter-sublattice exchange interaction.

In ferrimagnets the macroscopic damping, $\alpha_{\rm eff}$--defined as the rate of relaxation of transverse magnetisation oscillations--differs qualitatively from the microscopic damping parameters. The reason is the coupled dynamics of both sublattices. The collective oscillations, described as one effective ferromagnet, relax at a rate which is a combination of the damping parameters from different sublattices, weighted by their magnetisation. This rate of relaxation is what is experimentally observable\cite{Wangsness_1953_PhysRev_91_5}. In the macrospin rigid coupling approximation and not too close to the magnetisation compensation point the effective damping of the net magnetisation is

\begin{equation}
    \alpha_{\rm eff}(T) = \frac{\alpha_1(T) M_1(T)/\gamma_1 + \alpha_2(T) M_2(T)/\gamma_2}{M_1(T)/\gamma_1 - M_2(T)/\gamma_2}.
    \label{eq:effective_damping}
\end{equation}

At low temperatures (much below the Curie temperature, before the magnon-magnon interactions become significant), in a first step, $\alpha_1(T) = \lambda_1$ and $\alpha_2(T) = \lambda_2$ can be taken as constant. The effective damping scales inversely to the net magnetisation $M_1-M_2$, which has a more pronounced temperature dependence than in ferromagnets, with the striking difference that as the system approaches the magnetisation compensation temperature, $M_1-M_2 \rightarrow 0$, the macroscopic damping (Eq. \eqref{eq:effective_damping}) starts to diverge. Divergence of the damping parameter is unphysical, and points to the failure of the effective macrospin approach to describe ferromagnetic magnetisation dynamics.

The local gyromagnetic ratio $\gamma_i$ can also be site dependent in a ferrimagnet leading to an angular momentum compensation point where the sublattices no longer precess. For the macrospin model this is
\begin{equation}
    \gamma_{\rm eff}(T) = \frac{M_1(T) - M_2(T)}{M_1(T)/\gamma_1 - M_2(T)/\gamma_2}.
    \label{eq:effective_gamma}
\end{equation}
The frequency of the lowest energy mode (often called the ferromagnetic mode), when neglecting damping is
\begin{equation}
    \omega(T) = \gamma_{\rm eff}(T)  H.
    \label{eq:effective_omega}
\end{equation}
Approaching the magnetisation compensation point $\omega \rightarrow 0$. 
As demonstrated in Figure~\ref{fig:compensation_points}, at the angular momentum compensation point $M_1(T)/\gamma_1 - M_2(T)/\gamma_2=0$ so $\gamma_{\rm eff}$ and $\omega$ diverge. 
When the inclusion of the damping contribution is used in the calculations, $\omega = (\gamma_{\rm{eff}}/1+\alpha_{\rm eff}^2)H$. This correction removes the divergence of the frequency at the compensation point. 

\subsection{Conventional micromagnetism}

Length scales covered by atomistic spin dynamics simulations are limited to only several tens of nanometers of lateral size and macrospin approaches fail at describing magnetic textures, such as domain walls, vortices or skyrmions (Section~\ref{sec:textures}) which are a key aspect of many spintronic device functionalities. To model spintronic phenomena happening at the micrometer scale, such as magnetic domain motion, one needs to use the so-called micromagnetic models. Micromagnetics has become an pivotal tool for not only for the interpretation of experimental results but also for the design of spintronics devices and predicting novel effects. Originally micromagnetics was developed as a purely theoretical tool for the study of magnetisation processes, however nowadays computer software using finite difference or finite element methods to solve the equations for different geometries is common\cite{oommf, Vansteenkiste_2014_AIPAdv_4_10, Bisotti_2018_JOpenResSoftw_6, Lepadatu_2020_JApplPhys_128_24}.

Conventional micromagnetism was originally formulated within the 'ferromagnetic' framework. The basic underlying assumption of micromagnetism is that neighbouring atomic spins can be considered parallel over a certain scale given by the ratio between the anisotropy and exchange interaction. The magnetisation is therefore treated as a continuous field where at each point of the continuum space $\mathbf{r}$ the magnetisation is defined by one vector, $\mathbf{m}(\mathbf{r})$. Ferrimagnetism was first introduced in micromagnetics by assuming that the opposing sublattices are locked exactly antiparallel and so the coupled equations of motion can be re-written with a single Landau-Lifshitz-Gilbert equation of motion\cite{Oezelt_2015_JMagnMagnMater_381} using the effective parameters for the damping (Eq.~\eqref{eq:effective_damping}) and gyromagnetic ratio (Eq.~\eqref{eq:effective_gamma}). This is sometimes called an effective medium approach. It is useful in ferrimagnets with very strong exchange coupling between the sublattices and away from the compensation points which diverge in the effective parameter equations. Micromagnetism for ferrimagnets with explicit consideration of at least two sublattices is recent\cite{Martinez_2019_JMagnMagnMater_491, Alejos_2018_JApplPhys_123_1, Martinez_2020_AipAdv_10_1}. The assumption of a smoothly changing  magnetisation only holds in the spin subspace corresponding to each magnetic sublattice. At the local level, the spins are aligned antiparallel. Similarly to antiferromagnets, this causes an `exchange enhancement' of the spin dynamics. Exchange enhancement effects are important since allow for the speed up of the dynamics, for example the motion of domain walls at relativistic velocities--near to the maximum magnon velocity of the medium \cite{Caretta_2020_Science_370_6523}. Other magnetic textures, such as skyrmions, in ferrimagnets are also of great fundamental and technological interest. Micromagnetic models allow the investigation of low frequency dynamics of such magnetic textures at the device scale. However, conventional micromagnetism is unable to describe thermal effects. These are crucial in fields such as ultrafast spin dynamics and spincaloritronics and possibly for realistic device modelling due to contact heating. Solutions to the shortcoming of thermodynamic considerations have been proposed for example by rescaling the micromagnetic parameters to their values of at specific temperatures. Stochastic fields have been also added to the equations of motion, however since the finite size of the computational cells means that the full range of spin waves excitations can not be described. At intermediate-to-high temperatures the contribution of those excitations to the thermodynamic properties becomes large and they cannot be neglected. An elegant solution is the use of the Landau-Lifshitz-Bloch equation as a basis for micromagnetism, first derived for ferromagnets \cite{Garanin_1997_PhysRevB_55_5,Chubykalo-Fesenko_2006_PhysRevB_74_9} and later on for ferrimagnets \cite{Atxitia_2012_PhysRevB_86_10}.

\subsection{Landau-Lifshitz-Bloch (LLB) model} 
\label{sec:llb}
 
To include thermal effects into the equation of motion of magnetisation one has to either include a noise term, similar to atomistic spin dynamics methods (Section~\ref{sec:asd}), or to expand the equation of motion to take care of the effect of temperature on a mesoscopic level. 
This leads to the so-called Landau-Lifshitz-Bloch equation. 
The LLB equation describes magnetisation dynamics (rather than spin dynamics) at finite temperatures. It can be considered as an extension of already established micromagnetic methods with a comparable numerical effort.  Standard micromagnetic models are strictly zero temperature formalisms---the length of the magnetisation is fixed. In contrast, the LLB equation admits the relaxation of the magnetisation length. The higher order spin-spin interactions are accounted for in the derivation, producing an equation in which parameters have a temperature dependence. These parameters can be calculated within the mean-field approximation (MFA), higher order methods using Green's functions\cite{Bastardis_2012_PhysRevB_86_9}, atomistic spin dynamics, or parametrised from experimental measurements. 
This makes the LLB equation well suited for modelling scenarios where temperature also changes dynamically, such as the field of ultrafast magnetisation dynamics. Because of the increasing interest in antiferromagnetic and ferrimagnetic materials the LLB equation of motion for two sublattice magnets was recently derived\cite{Atxitia_2012_PhysRevB_86_10} and we briefly introduce it here.

The LLB equation in a two sublattice ferrimagnet is element specific, and for each sublattice averaged magnetisation at each site, $\mathbf{m}_{i,\nu}= \langle \mathbf{S}_{i,\nu} \rangle$,($\nu=$ 1,2), hence, its value is limited to 1. Moreover, in a first step it is common to assume a magnetically homogeneous state, $\mathbf{m}_{i,\nu} \rightarrow \mathbf{m}_{\nu}$ for all lattice sites at the same sublattice $\nu$, the dynamics equation  
reads as follows
\begin{eqnarray} 
        \frac{1}{\gamma_{\nu}}  \frac{\mathrm{d} \mathbf{m}_{\nu} }{\mathrm{d} t} =   &- [\mathbf{m}_{\nu}\times \mathbf{H}_{\rm{eff},\nu}]
           -\frac{\alpha_{\bot}^{\nu}}{m_{\nu}^{2}} [\mathbf{m}_{\nu}\times[ \mathbf{m}_{\nu}\times \mathbf{H}_{\rm{eff},\nu} ]] \nonumber \\
           &+\frac{\alpha_{\|}^{\nu}}{m_{\nu}^2} \left( \mathbf{m}_{\nu} \cdot \mathbf{H}_{\rm{eff},\nu} \right) \mathbf{m}_{\nu}.
\label{eq:ferrimagnetic-llbequation}
\end{eqnarray}
The relaxation of the magnetisation within the LLB equation depends on the longitudinal and transverse damping parameters,
$\alpha_{\|}^{\nu}$ and $\alpha_{\bot}^{\nu}$, and the effective field, $\mathbf{H}_{\rm{eff}}$. 
The damping and other input parameters for the two sublattice LLB equation are element specific. When the gyromagnetic ratio, $\gamma_{\nu}$, is different for each sublattice, owing to the particular relation between orbital and spin degrees of freedom at each sublattice, the total angular momentum compensation point differs from the magnetisation compensation point. 
Below $T_C$, the damping parameters $ \alpha_{\Vert}^{\nu}$ and $\alpha_{\bot}^{\nu}$ are 
\begin{equation}
           \alpha_{\Vert}^{\nu}  = \frac{2k_{\mathrm{B}} T\lambda_{\nu}} { \widetilde{J}_{0,\nu}},\quad \quad
           \alpha_{\bot}^{\nu}  =  \lambda_{\nu}\left(1-\frac{k_{\mathrm{B}} T}{ \widetilde{J}_{0,\nu}}\right).
\label{eq:FerriDampingParameters}
\end{equation}
where $ \widetilde{J}_{0,\nu}= J_{0,\nu}+ |J_{0,\nu\kappa}|m_{e,\kappa}/m_{e,\nu}$, $J_{0,\nu}$ is the intra-sublattice exchange and $J_{0,\nu\kappa}$ is the inter-sublattice exchange (note the sign of the second term does not depend on the sign of the inter-sublattice exchange interaction), $m_{e,\nu}$ is the length of the magnetisation at equilibrium for a given temperature. 
Above $T_C$ the longitudinal and transverse damping parameters are equal and coincide with the expression \cite{Garanin_1997_PhysRevB_55_5} for the classical LLB equation of a ferromagnet above $T_C$, $\alpha_{\bot}^{\nu} = \alpha_{\|}^{\nu} = 2\lambda_{\mu} T/3T_C$, and the equation of motion reduces to Bloch equation.
In Eqs. \eqref{eq:FerriDampingParameters}, the intrinsic damping parameters $\lambda_{\nu}$ depend on the particularities of the spin dissipation at the atomic level as discussed above in Section~\ref{sec:asd}. They can be the same or different for each sublattice. 
The effective field $\mathbf{H}_{\rm{eff},\nu}$ for sublattice $\nu$ is defined as
\begin{eqnarray} 
          &\mathbf{H}_{\rm{eff},\nu} &  =  \mathbf{H}+\mathbf{H}_{\rm{A},\nu}+\frac{J_{0,\nu\kappa}}{\mu_{\nu}}\boldsymbol{\Pi}_{\kappa} \nonumber 
          \\
         & & +
         \left[\frac{1}{2\Lambda_{\nu\nu}}\left(\frac{m_{\nu}^{2}}{m_{e,\nu}^{2}}-1\right)-
         \frac{1}{2\Lambda_{\nu\kappa}}\left(\frac{\tau_{\kappa}^2}{\tau_{e,\kappa}^2}-1\right)\right]\mathbf{m}_{\nu},
         \label{eq:effective-field-two-sublattice}
 \end{eqnarray}
where $\mathbf{H}$ is an applied field,
 $\boldsymbol{\Pi}_{\nu}=
-\left[ \mathbf{m}_{\kappa}\times\left[\mathbf{m}_{\kappa}
\times\mathbf{m}_{\nu}\right]\right]/m^ 2_{\kappa}$ is transverse to $\mathbf{m}_{\kappa}$, 
and $\boldsymbol{\tau}_{\nu}$ is the component of $\mathbf{m}_{\nu}$  parallel to $\mathbf{m}_{\kappa}$, 
in other words $\boldsymbol{\tau}_{\nu}=\mathbf{m}_{\kappa}\left(\mathbf{m}_{\nu}\cdot\mathbf{m}_{\kappa}\right)/m_{\kappa}^{2}$,
where  $\kappa\neq\nu$.
The anisotropy field, $\mathbf{H}_{\rm{A},\nu}$ is related to the zero-field transverse susceptibility or 
directly to the uniaxial anisotropy.
The temperature dependence of the parameters defining the longitudinal dynamics in Eq. \eqref{eq:effective-field-two-sublattice} is
\begin{equation}
\Lambda_{\nu\nu}= 
\frac{1}{ \widetilde{\chi}_{\|,\nu} } \left(1 +  
\frac{ J_{0,\nu \kappa}}{\mu_{0,\nu}}  \widetilde{\chi}_{\|,\kappa} \right) , \quad 
\Lambda_{\nu\kappa} =  \frac{J_{0,\nu \kappa}}{\mu_{0,\kappa}}  \frac{m_{e,\kappa}}{m_{e,\nu}}.
\end{equation}
Within the MFA, 
the equilibrium magnetisation of each sublattice can be obtained via the self-consistent solution of the Curie-Weiss equations $
 m_{e,\nu}=\mathrm{L}\left(( J_{0,\nu}
 m_{e,\nu}
 +|J_{0,\nu\kappa}| m_{e,\kappa} )/k_{\mathrm{B}}T
 \right)$, ($\mathrm{L}$ is the Langevin function)                 
 and the sublattice dependent longitudinal susceptibilities derived directly from them, $\widetilde{\chi^{\nu}_{\|}}=\partial m_{\nu}/\partial H$\cite{Atxitia_2012_PhysRevB_86_10}. 
 For temperatures above $T_C$, one can make use of the symmetry around $T_C$, $\widetilde{\chi}(\epsilon)=2\widetilde{\chi}(-\epsilon)$, 
where $\epsilon=1-T/T_C$ is small\cite{Atxitia_2012_PhysRevB_86_10}.
These parameters can be calculated directly from MFA, gained from  ASD simulations, or measured experimentally.

\subsection{Comparisons of ASD and LLB approaches}

There have been several studies comparing the ASD and ferromagnetic LLB approaches to show consistency\cite{Atxitia_2012_PhysRevB_86_10,Atxitia_2012_PhysRevB_86_10,Schlickeiser_2012_PhysRevB_86_21,Vogler_2019_PhysRevB_100_5}. 
Since the LLB model describes both the longitudinal and transverse relaxation of the magnetisation, two different types of comparison have been conducted so far. In the first case, transverse relaxation, which can also be described by the conventional micromagnetism, the LLB model provides also the temperature dependence of the damping parameters, magnetisation and effective fields. In the second case, longitudinal relaxation, this is described by the last term in Eq. \eqref{eq:ferrimagnetic-llbequation}, and accounts for the dynamics of the magnetisation length. For example, when the temperature of the heat-bath changes, the magnetisation length also changes. This cannot occur in conventional micromagnetic methods.

\begin{figure}
    \centering
    \includegraphics[width=0.48\textwidth]{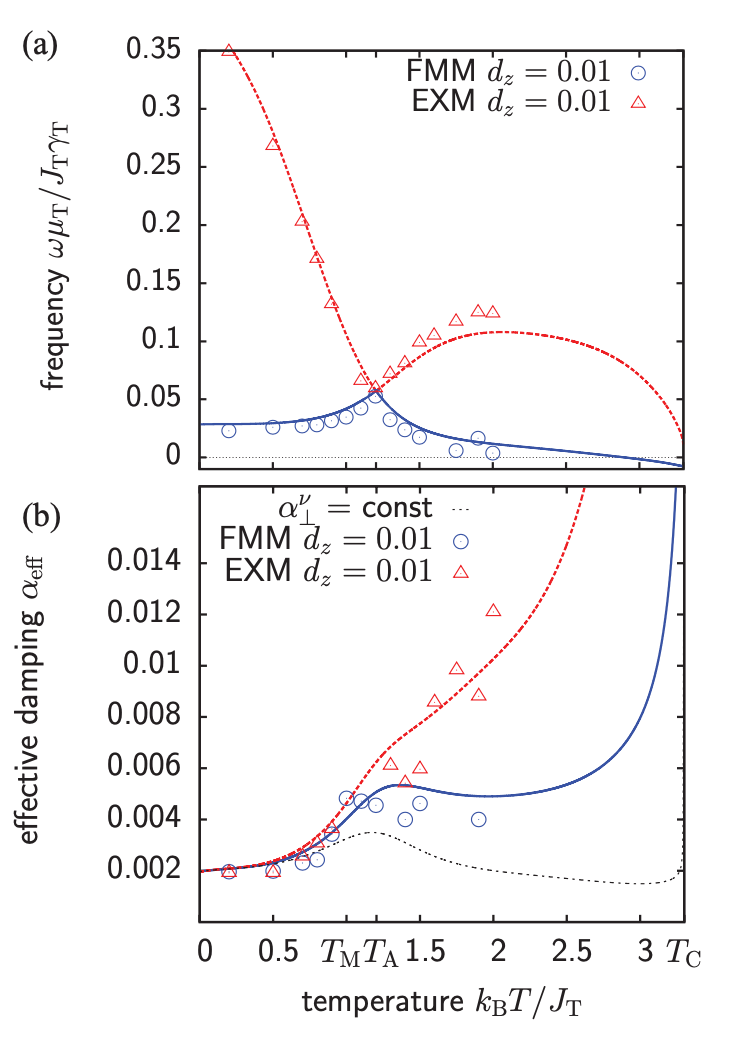}
    \caption{Comparison of ASD (points) and LLB (red and blue lines) calculations of the effective frequency and damping of the ferromagnetic (blue) and exchange (red) modes in a ferrimagnet. The black dashed line is the analytic solution when the temperature dependence of the microscopic damping parameters is not included. (Reprinted figure with permission from [F. Schlickeiser et al., Phys. Rev. B, \textbf{86}, 214416 (2012)] Copyright (2012) by the American Physical Society.)}
    \label{fig:schlickeiser_prb_2012}
\end{figure}

In transverse relaxation numerical experiments, the main difference between ferromagnets and ferrimagnets is found around the magnetisation and angular momentum compensation temperatures. As already described by Eqs. \eqref{eq:effective_damping} and \eqref{eq:effective_omega}, effective damping and frequency strongly depends on the net magnetisation and angular momentum, which in turn depend on temperature. Effective damping also depends on sublattice damping parameters, which are assumed to be constant in conventional micromagnetism, however within the LLB approach they depend on temperature.
 As highlighted above, features such as the temperature dependence of damping should obey  Eq.~(\ref{eq:effective_damping}). This emerges naturally in ASD but it is important the LLB also reproduces this physics. 
Direct comparison between ASD simulations and LLB predictions were made in order to validate the temperature dependence of the parameters involved in the transverse dynamics. Schlickeiser \textit{et al}. performed this ASD-LLB comparison \cite{Schlickeiser_2012_PhysRevB_86_21}. They showed good agreement between ASD and LLB results as shown in Fig.~\ref{fig:schlickeiser_prb_2012}. Their work makes it clear that the effective damping of ferrimagnets is higher than predicted by macroscopic approximations with a constant intrinsic damping parameter. As $T_C$ is approached, spin-spin interactions increase and so the intrinsic damping increases. The simulations also provide access to not only the effective Gilbert-like damping but also the damping of individual modes in the magnon spectrum. The exchange modes have a higher effective damping and a greater temperature dependence. These modes can play an important part in non-equilibrium and dynamical processes\cite{Barker_2013_SciRep_3_1}.

In longitudinal relaxation simulation, one aims to investigate the time scales for the magnetisation length to reach a new equilibrium when the temperature of the heat-bath is changed suddenly. A longitudinal relaxation term was already proposed by Baryaktar as a direct generalisation of the Landau-Lifshitz equation\cite{Baryakhtar_1988}. However, the theory does not answer what happens with phenomenological relaxation parameters. Later on Garanin solved this issue with the LLB equation\cite{Garanin_1997_PhysRevB_55_5}, and Atxitia et al. extended Garanin's ideas to ferrimagnets\cite{Atxitia_2012_PhysRevB_86_10}. In their work, Atxitia et al. already conducted a systematic ASD-LLB comparison of the element-specific longitudinal relaxation dynamics in a GdFeCo alloy. This study was restricted to relatively large intrinsic damping parameters, $\lambda_{\nu}=0.1$ and temperatures below $T_C$. Later, Nieves et al. \cite{Nieves_2015_LowTempPhys_41_9} extended the comparison to all temperatures. 
The comparison between ASD simulations and LLB prediction is very good. This good agreement opens the door to use the LLB model to derive analytical expression of the characteristic time scales of the longitudinal relaxation for situations where the deviations from equilibrium are relatively small. 
In ferrimagnets, the main characteristic of the longitudinal relaxation is that the time scales are element specific. For example, in GdFeCo, Gd relaxes slower than Fe. This is rooted in their different atomic magnetic moments. Interestingly, fluctuations act differently on Gd and Fe sublattices since the alloys are typical $>70\%$ Fe the Gd moments are like impurities inside this Fe ferromagnet, therefore Gd behaves as a polarised paramagnet. As a consequence, whilst magnetic fluctuations make Fe magnetisation dynamics experience critical slowing down near the Curie temperature, in Gd they do not, and the Gd sublattice relaxes faster than Fe at this point. This behaviour was predicted by the LLB model and it has recently been observed in femtosecond element-resolved experiments in GdFeCo\cite{Hennecke_2019_PhysRevLett_122_15}. 

\section{Materials}
\label{sec:materials}

Ferrimagnetic materials are everywhere, many of magnetic materials used in research are some kind of ferrimagnet. Actually the majority of ``ferromagnetic'' insulators are ferrimagnets\cite{Dionne_Book}. There are many ferrimagnets with high potential for technological applications. However two ferrimagnets have been of sustained interest in spintronics ; GdFeCo metallic alloys for ultrafast spintronics applications, and YIG, for insulator spintronics. From the computer modelling perspective, ASD is ideal for the modelling of those two ferrimagnets. First, GdFeCo is most commonly grown as an amorphous alloy, which makes it almost impossible for first principle calculation of macroscopic properties but it is a problem well suited for ASD and LLB models. Second, YIG presents a complex spin structure of the unit cell, and potential novel spintronic properties emerging from the atomic interactions between them can only be handled with atomistic models.   

\subsection{Amorphous GdFeCo}
\label{subsec:GdFeCo}

GdFeCo has long been used for magneto-optical based experiments but is now finding use in other spintronics areas due to its tuneable nature \cite{Yu_2018_NatMater_18_1, Kim_2019_NatMater_18_7, Sala_2021_NatCommun_12_1}. It is an amorphous alloy of Gd, Fe and Co where the content of each can be adjusted in the growth. 
In the Gd with partially filled 4f-shells a magnetic moment of $\sim 7 \mu_B$ arises from the localised electrons due to their coupling according to Hund’s rules. 
The valence band electrons are spin-polarised by the 4f electrons and give rise to an additional 5d6s atomic magnetic moment contribution of around $0.5 \mu_B$. They are strongly exchange coupled to the 4f electrons via a very strong intra-atomic exchange interaction of the order of 200 meV, thus one can fairly consider both kind of spins as one locked magnetic moment for most of the physics studied here.
The Gd moment is therefore much larger than the Fe and Co moments, but in alloys with only 15-30 \% of Gd the net magnetisation can be balanced forming the compensation points even though there are much fewer Gd moments. Being able to continuously adjust the composition allows fine tuning of the macroscopic quantities such as saturation magnetisation and the temperature of the magnetisation and angular momentum compensation points. In experiments, the Gd or FeCo composition are often varied in a systematic set of samples alongside the temperature or applied field. This has lead to important discoveries such as field free optical switching~\cite{Ostler_2012_NatCommun_3_1} but makes the parameter space to model much larger\cite{Ostler_2011_PhysRevB_84_2,Barker_2013_SciRep_3_1}

Although GdFeCo is amorphous it has generally been modelled on a lattice with random site occupancies \cite{Radu_2011_Nature_472_7342}. The exchange interactions are parametrised by nearest neighbour values chosen to reproduce the magnetisation compensation and Curie temperatures\cite{Ostler_2011_PhysRevB_84_2}. While the number of neighbours of each site is constant (due to the lattice) the number of specific Fe or Gd neighbours varies from site to site as would be expected in an amorphous system. Modelling the system as a truly amorphous arrangement of atoms is eminently possible but requires parameters such as the atomic radial distribution function and the distance dependent exchange which are unknown and hard to obtain. The lattice based random site modelling is therefore a crossover between attempts at quantitative modelling and toy models, where salient macroscopic features are reproduced using a somewhat generic base model.

\begin{figure}
    \centering
    \includegraphics[width=0.48\textwidth]{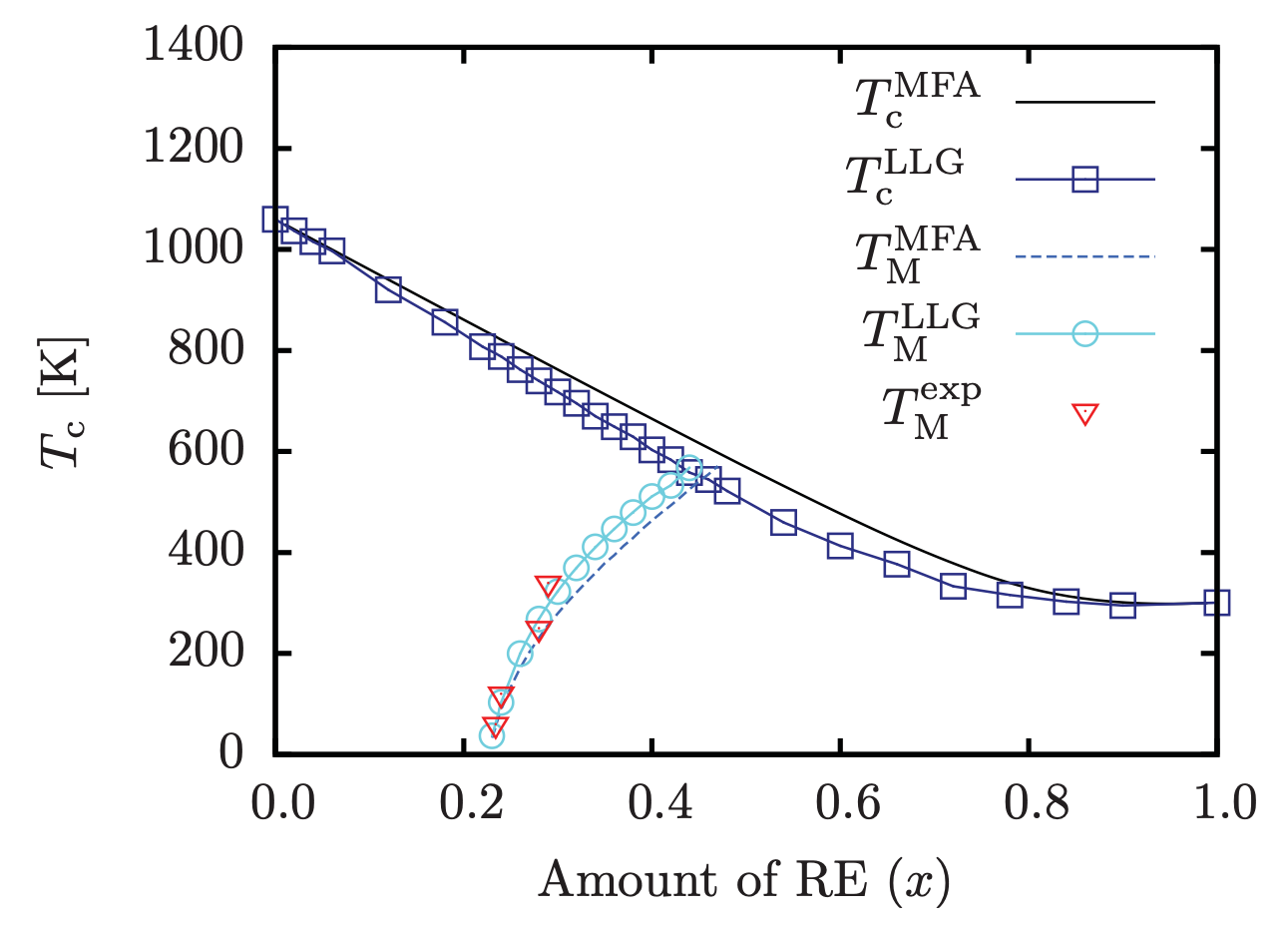}
    \caption{The Cuire temperature ($T_C$ - blue squares) and magnetisation compensation temperature ($T_M$ - blue circles) of the amorphous Gd$_{x}$(FeCo)$_{1-x}$ ASD model as a function of the fractional Gd content. Red triangles are experimental measurements of the magnetisation compensation temperature (Reprinted figure with permission from [T.A. Ostler et al., Phys. Rev. B, \textbf{84}, 024407 (2011)] Copyright (2011) by the American Physical Society.)}
    \label{fig:ostler_prb_2011}
\end{figure}

Ostler et al. performed the first extensive ASD modelling of GdFeCo for different compositions, parametrised from a series of experimental measurements \cite{Ostler_2011_PhysRevB_84_2}. Compensation points as well as critical temperatures were well reproduced by the model, showing an excellent agreement with both experimental and mean field modelling (Fig. \ref{fig:ostler_prb_2011}). They also parametrically change the Fe-Gd exchange coupling and calculate the thermal relaxation time of the element-specific magnetisation length, showing that stronger coupling leads to a faster thermalisation. Since then, the ASD model for GdFeCo proposed by Ostler et al. has been used for the description of ultrafast all-optical switching with great success, see Section~\ref{sec:ultrafast} for further details.

\subsection{Garnets}
Yttrium Iron Garnet (YIG)\cite{Cherepanov_1993_PhysRep_229_3} is one of the main ferrimagnetic materials used in insulator spintronics research. It's very low Gilbert damping, $\alpha\sim 1\times 10^{-5}$ means that spin waves and magnons have very long lifetimes. Recently it has become a platform for spincaloritronics\cite{Bauer_2012_NatMater_11_5} research where magnon spin and heat currents are passed through the YIG\cite{Cornelissen_2015_NatPhys_11_12}. Many other garnets can be formed by substitution, for example the yttrium is easily substituted by other rare-earth elements such as Gd, Tb or Er and the Fe can be substituted by gallium to reduce the magnetisation and introduce compensation points. 

Macroscopically the magnetic properties of YIG appear similar to a ferromagnet. Most analytic theories approximates YIG as a simple one magnon band ferromagnet. In fact YIG is a complex ferrimagnet with the primitive cell containing 20 Fe atoms. Much effort has been applied in building detailed, quantitative models of YIG and numerical modelling has played a role in understanding spintronic effects in YIG and other garnets such as gadolinium iron garnet (GdIG) beyond the ferromagnetic approximations.

The complexity of the crystal structures and magnetic spectra of ferrimagnetic insulators has meant that parametrising atomistic models can be difficult. It is common to infer Heisenberg exchange parameters from inelastic neutron scattering measurements of the magnon spectrum. Linear spin wave models are fitted to the measured dispersion\cite{Toth_2015_JPhysCondensMatter_27_16}. In these complex magnets only a few of the magnon modes can be identified in the neutron scattering making the fitting non-unique. For example YIG must contain 20 magnon modes but neutron measurements see only four or five modes clearly\cite{Princep_2017_npjQuantumMater_2_1}. Gd is also present in many ferrimagnets, particularly those used in magneto-optical based spintronics, as well as some insulating garnets. But it provides a particular difficulty for neutron scattering because its natural isotope has one of the largest neutron capture cross sections. Electronic structure calculations are therefore being increasingly used to establish the exchange interactions in material specific models.

The ferrimagnetic insulators like YIG have large electronic band gaps making the most common density functional theory (DFT) methods poorly suited for calculations due to the significant Coulomb interaction. Attempts have been made to calculate the magnetic properties of ferrimagnetic garnets using DFT+U but these could not simultaneously provide the correct electronic and magnetic properties \cite{Xie_2017_PhysRevB_95_1, Nakamoto_2017_PhysRevB_95_2, Iori_2019_PhysRevB_100_24, Campbell_2020_PhysRevB_102_14}.

Recently more advanced methods such as quasiparticle self-consistent GW \cite{Faleev_2004_PhysRevLett_93_12} have shown great promise for these materials with excellent results obtained for YIG \cite{Barker_2020_ElectronStruct_2_4}. This enables a truly `ab initio' parametrisation of atomic spin models, without even the choice of a Hubbard `U'. Progress in this area may allow accurate modelling of specific ferrimagnetic insulators even where it is difficult to measure the exchange interactions experimentally.

Recent advances in ASD calculations have introduced quantum statistics into the thermostat\cite{Barker_2019_PhysRevB_100_14,Anders_2020_arXiv_2009_00600}. This means the thermal magnons obey Planck rather than Rayleigh-Jeans statistics. The result is that quantitative calculations of thermodynamic properties can be made across almost the entire temperature range from zero Kelvin to the Curie point. The method has been verified for complex ferrimagnets such as YIG with excellent agreement for the temperature dependence of magnetisation, magnon heat capacity\cite{Barker_2019_PhysRevB_100_14} and magnon spectrum\cite{Nambu_2020_PhysRevLett_125_2}. Excellent agreement was found. The magnon heat capacity in particular is an important parameter for thermal and spin transport\cite{Cornelissen_2016_PhysRevB_94_1} but can only be measured below about 10~K in experiments due to the overwhelming contribution of the phonons \cite{Boona_2014_PhysRevB_90_6}. Classical equipartition overestimates its value orders of magnitude and using quantum statistic allowed accurate calculations of the room temperature value\cite{Barker_2019_PhysRevB_100_14}. The ASD calculations showed that at room temperature the presence of the higher terahertz modes means the magnon heat capacity is an order of magnitude larger than otherwise predicted when assuming a simple single magnon band model\cite{Rezende_2015_PhysRevB_91_10}.

\section{Research Domains}
\label{sec:survey}

\subsection{Ultrafast magnetisation dynamics}
\label{sec:ultrafast}

The field of ultrafast magnetisation dynamics began with the discovery of the sub-picosecond magnetic response of Nickel to femtosecond duration optical laser pulses \cite{Beaurepaire_1996_PhysRevLett_76_22}. The use of femtosecond laser pulses has enabled the ultrafast manipulation of the magnetic order, from femtosecond demagnetisation \cite{Beaurepaire_1996_PhysRevLett_76_22,Koopmans_2009_NatMater_9_3,Dornes_2019_Nature_565_7738}
to ultrafast spin currents \cite{Battiato_2010_PhysRevLett_105_2,Rudolf_2012_NatCommun_3_1} and switching of magnetic polarity \cite{Stanciu_2007_PhysRevLett_99_4,Radu_2011_Nature_472_7342,Ostler_2012_NatCommun_3_1,Stupakiewicz_2017_Nature_542_7639,Schlauderer_2019_Nature_569_7756}. Later, alternative rapid excitations have also been shown to drive magnetisation dynamics, including recent demonstrations using picosecond electric and spin currents \cite{Yang_2017_SciAdv_3_11,van_Hees_2020_NatCommun_11_1}.
 
Ferrimagnetic materials play a central role in this field. The most prominent material is amorphous GdFeCo alloys (see Sec. \ref{subsec:GdFeCo}) since it has long been the only material showing the unusual deterministic heat-induced magnetisation switching \cite{Ostler_2012_NatCommun_3_1}. Recently, another class of ferrimagnet, Mn$_2$Ru$_x$Ga \cite{Banerjee_2020_NatCommun_11_1}, has shown the same all thermal switching.
The theoretical description of laser induced all-optical switching (AOS)\cite{Stanciu_2007_PhysRevLett_99_4} of the magnetisation in GdFeCo ferrimagnetic alloys has remained a challenge
\cite{Stanciu_2007_PhysRevLett_99_4,Radu_2011_Nature_472_7342,Ostler_2011_PhysRevB_84_2,Le_Guyader_2012_ApplPhysLett_101_2,Graves_2013_NatMater_12_4,Baryakhtar_2013_LowTempPhys_39_12,Barker_2013_SciRep_3_1,Atxitia_2014_PhysRevB_89_22,Gridnev_2016_JPhysCondensMatter_28_47,Schellekens_2013_PhysRevB_87_2,Mangin_2014_NatMater_13_3,Mentink_2017_JPhysCondensMatter_29_45}
Experimental findings are mostly compared or interpreted in terms of atomistic spin dynamics simulations \cite{Ostler_2011_PhysRevB_84_2,Barker_2013_SciRep_3_1,Wienholdt_2013_PhysRevB_88_2,Gerlach_2017_PhysRevB_95_22} and the Landau-Lifshitz-Bloch equation \cite{Atxitia_2012_PhysRevB_86_10,Atxitia_2013_PhysRevB_87_22,Atxitia_2014_PhysRevB_89_22}.
 
Laser pulses can be as short as just a few femtoseconds, which can excite the electron system on timescales of the order of the exchange interaction allowing the investigation of the fundamental physics governing switching.  
When a metallic ferrimagnetic thin film is subjected to a near infrared laser pulse, only the electrons are accessible to the photon electric field. 
Initially, the absorbed energy is barely transferred to the lattice and consequently the electron system heats up. 
The electron and phonon temperatures, $T_{\rm{el}}$ and $T_{\rm{ph}}$, are decoupled for up to several picoseconds until the electron-phonon interaction equilibrates the two heat-baths. 
This phenomenology is well captured by the so-called two-temperature model (2TM) \cite{Kaganov1957,Chen_2006_IntJHeatMassTransf_49_1-2} which can be written as two coupled differential equations:
\begin{align}
C_{\rm{el}} \frac{\partial T_{\rm{el}}}{\partial t} &= -g_{\rm{ep}}\left( T_{\rm{el}} - T_{\rm{ph}} \right) + P_{l}(t)
\label{eq:2TM-el}
\\
C_{\rm{ph}} \frac{\partial T_{\rm{ph}}}{\partial t} &= +g_{\rm{ep}}\left( T_{\rm{el}} - T_{\rm{ph}} \right).
\label{eq:2TM-ph}
\end{align}
The parameters entering Eqs. \eqref{eq:2TM-el} and \eqref{eq:2TM-ph} are material dependent. For instance, for GdFeCo alloys, $C_{\rm{el}}=\gamma_{\rm{el}} T_{\rm{el}}$ where  $\gamma_{\rm{el}}=6\times 10^2$ J/m$^3$K$^2$, and $C_\text{ph}=3.8\times 10^6$ J/m$^3$K represent the specific heat of the electron- and phonon system. The electron-phonon coupling, $g_{\rm{ep}}=7 \times 10^{17}$ J/m$^3$K. 
Here, $P_l(t)=P_0 \exp( -t^2/t_p^2)$ represents the absorbed energy by the electron system, coming from the laser. The laser duration is $t_p$. 
 
\begin{figure}
    \centering
    \includegraphics[width=0.48\textwidth]{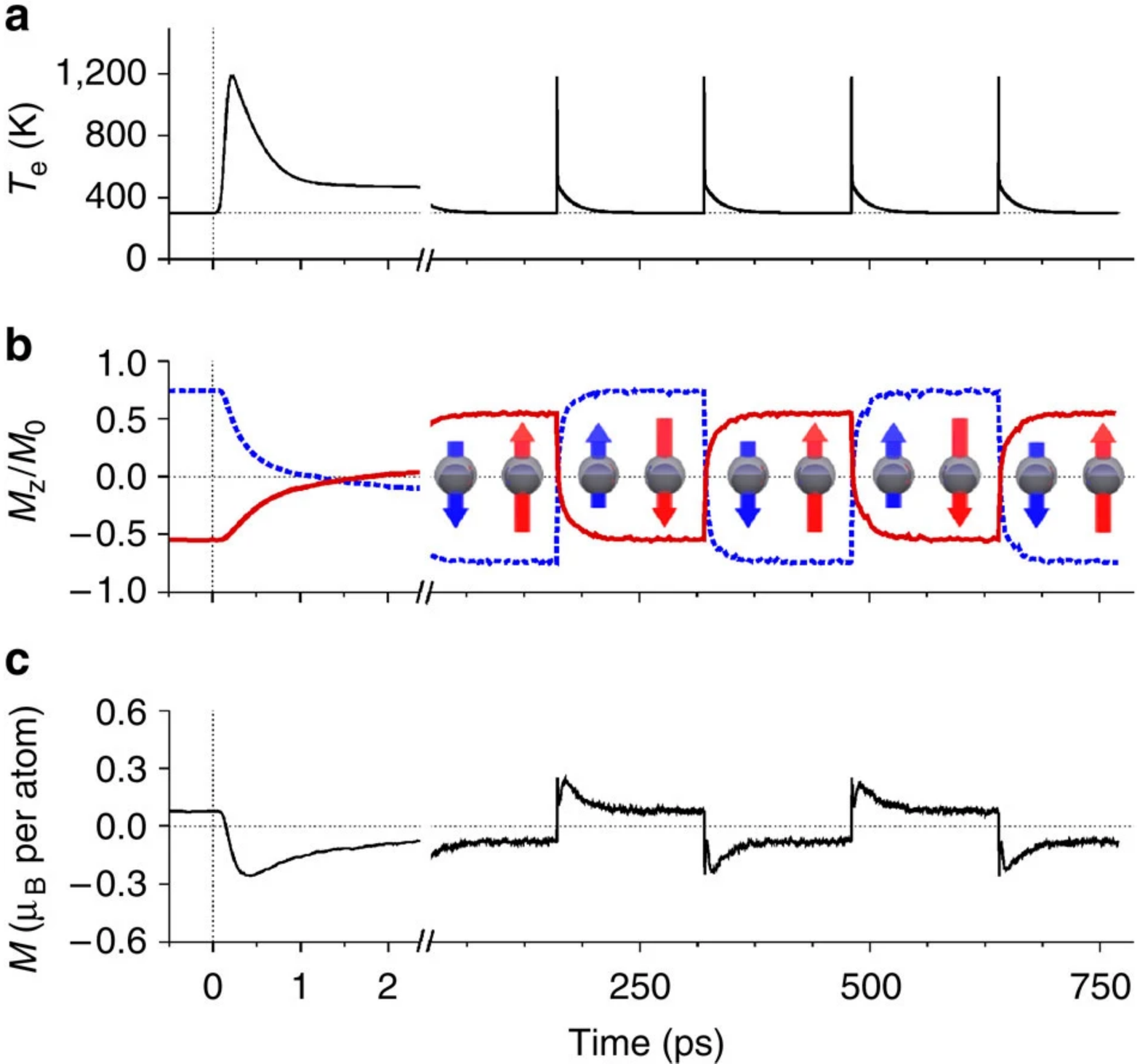}
    \caption{(a) Electron temperature in the two temperature model simulating the thermal laser pulses. (b) The sublattice magnetisation dynamics from ASD, Fe (dashed blue line) and Gd (solid red line). The sublattices switch deterministically with only a thermal pulse and no applied magnetic field. (c) Net magnetisation as a function of time through the switching events. (Reprinted figure from [T.A. Ostler et al., Nat. Commun., \textbf{3}, 666 (2012)])}
    \label{fig:ostler_ncomms_2012}
\end{figure}

ASD simulations coupled to the 2TM predicted and experiments subsequently confirmed, that the heat loaded by a femtosecond laser pulse into these particular systems is sufficient to toggle switch the magnetisation polarity \cite{Ostler_2012_NatCommun_3_1} as shown in the simulation results of Fig.~\ref{fig:ostler_ncomms_2012}.
Despite the Gd and FeCo spin sublattices being coupled antiferromagnetically,
during the switching process, and for only a few picoseconds, the Gd and FeCo spins show a parallel alignment, dubbed a `transient 
ferromagnetic-like state'. 
These insights were only possible thanks to element-specific time resolved femtosecond x-ray magnetic circular dichroism experiments conducted by Radu and co-workers\cite{Radu_2011_Nature_472_7342}.
This class of magnetic switching offers great promise for future applications as picosecond writing times have already been demonstrated in micro\cite{Le_Guyader_2012_ApplPhysLett_101_2} and nanoscale magnetic dots\cite{El-Ghazaly_2019_ApplPhysLett_114_23}. 
Recent experiments have brought this closer to more conventional spintronics paradigms, showing that 
besides using femtosecond laser pulses, GdFeCo nanostructures can be switched by the heat provided by picosecond electric currents\cite{Yang_2017_SciAdv_3_11} and picosecond optical pulses\cite{Gorchon_2016_PhysRevB_94_18}.
Thanks to ASD 
simulations it has been recently shown that switching using femtosecond to picosecond heating follows the same switching path.\cite{Jakobs_arxiv_2020}

Full micromagnetic models based in the LLB equation have been able to predict the ultrafast antiferromagnetic skyrmions creation after ultrafast demagnetisation\cite{Lepadatu_2020_JApplPhys_128_24}.  Micromagnetic simulations based on the LLB have been also used for the description of all-optical switching of the magnetisation using multiple-shot technique in cross devices made of synthetic ferrimagnets composed of rare-earth free materials \cite{Liao_2019_AdvSci_6_24}. Switching behaviours can be controlled by the Curie temperature of each layer in the bilayer, which in turn is controlled by their thickness.
 
\subsection{Spincaloritronics}
\label{sec:spincaloritronics}

The field of spincaloritronics studies the physical effects coming out from the coupling between spin and heat\cite{Bauer_2012_NatMater_11_5}. Spincaloritronics phenomena include the spin Seebeck effect\cite{Uchida_2008_Nature_455_7214,Uchida_2010_NatMater_9_11,Xiao_2010_PhysRevB_81_21}, spin Peltier effect\cite{Flipse_2014_PhysRevLett_113_2,Daimon_2016_NatCommun_7_1}, spin Nernst effect\cite{Meyer_2017_NatMater_16_10}, in metallic, semiconductor, non-magnetic and magnetic insulators. Magnetic insulators are extensively used here because they allow the study of purely magnon driven effects without charge transport convoluting results. Ideally, experiments would study the magnon transport in ferromagnets before branching into ferri- and antiferromagnets but, as mentioned before, most ``ferromagnetic'' magnetic insulators are in-fact ferrimagnets. Many theoretical works in this field neglect the complications associated to the existence of at least two antiparallel sublattices and approximate the materials as a ferromagnet by assuming a single magnon band with a $\hbar\omega\sim k^2$ dispersion. To explore what such approximations may miss in a true ferrimagnetic material ASD modelling has been used extensively. 

The magnonic spin Seebeck effect in a ferrimagnet--the response of the magnetisation under a temperature gradient--has been studied using ASD simulations considering the simplified situation of a temperature step instead of a gradient. In this context, ASD modelling of a generic two sublattice ferrimagnet in a temperature step provided information about the spatial distribution of sublattice magnetisation\cite{Ritzmann_2017_PhysRevB_95_5}.
In this numerical experiment, a stationary non-equilibrium magnon accumulation arises and the shape corresponds to the derivative of the equilibrium magnetisation $\partial m_z / \partial T$ in the absence of temperature step. In ferromagnets, $\partial m_z / \partial T < 0$, but in ferrimagnets showing compensation points, the sign of it can change, hence, the existence of a temperature at which magnetisation accumulation vanishes at the step site.

Modelling in this field has also drawn attention to the magnon polarisation in ferrimagnets\cite{Barker_2016_PhysRevLett_116_14,Princep_2017_npjQuantumMater_2_1,Nambu_2020_PhysRevLett_125_2}. Here polarisation is related to the clockwise versus anticlockwise rotation of the spins or alternatively whether a magnon mode carries $+\hbar$ or $-\hbar$ spin angular momentum.  In a ferromagnet, magnons have a single circular polarisation (or elliptical with dipolar interactions) corresponding to the anticlockwise rotation of the magnetic moments in a field. In a uniaxial antiferromagnet there are both anticlockwise and clockwise polarisations but the magnon modes degenerate so the polarisation cannot easily be measured. In a ferrimagnet the anticlockwise and clockwise polarisations also exist due to the opposing sublattices, but the exchange field between the sublattices splits the modes so that clockwise magnons are higher in energy than anticlockwise.

Experimental measurements and ASD modelling of the spin Seebeck effect in Gadolinium iron garnet (GdIG) showed that magnon polarisation has observable effects in spintronics\cite{Geprags_2016_NatCommun_7_1}. Two changes of sign in the spin Seebeck effect were found as the temperature is increased. One is expected across the magnetisation compensation point as the sublattices reverse in the applied field across this point. The change in sign of the spin Seebeck effect at low temperatures does not correspond to any macroscopic changes of the ferrimagnet. Theory and ASD modelling showed that the changing thermal occupation of magnon modes with different polarisation causes the sign change at low temperature. 

\begin{figure}
    \centering
    \includegraphics[width=0.48\textwidth]{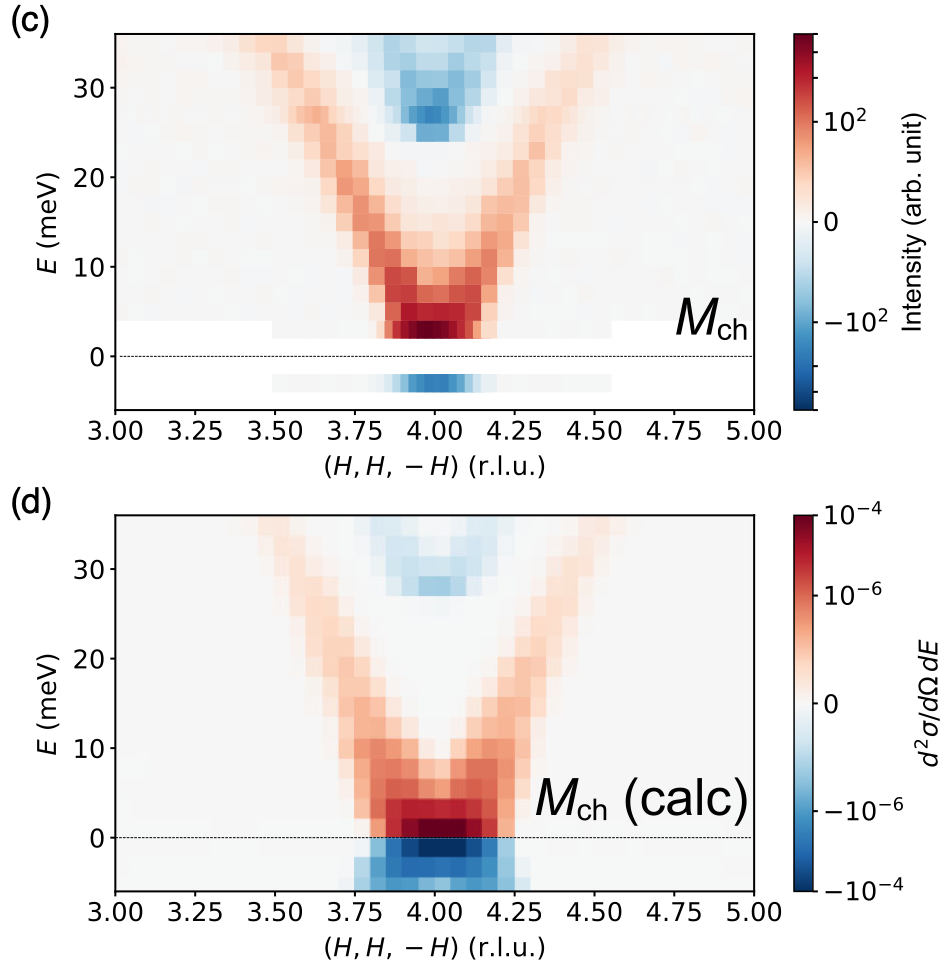}
    \caption{Upper (c) experimental measurements of the magnetic chiral part of polarised inelastic neutron scattering from YIG. Lower (d) ASD calculations of the polarised inelastic neutron scattering cross section from a parametrised model of YIG. The ASD results have been convoluted with the experimental measurement resolution. (Reprinted figure with permission from [Y. Nambu et al., Phys. Rev. Lett., \textbf{125}, 027201 (2020)] Copyright (2020) by the American Physical Society.)}
    \label{fig:nambu_prl_2020}
\end{figure}

Although the polarisation of magnons has been described by theory from the early understanding of magnons it had never been directly measured, possibly due to the lack of apparent physical consequence until recently. Polarised inelastic neutron scattering was used to measure the polarisation of magnons for the first time\cite{Nambu_2020_PhysRevLett_125_2}, supported by quantitative ASD modelling. The scattering cross section in the required experimental geometry is very small making the experiments difficult. The first attempt was only partially successful which was identified because of the prior calculation of the polarised neutron scattering cross section using ASD modelling. A second measurement was successful and very good agreement was found between the ASD calculations and the experimental measurements as shown in Fig.~\ref{fig:nambu_prl_2020}. The polarisation of different magnon modes may be useful in spintronics applications as the magnon transport appears to be sensitive to the polarisation in experiments\cite{Cramer_2017_NanoLett_17_6}. Theoretical suggestions of how to use the polarisation are also being made\cite{Kamra_2017_PhysRevB_96_2}.

\begin{figure}
    \centering
    \includegraphics[width=0.48\textwidth]{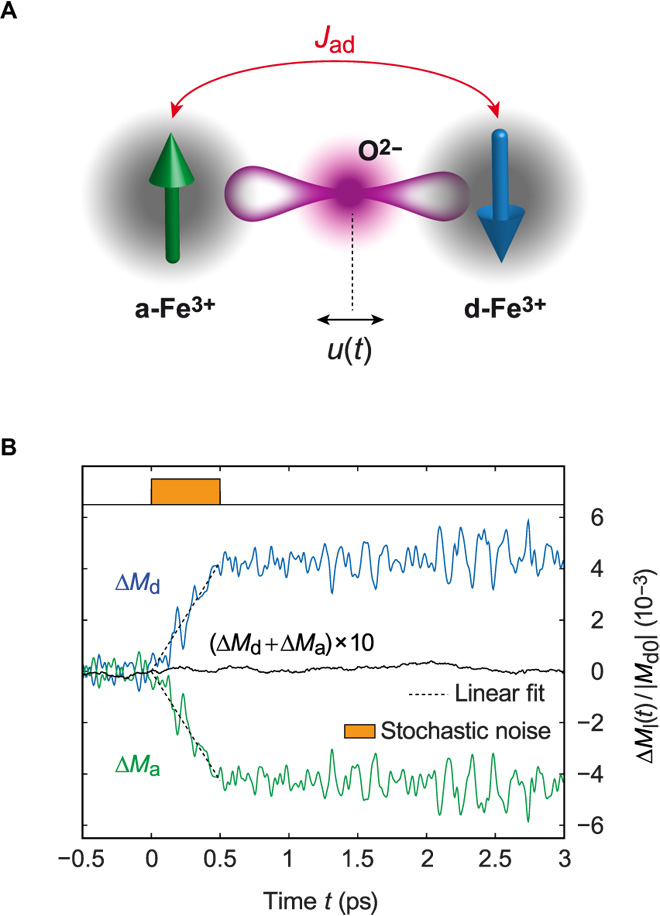}
    \caption{(A) Superexchange model of the inter-sublattice coupling in YIG. The oxygen displacement ($u(t)$) caused by phonon excitations at 20 THz causes fluctuations in the superexchange $J_{ad}$. (B) ASD model with stochastic exchange for 0.5~ps. The change in magnetisation of each sublattice is shown in blue and green respectively. The total change in magnetisation (black line) is zero. (Reprinted figure from [Maehrlein et al., Sci. Adv., \textbf{4}, eaar5164 (2018)])}
    \label{fig:maehrlein_sciadv_2018}
\end{figure}

The high frequency magnon modes of the YIG has motivated experiments to excite these high frequency modes\cite{Maehrlein_2018_SciAdv_4_7,Seifert_2018_NatCommun_9_1,Khan_2019_JPhysCondensMatter_31_27,Hsu_2020_PhysRevB_102_17}.
Maehrlein et al.\cite{Maehrlein_2018_SciAdv_4_7} resonantly excited long-wavelength THz phonons in YIG using THz laser pulses. Changes in the magnetisation were measured via the magneto-optical Faraday effect with an optical probe pulse. A reduction in the sublattice magnetisation was observed on a picosecond timescale\cite{Maehrlein_2018_SciAdv_4_7}. The change in magnetisation persists for microseconds before recovering. The excitation of the infra-red active THz phonons predominantly makes the light oxygen ions to move, displacing them from their equilibrium positions. It is assumed to cause fluctuations in the superexchange between Fe atoms due to the changes in the bond distances and angles, shown pictorially in Fig.~\ref{fig:maehrlein_sciadv_2018}A. To confirm the experimental findings ASD calculations were performed with a modified Heisenberg exchange which included dynamical stochastic fluctuations of the intersublattice exchange parameters. From the model it was determined that the net magnetisation remains constant, even though the sublattice magnetisation is decreasing (Fig.~\ref{fig:maehrlein_sciadv_2018}B). This is because the isotropic exchange interaction conserves the the spin angular momentum. 

The fields of ultrafast magnetisation dynamics and spincaloritronics sometimes combine and effects such as the spin Seebeck effect are probed on the sub-picosecond timescale to understand the timescales of the fundamental processes. Seifert et al. used a femtosecond laser to suddenly heat the Pt of a YIG-Pt bilayer \cite{Seifert_2018_NatCommun_9_1}. The YIG is transparent to the laser and so there is a significant temperature difference between the hot electrons in the Pt and the magnons and phonons of the YIG. The aim was to understand how quickly the spin current across the YIG-Pt interface is established due to electron-magnon scattering. The spin current (measured from the THz emission due to the current formed by the inverse spin Hall effect) emerges almost instantaneously and peaks within 500 femtoseconds of the laser excitation. A dynamical theory was created which required parameters such as the frequency dependent spin susceptibility of the YIG. This was calculated using ASD with a detailed model of YIG and showed that the magnetic system can respond instantly due to the lack of inertia. The quantitative theory allowed an estimate of the interfacial $sd$-exchange coupling and spin mixing conductance in good agreement with other values in the literature.

\subsection{Magnetic textures}
\label{sec:textures}

The motion of spin textures such as domain walls and skyrmions is a large area of study because of proposed device concepts using the textures for information storage and processing. Much work has been done on ferromagnetic textures but ferrimagnets are relatively unexplored beyond treating a ferrimagnet as a ferromagnet. The main difference between ferro and ferrimagnets that can be expected is in how the dynamics of textures changes due to the presence of the angular momentum and magnetisation compensation points. It is expected that at the angular compensation point, ferrimagnets behave like antiferromagnets. Antiferromagnetic textures can move at high velocities due to the compensation of torques acting at each sublattice and responsible of the so-called Walker breakdown common for ferromagnets. Analytical theory based on existing macrospin approaches have been used to predict and interpret domain wall dynamics under external field $H$\cite{Kim_2017_NatMater_16_12}. The velocity, $v_{\rm{DW}}$ and precession, $\Omega_{\rm{DW}}$ of the ferrimagnetic domain wall of width $\Delta_{\rm{DW}}$ can be expressed in terms of effective damping (Eq. \eqref{eq:effective_damping} and gyromagnetic ratio (Eq. \eqref{eq:effective_gamma}: 
\begin{equation}
         v_{\rm{DW}}  =   \Delta_{\rm{DW}}\frac{\alpha_{\rm{eff}} }{1+\alpha_{\rm{eff}}^2} \gamma_{\rm{eff}}H \quad ;\quad 
         \Omega_{\rm{DW}}  =  \frac{ 1}{1+\alpha_{\rm{eff}}^2} \gamma_{\rm{eff}}H. 
         \label{eq:ferri-DW-eff}
\end{equation}

\begin{figure*}
    \centering
    \includegraphics[width=0.96\textwidth]{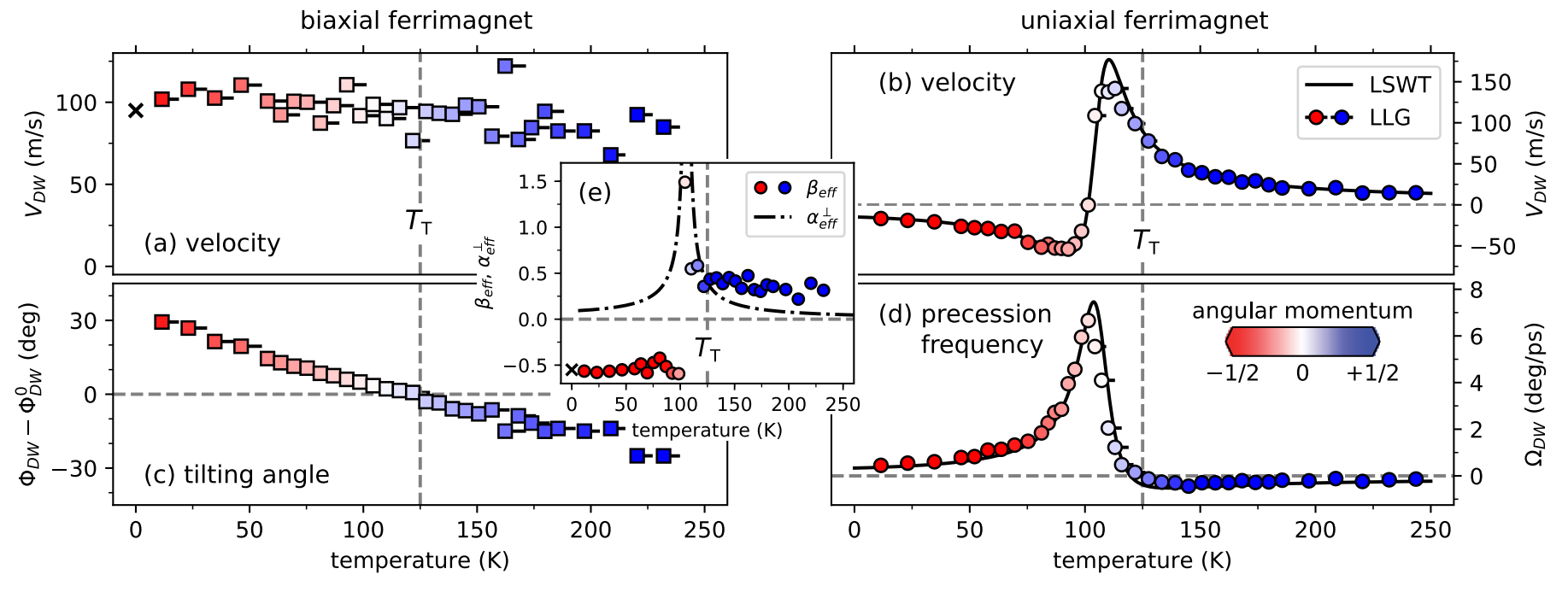}
    \caption{
    Velocity, precession and tilting of ferrimagnetic domain walls (DW) under thermal gradients from ASD simulations. (a) Biaxial anisotropy ferrimagnetic DWs dynamics can be described as an effective ferromagnet, velocity scales with temperature gradient. (b) Tilting of the DW vanishes at the torque compensation temperature $T_{\rm{t}}$ at which the DW dynamics shows similar characteristics to antiferromagnets, namely, quasi-inertia-free motion and the absence of Walker breakdown. (c) Uniaxial anisotropy. Above Walker breakdown, the ferrimagnetic DW can show the opposite, counter intuitive behaviour of moving toward the cold end. Below the compensation temperature the wall moves toward the cold end, whereas above it toward the hot end. This behaviour is driven by angular momentum transfer and therefore strongly related to the angular momentum compensation temperature. (e) The inset shows the temperature dependence of the non-adiabaticy parameter $\beta_{\rm{eff}}$, defined as the ratio between adiabatic and non-adiabatic torque created by a spin current. (Reprinted figure from [Donges et al., Phys. Rev. Research, \textbf{2}, 013293 (2020)])
    }
    \label{fig:donges_prr_2020}
\end{figure*}

Still, thermal effects, such as motion of magnetic textures under thermal gradients require computational models. ASD modelling has been useful in this area. Donges et al. performed an extensive study of how domain walls in ferrimagnets move when thermal gradients are applied~\cite{Donges_2020_PhysRevRes_2_1}. In ferro- and antiferromagnets, domain walls always move towards hotter regions \cite{Hinzke_2011_PhysRevLett_107_2,Schlickeiser_2014_PhysRevLett_113_9,Selzer_2016_PhysRevLett_117_10}. In ferrimagnets they found the same behaviour below the Walker breakdown--where domain walls behave like an effective ferromagnetic DW under thermal gradient. But above the Walker breakdown the domain walls can move towards colder regions if the temperature is less than the angular momentum compensation point as shown in Fig.~\ref{fig:donges_prr_2020}. Their results explain anomalies in experimental measurements of domain wall motion in ultrafast laser experiments on GdFeCo where domain walls have been observed to move away from the heated region~\cite{Shokr_2019_PhysRevB_99_21}--the opposite of previous expectations. This highlights the need for ASD modelling of ferrimagnets as these features cannot be seen in ferromagnetic models such as Eqs. \eqref{eq:ferri-DW-eff} and are difficult to isolate in experiments from many other possible effects. They also find a `torque compensation point' where the ferrimagnetic domain wall moves similarly to an antiferromagnet domain wall with inertia free motion and no Walker breakdown.

In the magnetic skyrmion field there is an effort to remove or control the skyrmion Hall effect--the transverse motion of a skyrmion when it is pushed by a directional stimulus. In antiferromagnets it is known that skyrmions have no hall effect due to the exact cancellation of the Magnus force from the two sublattices\cite{Barker_2016_PhysRevLett_116_14, Zhang_2016_NatCommun_7_1}. Recent experiments in GdFeCo have demonstrated that the skyrmion Hall effect can also be zero in ferrimagnets at the angular momentum compensation point\cite{Hirata_2019_NatNanotechnol_14_3}
. The experimental evidence comes from MOKE imaging and were compared with ASD simulations showing a good agreement confirming the absence of a skyrmion Hall effect when the next spin density is zero.

\section{Summary and Outlook}
\label{sec:outlook}

The modelling techniques presented in this review have already played a central role in uncovering material properties and physical phenomena in ferrimagnetic spintronics which are absent in the commonly used effective `ferromagnetic' models. Even with the rise of antiferromagntic spintronics\cite{Jungwirth_2016_NatNanotechnol_11_3}, we believe ferrimagnetic spintronics will grow due to the manipulations that can be induced by use of the compensation points. One of the most important aspects of computer modelling is the close collaboration with experimental partners. This allows not only to interpret experimental results but also to validate theories and predict new physical phenomena. Paradigmatic examples have been covered in this review. For example, in the field of ultrafast dynamics, modelling techniques have been pivotal to the interpretation of the transient ferromagnetic-like state in GdFeCo, validation of the femtosecond formation of the Spin Seebeck effect in YIG, and the prediction of thermally-induced ultrafast magnetisation switching in GdFeCo. One characteristic of ferrimagnets is that interesting physics appears when the energy input is of the same order as the antiferromagnetic coupling between distinct sublattices. For example, by changing temperature (thermal energy input), magnetisation and angular momentum compensation point can be crossed at which some dynamical properties are similar to antiferromagnets. Lately, models which include thermal effects have dominated this field. Improvements in the modelling of temperature by inclusion of quantum statistics are likely to become more prevalent and provide more predictive weight to multiscale simulations. Efforts to directly model the heat bath subsystems are also likely to increase as there is much interest in the strong coupling between systems such as phonons and magnons. As space and timescales are pushed further models at the atomic level must be further developed beyond the localised rigid spin approximation. There is also a notable absence of models for elements with significant orbital moments which may be required for rare-earth elements which are found in many ferrimagnets.

Going forward we anticipate ferrimagnets to be ever more investigated for spintronic applications. Ferrimagnets are the door for fast control of magnetic functionalities as given by their antiferromagnetic character while conserving the advantage of ferromagnets in terms of measuring and controlling by conventional means such as magnetic fields. Recent examples, include the field of skyrmionics and spin-orbit torque switching. Ferrimagnetic insulators in particular are likely to continue to be used widely in experiments as they provide an important spin source for fundamental and device inspired experiments which attempt to modify spin current transmission. Recently, practical realisation of relativistic kinematics  in isolated  magnetic  solitons has been  demonstrated \cite{Caretta_2020_Science_370_6523} in ferrimagnets, establishing an experimental framework to study relativistic solitonic physics.

So far ferrimagnets have hardly featured in some popular topics such as topological magnetism where the additional symmetries allowed by a two sublattice magnet with a net magnetisation may be of interest. A similarly open avenue to be explored is the investigation of two-dimensional ferrimagnets. A promising route to realise this class of 2D ferrimagnets could be by exploiting the highly tunable superlattice patterns in twisted (or moir\'{e}) bilayers of atomically thin materials. Atomic modelling in these areas would be timely to predict if there is any novel physics to be found before embarking on undoubtedly difficult experiments.

\section*{Acknowledgements}
J.B. acknowledges support from the Royal Society through a University Research Fellowship.
U.A. acknowledges support from the Deutsche Forschungsgemeinschaft through SFB/TRR 227  "Ultrafast Spin Dynamics", Project A08.

\bibliographystyle{apsrev4-1}
\bibliography{biblio_clean.bib}

\begin{thebibliography}{130}%
\makeatletter
\providecommand \@ifxundefined [1]{%
 \@ifx{#1\undefined}
}%
\providecommand \@ifnum [1]{%
 \ifnum #1\expandafter \@firstoftwo
 \else \expandafter \@secondoftwo
 \fi
}%
\providecommand \@ifx [1]{%
 \ifx #1\expandafter \@firstoftwo
 \else \expandafter \@secondoftwo
 \fi
}%
\providecommand \natexlab [1]{#1}%
\providecommand \enquote  [1]{``#1''}%
\providecommand \bibnamefont  [1]{#1}%
\providecommand \bibfnamefont [1]{#1}%
\providecommand \citenamefont [1]{#1}%
\providecommand \href@noop [0]{\@secondoftwo}%
\providecommand \href [0]{\begingroup \@sanitize@url \@href}%
\providecommand \@href[1]{\@@startlink{#1}\@@href}%
\providecommand \@@href[1]{\endgroup#1\@@endlink}%
\providecommand \@sanitize@url [0]{\catcode `\\12\catcode `\$12\catcode
  `\&12\catcode `\#12\catcode `\^12\catcode `\_12\catcode `\%12\relax}%
\providecommand \@@startlink[1]{}%
\providecommand \@@endlink[0]{}%
\providecommand \url  [0]{\begingroup\@sanitize@url \@url }%
\providecommand \@url [1]{\endgroup\@href {#1}{\urlprefix }}%
\providecommand \urlprefix  [0]{URL }%
\providecommand \Eprint [0]{\href }%
\providecommand \doibase [0]{http://dx.doi.org/}%
\providecommand \selectlanguage [0]{\@gobble}%
\providecommand \bibinfo  [0]{\@secondoftwo}%
\providecommand \bibfield  [0]{\@secondoftwo}%
\providecommand \translation [1]{[#1]}%
\providecommand \BibitemOpen [0]{}%
\providecommand \bibitemStop [0]{}%
\providecommand \bibitemNoStop [0]{.\EOS\space}%
\providecommand \EOS [0]{\spacefactor3000\relax}%
\providecommand \BibitemShut  [1]{\csname bibitem#1\endcsname}%
\let\auto@bib@innerbib\@empty
\bibitem [{\citenamefont {Bader}\ and\ \citenamefont
  {Parkin}(2010)}]{Bader_2010_AnnRevCondensMatterPhys_1_1}%
  \BibitemOpen
  \bibfield  {author} {\bibinfo {author} {\bibfnamefont {S.}~\bibnamefont
  {Bader}}\ and\ \bibinfo {author} {\bibfnamefont {S.}~\bibnamefont {Parkin}},\
  }\href {\doibase 10.1146/annurev-conmatphys-070909-104123} {\bibfield
  {journal} {\bibinfo  {journal} {Ann. Rev. Condens. Matter Phys.}\ }\textbf
  {\bibinfo {volume} {1}},\ \bibinfo {pages} {71} (\bibinfo {year}
  {2010})}\BibitemShut {NoStop}%
\bibitem [{\citenamefont {Otani}\ \emph {et~al.}(2017)\citenamefont {Otani},
  \citenamefont {Shiraishi}, \citenamefont {Oiwa}, \citenamefont {Saitoh},\
  and\ \citenamefont {Murakami}}]{Otani_2017_NatPhys_13_9}%
  \BibitemOpen
  \bibfield  {author} {\bibinfo {author} {\bibfnamefont {Y.}~\bibnamefont
  {Otani}}, \bibinfo {author} {\bibfnamefont {M.}~\bibnamefont {Shiraishi}},
  \bibinfo {author} {\bibfnamefont {A.}~\bibnamefont {Oiwa}}, \bibinfo {author}
  {\bibfnamefont {E.}~\bibnamefont {Saitoh}}, \ and\ \bibinfo {author}
  {\bibfnamefont {S.}~\bibnamefont {Murakami}},\ }\href {\doibase
  10.1038/nphys4192} {\bibfield  {journal} {\bibinfo  {journal} {Nat. Phys.}\
  }\textbf {\bibinfo {volume} {13}},\ \bibinfo {pages} {829} (\bibinfo {year}
  {2017})}\BibitemShut {NoStop}%
\bibitem [{\citenamefont {Finocchio}\ \emph {et~al.}(2021)\citenamefont
  {Finocchio}, \citenamefont {Di~Ventra}, \citenamefont {Camsari},
  \citenamefont {Everschor-Sitte}, \citenamefont {Khalili~Amiri},\ and\
  \citenamefont {Zeng}}]{Finocchio_2021_JMagnMagnMater_521}%
  \BibitemOpen
  \bibfield  {author} {\bibinfo {author} {\bibfnamefont {G.}~\bibnamefont
  {Finocchio}}, \bibinfo {author} {\bibfnamefont {M.}~\bibnamefont
  {Di~Ventra}}, \bibinfo {author} {\bibfnamefont {K.~Y.}\ \bibnamefont
  {Camsari}}, \bibinfo {author} {\bibfnamefont {K.}~\bibnamefont
  {Everschor-Sitte}}, \bibinfo {author} {\bibfnamefont {P.}~\bibnamefont
  {Khalili~Amiri}}, \ and\ \bibinfo {author} {\bibfnamefont {Z.}~\bibnamefont
  {Zeng}},\ }\href {\doibase 10.1016/j.jmmm.2020.167506} {\bibfield  {journal}
  {\bibinfo  {journal} {J. Magn. Magn. Mater.}\ }\textbf {\bibinfo {volume}
  {521}},\ \bibinfo {pages} {167506} (\bibinfo {year} {2021})}\BibitemShut
  {NoStop}%
\bibitem [{\citenamefont {Parkin}(1995)}]{Parkin_1995_AnnRevMaterSci_25_1}%
  \BibitemOpen
  \bibfield  {author} {\bibinfo {author} {\bibfnamefont {S.~S.~P.}\
  \bibnamefont {Parkin}},\ }\href {\doibase
  10.1146/annurev.ms.25.080195.002041} {\bibfield  {journal} {\bibinfo
  {journal} {Ann. Rev. Mater. Sci.}\ }\textbf {\bibinfo {volume} {25}},\
  \bibinfo {pages} {357} (\bibinfo {year} {1995})}\BibitemShut {NoStop}%
\bibitem [{\citenamefont {Baibich}\ \emph {et~al.}(1988)\citenamefont
  {Baibich}, \citenamefont {Broto}, \citenamefont {Fert}, \citenamefont
  {Van~Dau}, \citenamefont {Petroff}, \citenamefont {Etienne}, \citenamefont
  {Creuzet}, \citenamefont {Friederich},\ and\ \citenamefont
  {Chazelas}}]{Baibich_1988_PhysRevLett_61_21}%
  \BibitemOpen
  \bibfield  {author} {\bibinfo {author} {\bibfnamefont {M.~N.}\ \bibnamefont
  {Baibich}}, \bibinfo {author} {\bibfnamefont {J.~M.}\ \bibnamefont {Broto}},
  \bibinfo {author} {\bibfnamefont {A.}~\bibnamefont {Fert}}, \bibinfo {author}
  {\bibfnamefont {F.~N.}\ \bibnamefont {Van~Dau}}, \bibinfo {author}
  {\bibfnamefont {F.}~\bibnamefont {Petroff}}, \bibinfo {author} {\bibfnamefont
  {P.}~\bibnamefont {Etienne}}, \bibinfo {author} {\bibfnamefont
  {G.}~\bibnamefont {Creuzet}}, \bibinfo {author} {\bibfnamefont
  {A.}~\bibnamefont {Friederich}}, \ and\ \bibinfo {author} {\bibfnamefont
  {J.}~\bibnamefont {Chazelas}},\ }\href {\doibase 10.1103/physrevlett.61.2472}
  {\bibfield  {journal} {\bibinfo  {journal} {Phys. Rev. Lett.}\ }\textbf
  {\bibinfo {volume} {61}},\ \bibinfo {pages} {2472} (\bibinfo {year}
  {1988})}\BibitemShut {NoStop}%
\bibitem [{\citenamefont {Binasch}\ \emph {et~al.}(1989)\citenamefont
  {Binasch}, \citenamefont {Gr\"{u}nberg}, \citenamefont {Saurenbach},\ and\
  \citenamefont {Zinn}}]{Binasch_1989_PhysRevB_39_7}%
  \BibitemOpen
  \bibfield  {author} {\bibinfo {author} {\bibfnamefont {G.}~\bibnamefont
  {Binasch}}, \bibinfo {author} {\bibfnamefont {P.}~\bibnamefont
  {Gr\"{u}nberg}}, \bibinfo {author} {\bibfnamefont {F.}~\bibnamefont
  {Saurenbach}}, \ and\ \bibinfo {author} {\bibfnamefont {W.}~\bibnamefont
  {Zinn}},\ }\href {\doibase 10.1103/physrevb.39.4828} {\bibfield  {journal}
  {\bibinfo  {journal} {Phys. Rev. B}\ }\textbf {\bibinfo {volume} {39}},\
  \bibinfo {pages} {4828} (\bibinfo {year} {1989})}\BibitemShut {NoStop}%
\bibitem [{\citenamefont {Dyakonov}\ and\ \citenamefont
  {Perel}(1971{\natexlab{a}})}]{Dyakonov_1971_SovPhysJETP_13}%
  \BibitemOpen
  \bibfield  {author} {\bibinfo {author} {\bibfnamefont {M.}~\bibnamefont
  {Dyakonov}}\ and\ \bibinfo {author} {\bibfnamefont {V.}~\bibnamefont
  {Perel}},\ }\href {http://www.jetpletters.ac.ru/ps/1587/article_24366.shtml}
  {\bibfield  {journal} {\bibinfo  {journal} {Sov. Phys. JETP Lett.}\ }\textbf
  {\bibinfo {volume} {13}},\ \bibinfo {pages} {467} (\bibinfo {year}
  {1971}{\natexlab{a}})}\BibitemShut {NoStop}%
\bibitem [{\citenamefont {Dyakonov}\ and\ \citenamefont
  {Perel}(1971{\natexlab{b}})}]{Dyakonov_1971_PhysLettA_35_6}%
  \BibitemOpen
  \bibfield  {author} {\bibinfo {author} {\bibfnamefont {M.}~\bibnamefont
  {Dyakonov}}\ and\ \bibinfo {author} {\bibfnamefont {V.}~\bibnamefont
  {Perel}},\ }\href {\doibase 10.1016/0375-9601(71)90196-4} {\bibfield
  {journal} {\bibinfo  {journal} {Phys. Lett. A}\ }\textbf {\bibinfo {volume}
  {35}},\ \bibinfo {pages} {459} (\bibinfo {year}
  {1971}{\natexlab{b}})}\BibitemShut {NoStop}%
\bibitem [{\citenamefont {Hirsch}(1999)}]{Hirsch_1999_PhysRevLett_83_9}%
  \BibitemOpen
  \bibfield  {author} {\bibinfo {author} {\bibfnamefont {J.~E.}\ \bibnamefont
  {Hirsch}},\ }\href {\doibase 10.1103/physrevlett.83.1834} {\bibfield
  {journal} {\bibinfo  {journal} {Phys. Rev. Lett.}\ }\textbf {\bibinfo
  {volume} {83}},\ \bibinfo {pages} {1834} (\bibinfo {year}
  {1999})}\BibitemShut {NoStop}%
\bibitem [{\citenamefont {Kato}\ \emph {et~al.}(2004)\citenamefont {Kato},
  \citenamefont {Myers}, \citenamefont {Gossard},\ and\ \citenamefont
  {Awschalom}}]{Kato_2004_Science_306_5703}%
  \BibitemOpen
  \bibfield  {author} {\bibinfo {author} {\bibfnamefont {Y.~K.}\ \bibnamefont
  {Kato}}, \bibinfo {author} {\bibfnamefont {R.~C.}\ \bibnamefont {Myers}},
  \bibinfo {author} {\bibfnamefont {A.~C.}\ \bibnamefont {Gossard}}, \ and\
  \bibinfo {author} {\bibfnamefont {D.~D.}\ \bibnamefont {Awschalom}},\ }\href
  {\doibase 10.1126/science.1105514} {\bibfield  {journal} {\bibinfo  {journal}
  {Science}\ }\textbf {\bibinfo {volume} {306}},\ \bibinfo {pages} {1910}
  (\bibinfo {year} {2004})}\BibitemShut {NoStop}%
\bibitem [{\citenamefont {Wunderlich}\ \emph {et~al.}(2005)\citenamefont
  {Wunderlich}, \citenamefont {Kaestner}, \citenamefont {Sinova},\ and\
  \citenamefont {Jungwirth}}]{Wunderlich_2005_PhysRevLett_94_4}%
  \BibitemOpen
  \bibfield  {author} {\bibinfo {author} {\bibfnamefont {J.}~\bibnamefont
  {Wunderlich}}, \bibinfo {author} {\bibfnamefont {B.}~\bibnamefont
  {Kaestner}}, \bibinfo {author} {\bibfnamefont {J.}~\bibnamefont {Sinova}}, \
  and\ \bibinfo {author} {\bibfnamefont {T.}~\bibnamefont {Jungwirth}},\ }\href
  {\doibase 10.1103/physrevlett.94.047204} {\bibfield  {journal} {\bibinfo
  {journal} {Phys. Rev. Lett.}\ }\textbf {\bibinfo {volume} {94}},\ \bibinfo
  {pages} {047204} (\bibinfo {year} {2005})}\BibitemShut {NoStop}%
\bibitem [{\citenamefont
  {Edelstein}(1990)}]{Edelstein_1990_SolidStateCommun_73_3}%
  \BibitemOpen
  \bibfield  {author} {\bibinfo {author} {\bibfnamefont {V.}~\bibnamefont
  {Edelstein}},\ }\href {\doibase 10.1016/0038-1098(90)90963-c} {\bibfield
  {journal} {\bibinfo  {journal} {Solid State Commun.}\ }\textbf {\bibinfo
  {volume} {73}},\ \bibinfo {pages} {233} (\bibinfo {year} {1990})}\BibitemShut
  {NoStop}%
\bibitem [{\citenamefont {S\'{a}nchez}\ \emph {et~al.}(2013)\citenamefont
  {S\'{a}nchez}, \citenamefont {Vila}, \citenamefont {Desfonds}, \citenamefont
  {Gambarelli}, \citenamefont {Attan\'{e}}, \citenamefont {De~Teresa},
  \citenamefont {Mag\'{e}n},\ and\ \citenamefont
  {Fert}}]{Sanchez_2013_NatCommun_4_1}%
  \BibitemOpen
  \bibfield  {author} {\bibinfo {author} {\bibfnamefont {J.~C.~R.}\
  \bibnamefont {S\'{a}nchez}}, \bibinfo {author} {\bibfnamefont
  {L.}~\bibnamefont {Vila}}, \bibinfo {author} {\bibfnamefont {G.}~\bibnamefont
  {Desfonds}}, \bibinfo {author} {\bibfnamefont {S.}~\bibnamefont
  {Gambarelli}}, \bibinfo {author} {\bibfnamefont {J.~P.}\ \bibnamefont
  {Attan\'{e}}}, \bibinfo {author} {\bibfnamefont {J.~M.}\ \bibnamefont
  {De~Teresa}}, \bibinfo {author} {\bibfnamefont {C.}~\bibnamefont
  {Mag\'{e}n}}, \ and\ \bibinfo {author} {\bibfnamefont {A.}~\bibnamefont
  {Fert}},\ }\href {\doibase 10.1038/ncomms3944} {\bibfield  {journal}
  {\bibinfo  {journal} {Nat. Commun.}\ }\textbf {\bibinfo {volume} {4}},\
  \bibinfo {pages} {2944} (\bibinfo {year} {2013})}\BibitemShut {NoStop}%
\bibitem [{\citenamefont {Kirilyuk}\ \emph {et~al.}(2010)\citenamefont
  {Kirilyuk}, \citenamefont {Kimel},\ and\ \citenamefont
  {Rasing}}]{Kirilyuk_2010_RevModPhys_82_3}%
  \BibitemOpen
  \bibfield  {author} {\bibinfo {author} {\bibfnamefont {A.}~\bibnamefont
  {Kirilyuk}}, \bibinfo {author} {\bibfnamefont {A.~V.}\ \bibnamefont {Kimel}},
  \ and\ \bibinfo {author} {\bibfnamefont {T.}~\bibnamefont {Rasing}},\ }\href
  {\doibase 10.1103/revmodphys.82.2731} {\bibfield  {journal} {\bibinfo
  {journal} {Rev. Mod. Phys.}\ }\textbf {\bibinfo {volume} {82}},\ \bibinfo
  {pages} {2731} (\bibinfo {year} {2010})}\BibitemShut {NoStop}%
\bibitem [{\citenamefont {Walowski}\ and\ \citenamefont
  {M\"{u}nzenberg}(2016)}]{Walowski_2016_JApplPhys_120_14}%
  \BibitemOpen
  \bibfield  {author} {\bibinfo {author} {\bibfnamefont {J.}~\bibnamefont
  {Walowski}}\ and\ \bibinfo {author} {\bibfnamefont {M.}~\bibnamefont
  {M\"{u}nzenberg}},\ }\href {\doibase 10.1063/1.4958846} {\bibfield  {journal}
  {\bibinfo  {journal} {J. Appl. Phys.}\ }\textbf {\bibinfo {volume} {120}},\
  \bibinfo {pages} {140901} (\bibinfo {year} {2016})}\BibitemShut {NoStop}%
\bibitem [{\citenamefont {Cornelissen}\ \emph {et~al.}(2015)\citenamefont
  {Cornelissen}, \citenamefont {Liu}, \citenamefont {Duine}, \citenamefont
  {Youssef},\ and\ \citenamefont {van Wees}}]{Cornelissen_2015_NatPhys_11_12}%
  \BibitemOpen
  \bibfield  {author} {\bibinfo {author} {\bibfnamefont {L.~J.}\ \bibnamefont
  {Cornelissen}}, \bibinfo {author} {\bibfnamefont {J.}~\bibnamefont {Liu}},
  \bibinfo {author} {\bibfnamefont {R.~A.}\ \bibnamefont {Duine}}, \bibinfo
  {author} {\bibfnamefont {J.~B.}\ \bibnamefont {Youssef}}, \ and\ \bibinfo
  {author} {\bibfnamefont {B.~J.}\ \bibnamefont {van Wees}},\ }\href {\doibase
  10.1038/nphys3465} {\bibfield  {journal} {\bibinfo  {journal} {Nat. Phys.}\
  }\textbf {\bibinfo {volume} {11}},\ \bibinfo {pages} {1022} (\bibinfo {year}
  {2015})}\BibitemShut {NoStop}%
\bibitem [{\citenamefont {Nowak}(2001)}]{Nowak_Book_ASD}%
  \BibitemOpen
  \bibfield  {author} {\bibinfo {author} {\bibfnamefont {U.}~\bibnamefont
  {Nowak}},\ }\enquote {\bibinfo {title} {Thermally activated reversal in
  magnetic nanostructures},}\ in\ \href {\doibase
  https://doi.org/10.1142/9789812811578_0002} {\emph {\bibinfo {booktitle}
  {Annual Reviews of Computational Physics IX}}}\ (\bibinfo  {publisher} {World
  Scientific},\ \bibinfo {year} {2001})\ pp.\ \bibinfo {pages}
  {105--151}\BibitemShut {NoStop}%
\bibitem [{\citenamefont {Nowak}\ \emph {et~al.}(2005)\citenamefont {Nowak},
  \citenamefont {Mryasov}, \citenamefont {Wieser}, \citenamefont {Guslienko},\
  and\ \citenamefont {Chantrell}}]{Nowak_2005_PhysRevB_72_17}%
  \BibitemOpen
  \bibfield  {author} {\bibinfo {author} {\bibfnamefont {U.}~\bibnamefont
  {Nowak}}, \bibinfo {author} {\bibfnamefont {O.~N.}\ \bibnamefont {Mryasov}},
  \bibinfo {author} {\bibfnamefont {R.}~\bibnamefont {Wieser}}, \bibinfo
  {author} {\bibfnamefont {K.}~\bibnamefont {Guslienko}}, \ and\ \bibinfo
  {author} {\bibfnamefont {R.~W.}\ \bibnamefont {Chantrell}},\ }\href {\doibase
  10.1103/physrevb.72.172410} {\bibfield  {journal} {\bibinfo  {journal} {Phys.
  Rev. B}\ }\textbf {\bibinfo {volume} {72}},\ \bibinfo {pages} {172410}
  (\bibinfo {year} {2005})}\BibitemShut {NoStop}%
\bibitem [{\citenamefont {Skubic}\ \emph {et~al.}(2008)\citenamefont {Skubic},
  \citenamefont {Hellsvik}, \citenamefont {Nordstr\"{o}m},\ and\ \citenamefont
  {Eriksson}}]{Skubic_2008_JPhysCondensMatter_20_31}%
  \BibitemOpen
  \bibfield  {author} {\bibinfo {author} {\bibfnamefont {B.}~\bibnamefont
  {Skubic}}, \bibinfo {author} {\bibfnamefont {J.}~\bibnamefont {Hellsvik}},
  \bibinfo {author} {\bibfnamefont {L.}~\bibnamefont {Nordstr\"{o}m}}, \ and\
  \bibinfo {author} {\bibfnamefont {O.}~\bibnamefont {Eriksson}},\ }\href
  {\doibase 10.1088/0953-8984/20/31/315203} {\bibfield  {journal} {\bibinfo
  {journal} {J. Phys.: Condens. Matter}\ }\textbf {\bibinfo {volume} {20}},\
  \bibinfo {pages} {315203} (\bibinfo {year} {2008})}\BibitemShut {NoStop}%
\bibitem [{\citenamefont {Ma}\ and\ \citenamefont
  {Dudarev}(2011)}]{Ma_2011_PhysRevB_83_13}%
  \BibitemOpen
  \bibfield  {author} {\bibinfo {author} {\bibfnamefont {P.-W.}\ \bibnamefont
  {Ma}}\ and\ \bibinfo {author} {\bibfnamefont {S.~L.}\ \bibnamefont
  {Dudarev}},\ }\href {\doibase 10.1103/physrevb.83.134418} {\bibfield
  {journal} {\bibinfo  {journal} {Phys. Rev. B}\ }\textbf {\bibinfo {volume}
  {83}},\ \bibinfo {pages} {134418} (\bibinfo {year} {2011})}\BibitemShut
  {NoStop}%
\bibitem [{\citenamefont {Evans}\ \emph {et~al.}(2014)\citenamefont {Evans},
  \citenamefont {Fan}, \citenamefont {Chureemart}, \citenamefont {Ostler},
  \citenamefont {Ellis},\ and\ \citenamefont
  {Chantrell}}]{Evans_2014_JPhysCondensMatter_26_10}%
  \BibitemOpen
  \bibfield  {author} {\bibinfo {author} {\bibfnamefont {R.~F.~L.}\
  \bibnamefont {Evans}}, \bibinfo {author} {\bibfnamefont {W.~J.}\ \bibnamefont
  {Fan}}, \bibinfo {author} {\bibfnamefont {P.}~\bibnamefont {Chureemart}},
  \bibinfo {author} {\bibfnamefont {T.~A.}\ \bibnamefont {Ostler}}, \bibinfo
  {author} {\bibfnamefont {M.~O.~A.}\ \bibnamefont {Ellis}}, \ and\ \bibinfo
  {author} {\bibfnamefont {R.~W.}\ \bibnamefont {Chantrell}},\ }\href {\doibase
  10.1088/0953-8984/26/10/103202} {\bibfield  {journal} {\bibinfo  {journal}
  {J. Phys.: Condens. Matter}\ }\textbf {\bibinfo {volume} {26}},\ \bibinfo
  {pages} {103202} (\bibinfo {year} {2014})}\BibitemShut {NoStop}%
\bibitem [{\citenamefont {Garanin}(1997)}]{Garanin_1997_PhysRevB_55_5}%
  \BibitemOpen
  \bibfield  {author} {\bibinfo {author} {\bibfnamefont {D.~A.}\ \bibnamefont
  {Garanin}},\ }\href {\doibase 10.1103/physrevb.55.3050} {\bibfield  {journal}
  {\bibinfo  {journal} {Phys. Rev. B}\ }\textbf {\bibinfo {volume} {55}},\
  \bibinfo {pages} {3050} (\bibinfo {year} {1997})}\BibitemShut {NoStop}%
\bibitem [{\citenamefont {Garanin}\ and\ \citenamefont
  {Chubykalo-Fesenko}(2004)}]{Garanin_2004_PhysRevB_70_21}%
  \BibitemOpen
  \bibfield  {author} {\bibinfo {author} {\bibfnamefont {D.~A.}\ \bibnamefont
  {Garanin}}\ and\ \bibinfo {author} {\bibfnamefont {O.}~\bibnamefont
  {Chubykalo-Fesenko}},\ }\href {\doibase 10.1103/physrevb.70.212409}
  {\bibfield  {journal} {\bibinfo  {journal} {Phys. Rev. B}\ }\textbf {\bibinfo
  {volume} {70}},\ \bibinfo {pages} {212409} (\bibinfo {year}
  {2004})}\BibitemShut {NoStop}%
\bibitem [{\citenamefont {Chubykalo-Fesenko}\ \emph {et~al.}(2006)\citenamefont
  {Chubykalo-Fesenko}, \citenamefont {Nowak}, \citenamefont {Chantrell},\ and\
  \citenamefont {Garanin}}]{Chubykalo-Fesenko_2006_PhysRevB_74_9}%
  \BibitemOpen
  \bibfield  {author} {\bibinfo {author} {\bibfnamefont {O.}~\bibnamefont
  {Chubykalo-Fesenko}}, \bibinfo {author} {\bibfnamefont {U.}~\bibnamefont
  {Nowak}}, \bibinfo {author} {\bibfnamefont {R.~W.}\ \bibnamefont
  {Chantrell}}, \ and\ \bibinfo {author} {\bibfnamefont {D.}~\bibnamefont
  {Garanin}},\ }\href {\doibase 10.1103/physrevb.74.094436} {\bibfield
  {journal} {\bibinfo  {journal} {Phys. Rev. B}\ }\textbf {\bibinfo {volume}
  {74}},\ \bibinfo {pages} {094436} (\bibinfo {year} {2006})}\BibitemShut
  {NoStop}%
\bibitem [{\citenamefont {Evans}\ \emph {et~al.}(2012)\citenamefont {Evans},
  \citenamefont {Hinzke}, \citenamefont {Atxitia}, \citenamefont {Nowak},
  \citenamefont {Chantrell},\ and\ \citenamefont
  {Chubykalo-Fesenko}}]{Evans_2012_PhysRevB_85_1}%
  \BibitemOpen
  \bibfield  {author} {\bibinfo {author} {\bibfnamefont {R.~F.~L.}\
  \bibnamefont {Evans}}, \bibinfo {author} {\bibfnamefont {D.}~\bibnamefont
  {Hinzke}}, \bibinfo {author} {\bibfnamefont {U.}~\bibnamefont {Atxitia}},
  \bibinfo {author} {\bibfnamefont {U.}~\bibnamefont {Nowak}}, \bibinfo
  {author} {\bibfnamefont {R.~W.}\ \bibnamefont {Chantrell}}, \ and\ \bibinfo
  {author} {\bibfnamefont {O.}~\bibnamefont {Chubykalo-Fesenko}},\ }\href
  {\doibase 10.1103/physrevb.85.014433} {\bibfield  {journal} {\bibinfo
  {journal} {Phys. Rev. B}\ }\textbf {\bibinfo {volume} {85}},\ \bibinfo
  {pages} {014433} (\bibinfo {year} {2012})}\BibitemShut {NoStop}%
\bibitem [{\citenamefont {Atxitia}\ \emph {et~al.}(2012)\citenamefont
  {Atxitia}, \citenamefont {Nieves},\ and\ \citenamefont
  {Chubykalo-Fesenko}}]{Atxitia_2012_PhysRevB_86_10}%
  \BibitemOpen
  \bibfield  {author} {\bibinfo {author} {\bibfnamefont {U.}~\bibnamefont
  {Atxitia}}, \bibinfo {author} {\bibfnamefont {P.}~\bibnamefont {Nieves}}, \
  and\ \bibinfo {author} {\bibfnamefont {O.}~\bibnamefont
  {Chubykalo-Fesenko}},\ }\href {\doibase 10.1103/physrevb.86.104414}
  {\bibfield  {journal} {\bibinfo  {journal} {Phys. Rev. B}\ }\textbf {\bibinfo
  {volume} {86}},\ \bibinfo {pages} {104414} (\bibinfo {year}
  {2012})}\BibitemShut {NoStop}%
\bibitem [{\citenamefont {Atxitia}\ \emph {et~al.}(2016)\citenamefont
  {Atxitia}, \citenamefont {Hinzke},\ and\ \citenamefont
  {Nowak}}]{Atxitia_2016_JPhysDApplPhys_50_3}%
  \BibitemOpen
  \bibfield  {author} {\bibinfo {author} {\bibfnamefont {U.}~\bibnamefont
  {Atxitia}}, \bibinfo {author} {\bibfnamefont {D.}~\bibnamefont {Hinzke}}, \
  and\ \bibinfo {author} {\bibfnamefont {U.}~\bibnamefont {Nowak}},\ }\href
  {\doibase 10.1088/1361-6463/50/3/033003} {\bibfield  {journal} {\bibinfo
  {journal} {J. Phys. D: Appl. Phys.}\ }\textbf {\bibinfo {volume} {50}},\
  \bibinfo {pages} {033003} (\bibinfo {year} {2016})}\BibitemShut {NoStop}%
\bibitem [{\citenamefont {Woo}\ \emph {et~al.}(2018)\citenamefont {Woo},
  \citenamefont {Song}, \citenamefont {Zhang}, \citenamefont {Zhou},
  \citenamefont {Ezawa}, \citenamefont {Liu}, \citenamefont {Finizio},
  \citenamefont {Raabe}, \citenamefont {Lee}, \citenamefont {Kim},
  \citenamefont {Park}, \citenamefont {Kim}, \citenamefont {Kim}, \citenamefont
  {Lee}, \citenamefont {Lee}, \citenamefont {Choi}, \citenamefont {Min},
  \citenamefont {Koo},\ and\ \citenamefont {Chang}}]{Woo_2018_NatCommun_9_1}%
  \BibitemOpen
  \bibfield  {author} {\bibinfo {author} {\bibfnamefont {S.}~\bibnamefont
  {Woo}}, \bibinfo {author} {\bibfnamefont {K.~M.}\ \bibnamefont {Song}},
  \bibinfo {author} {\bibfnamefont {X.}~\bibnamefont {Zhang}}, \bibinfo
  {author} {\bibfnamefont {Y.}~\bibnamefont {Zhou}}, \bibinfo {author}
  {\bibfnamefont {M.}~\bibnamefont {Ezawa}}, \bibinfo {author} {\bibfnamefont
  {X.}~\bibnamefont {Liu}}, \bibinfo {author} {\bibfnamefont {S.}~\bibnamefont
  {Finizio}}, \bibinfo {author} {\bibfnamefont {J.}~\bibnamefont {Raabe}},
  \bibinfo {author} {\bibfnamefont {N.~J.}\ \bibnamefont {Lee}}, \bibinfo
  {author} {\bibfnamefont {S.-I.}\ \bibnamefont {Kim}}, \bibinfo {author}
  {\bibfnamefont {S.-Y.}\ \bibnamefont {Park}}, \bibinfo {author}
  {\bibfnamefont {Y.}~\bibnamefont {Kim}}, \bibinfo {author} {\bibfnamefont
  {J.-Y.}\ \bibnamefont {Kim}}, \bibinfo {author} {\bibfnamefont
  {D.}~\bibnamefont {Lee}}, \bibinfo {author} {\bibfnamefont {O.}~\bibnamefont
  {Lee}}, \bibinfo {author} {\bibfnamefont {J.~W.}\ \bibnamefont {Choi}},
  \bibinfo {author} {\bibfnamefont {B.-C.}\ \bibnamefont {Min}}, \bibinfo
  {author} {\bibfnamefont {H.~C.}\ \bibnamefont {Koo}}, \ and\ \bibinfo
  {author} {\bibfnamefont {J.}~\bibnamefont {Chang}},\ }\href {\doibase
  10.1038/s41467-018-03378-7} {\bibfield  {journal} {\bibinfo  {journal} {Nat.
  Commun.}\ }\textbf {\bibinfo {volume} {9}},\ \bibinfo {pages} {959} (\bibinfo
  {year} {2018})}\BibitemShut {NoStop}%
\bibitem [{\citenamefont {Caretta}\ \emph {et~al.}(2018)\citenamefont
  {Caretta}, \citenamefont {Mann}, \citenamefont {B\"{u}ttner}, \citenamefont
  {Ueda}, \citenamefont {Pfau}, \citenamefont {G\"{u}nther}, \citenamefont
  {Hessing}, \citenamefont {Churikova}, \citenamefont {Klose}, \citenamefont
  {Schneider}, \citenamefont {Engel}, \citenamefont {Marcus}, \citenamefont
  {Bono}, \citenamefont {Bagschik}, \citenamefont {Eisebitt},\ and\
  \citenamefont {Beach}}]{Caretta_2018_NatNanotechnol_13_12}%
  \BibitemOpen
  \bibfield  {author} {\bibinfo {author} {\bibfnamefont {L.}~\bibnamefont
  {Caretta}}, \bibinfo {author} {\bibfnamefont {M.}~\bibnamefont {Mann}},
  \bibinfo {author} {\bibfnamefont {F.}~\bibnamefont {B\"{u}ttner}}, \bibinfo
  {author} {\bibfnamefont {K.}~\bibnamefont {Ueda}}, \bibinfo {author}
  {\bibfnamefont {B.}~\bibnamefont {Pfau}}, \bibinfo {author} {\bibfnamefont
  {C.~M.}\ \bibnamefont {G\"{u}nther}}, \bibinfo {author} {\bibfnamefont
  {P.}~\bibnamefont {Hessing}}, \bibinfo {author} {\bibfnamefont
  {A.}~\bibnamefont {Churikova}}, \bibinfo {author} {\bibfnamefont
  {C.}~\bibnamefont {Klose}}, \bibinfo {author} {\bibfnamefont
  {M.}~\bibnamefont {Schneider}}, \bibinfo {author} {\bibfnamefont
  {D.}~\bibnamefont {Engel}}, \bibinfo {author} {\bibfnamefont
  {C.}~\bibnamefont {Marcus}}, \bibinfo {author} {\bibfnamefont
  {D.}~\bibnamefont {Bono}}, \bibinfo {author} {\bibfnamefont {K.}~\bibnamefont
  {Bagschik}}, \bibinfo {author} {\bibfnamefont {S.}~\bibnamefont {Eisebitt}},
  \ and\ \bibinfo {author} {\bibfnamefont {G.~S.~D.}\ \bibnamefont {Beach}},\
  }\href {\doibase 10.1038/s41565-018-0255-3} {\bibfield  {journal} {\bibinfo
  {journal} {Nat. Nanotechnol.}\ }\textbf {\bibinfo {volume} {13}},\ \bibinfo
  {pages} {1154} (\bibinfo {year} {2018})}\BibitemShut {NoStop}%
\bibitem [{\citenamefont {Liechtenstein}\ \emph {et~al.}(1984)\citenamefont
  {Liechtenstein}, \citenamefont {Katsnelson},\ and\ \citenamefont
  {Gubanov}}]{Liechtenstein_1984_JPhysFMetalPhys_14_7}%
  \BibitemOpen
  \bibfield  {author} {\bibinfo {author} {\bibfnamefont {A.~I.}\ \bibnamefont
  {Liechtenstein}}, \bibinfo {author} {\bibfnamefont {M.~I.}\ \bibnamefont
  {Katsnelson}}, \ and\ \bibinfo {author} {\bibfnamefont {V.~A.}\ \bibnamefont
  {Gubanov}},\ }\href {\doibase 10.1088/0305-4608/14/7/007} {\bibfield
  {journal} {\bibinfo  {journal} {J. Phys. F: Metal Phys.}\ }\textbf {\bibinfo
  {volume} {14}},\ \bibinfo {pages} {L125} (\bibinfo {year}
  {1984})}\BibitemShut {NoStop}%
\bibitem [{\citenamefont {Nowak}(2007)}]{Nowak_Book}%
  \BibitemOpen
  \bibfield  {author} {\bibinfo {author} {\bibfnamefont {U.}~\bibnamefont
  {Nowak}},\ }\enquote {\bibinfo {title} {Classical spin models},}\ in\ \href
  {\doibase https://doi.org/10.1002/9780470022184.hmm205} {\emph {\bibinfo
  {booktitle} {Handbook of Magnetism and Advanced Magnetic Materials}}}\
  (\bibinfo  {publisher} {Wiley},\ \bibinfo {year} {2007})\ \Eprint
  {http://arxiv.org/abs/https://onlinelibrary.wiley.com/doi/pdf/10.1002/9780470022184.hmm205}
  {https://onlinelibrary.wiley.com/doi/pdf/10.1002/9780470022184.hmm205}
  \BibitemShut {NoStop}%
\bibitem [{\citenamefont {Brown}(1963)}]{Brown_1963_PhysRev_130_5}%
  \BibitemOpen
  \bibfield  {author} {\bibinfo {author} {\bibfnamefont {W.~F.}\ \bibnamefont
  {Brown}},\ }\href {\doibase 10.1103/physrev.130.1677} {\bibfield  {journal}
  {\bibinfo  {journal} {Phys. Rev.}\ }\textbf {\bibinfo {volume} {130}},\
  \bibinfo {pages} {1677} (\bibinfo {year} {1963})}\BibitemShut {NoStop}%
\bibitem [{\citenamefont {Atxitia}\ \emph {et~al.}(2009)\citenamefont
  {Atxitia}, \citenamefont {Chubykalo-Fesenko}, \citenamefont {Chantrell},
  \citenamefont {Nowak},\ and\ \citenamefont
  {Rebei}}]{Atxitia_2009_PhysRevLett_102_5}%
  \BibitemOpen
  \bibfield  {author} {\bibinfo {author} {\bibfnamefont {U.}~\bibnamefont
  {Atxitia}}, \bibinfo {author} {\bibfnamefont {O.}~\bibnamefont
  {Chubykalo-Fesenko}}, \bibinfo {author} {\bibfnamefont {R.~W.}\ \bibnamefont
  {Chantrell}}, \bibinfo {author} {\bibfnamefont {U.}~\bibnamefont {Nowak}}, \
  and\ \bibinfo {author} {\bibfnamefont {A.}~\bibnamefont {Rebei}},\ }\href
  {\doibase 10.1103/physrevlett.102.057203} {\bibfield  {journal} {\bibinfo
  {journal} {Phys. Rev. Lett.}\ }\textbf {\bibinfo {volume} {102}},\ \bibinfo
  {pages} {057203} (\bibinfo {year} {2009})}\BibitemShut {NoStop}%
\bibitem [{\citenamefont {Barker}\ and\ \citenamefont
  {Bauer}(2019)}]{Barker_2019_PhysRevB_100_14}%
  \BibitemOpen
  \bibfield  {author} {\bibinfo {author} {\bibfnamefont {J.}~\bibnamefont
  {Barker}}\ and\ \bibinfo {author} {\bibfnamefont {G.~E.~W.}\ \bibnamefont
  {Bauer}},\ }\href {\doibase 10.1103/physrevb.100.140401} {\bibfield
  {journal} {\bibinfo  {journal} {Phys. Rev. B}\ }\textbf {\bibinfo {volume}
  {100}},\ \bibinfo {pages} {140401} (\bibinfo {year} {2019})}\BibitemShut
  {NoStop}%
\bibitem [{\citenamefont {Anders}\ \emph {et~al.}(2020)\citenamefont {Anders},
  \citenamefont {Sait},\ and\ \citenamefont
  {Horsley}}]{Anders_2020_arXiv_2009_00600}%
  \BibitemOpen
  \bibfield  {author} {\bibinfo {author} {\bibfnamefont {J.}~\bibnamefont
  {Anders}}, \bibinfo {author} {\bibfnamefont {C.~R.~J.}\ \bibnamefont {Sait}},
  \ and\ \bibinfo {author} {\bibfnamefont {S.~A.~R.}\ \bibnamefont {Horsley}},\
  }\href {https://arxiv.org/abs/2009.00600} {\bibfield  {journal} {\bibinfo
  {journal} {arXiv:2009.00600}\ } (\bibinfo {year} {2020})}\BibitemShut
  {NoStop}%
\bibitem [{\citenamefont {Ito}\ \emph {et~al.}(2019)\citenamefont {Ito},
  \citenamefont {Kikkawa}, \citenamefont {Barker}, \citenamefont {Hirobe},
  \citenamefont {Shiomi},\ and\ \citenamefont
  {Saitoh}}]{Ito_2019_PhysRevB_100_6}%
  \BibitemOpen
  \bibfield  {author} {\bibinfo {author} {\bibfnamefont {N.}~\bibnamefont
  {Ito}}, \bibinfo {author} {\bibfnamefont {T.}~\bibnamefont {Kikkawa}},
  \bibinfo {author} {\bibfnamefont {J.}~\bibnamefont {Barker}}, \bibinfo
  {author} {\bibfnamefont {D.}~\bibnamefont {Hirobe}}, \bibinfo {author}
  {\bibfnamefont {Y.}~\bibnamefont {Shiomi}}, \ and\ \bibinfo {author}
  {\bibfnamefont {E.}~\bibnamefont {Saitoh}},\ }\href {\doibase
  10.1103/physrevb.100.060402} {\bibfield  {journal} {\bibinfo  {journal}
  {Phys. Rev. B}\ }\textbf {\bibinfo {volume} {100}},\ \bibinfo {pages}
  {060402} (\bibinfo {year} {2019})}\BibitemShut {NoStop}%
\bibitem [{\citenamefont {Bergman}\ \emph {et~al.}(2010)\citenamefont
  {Bergman}, \citenamefont {Taroni}, \citenamefont {Bergqvist}, \citenamefont
  {Hellsvik}, \citenamefont {Hj\"{o}rvarsson},\ and\ \citenamefont
  {Eriksson}}]{Bergman_2010_PhysRevB_81_14}%
  \BibitemOpen
  \bibfield  {author} {\bibinfo {author} {\bibfnamefont {A.}~\bibnamefont
  {Bergman}}, \bibinfo {author} {\bibfnamefont {A.}~\bibnamefont {Taroni}},
  \bibinfo {author} {\bibfnamefont {L.}~\bibnamefont {Bergqvist}}, \bibinfo
  {author} {\bibfnamefont {J.}~\bibnamefont {Hellsvik}}, \bibinfo {author}
  {\bibfnamefont {B.}~\bibnamefont {Hj\"{o}rvarsson}}, \ and\ \bibinfo {author}
  {\bibfnamefont {O.}~\bibnamefont {Eriksson}},\ }\href {\doibase
  10.1103/physrevb.81.144416} {\bibfield  {journal} {\bibinfo  {journal} {Phys.
  Rev. B}\ }\textbf {\bibinfo {volume} {81}},\ \bibinfo {pages} {144416}
  (\bibinfo {year} {2010})}\BibitemShut {NoStop}%
\bibitem [{\citenamefont {Ostler}\ \emph {et~al.}(2012)\citenamefont {Ostler},
  \citenamefont {Barker}, \citenamefont {Evans}, \citenamefont {Chantrell},
  \citenamefont {Atxitia}, \citenamefont {Chubykalo-Fesenko}, \citenamefont
  {El~Moussaoui}, \citenamefont {Le~Guyader}, \citenamefont {Mengotti},
  \citenamefont {Heyderman}, \citenamefont {Nolting}, \citenamefont
  {Tsukamoto}, \citenamefont {Itoh}, \citenamefont {Afanasiev}, \citenamefont
  {Ivanov}, \citenamefont {Kalashnikova}, \citenamefont {Vahaplar},
  \citenamefont {Mentink}, \citenamefont {Kirilyuk}, \citenamefont {Rasing},\
  and\ \citenamefont {Kimel}}]{Ostler_2012_NatCommun_3_1}%
  \BibitemOpen
  \bibfield  {author} {\bibinfo {author} {\bibfnamefont {T.}~\bibnamefont
  {Ostler}}, \bibinfo {author} {\bibfnamefont {J.}~\bibnamefont {Barker}},
  \bibinfo {author} {\bibfnamefont {R.}~\bibnamefont {Evans}}, \bibinfo
  {author} {\bibfnamefont {R.}~\bibnamefont {Chantrell}}, \bibinfo {author}
  {\bibfnamefont {U.}~\bibnamefont {Atxitia}}, \bibinfo {author} {\bibfnamefont
  {O.}~\bibnamefont {Chubykalo-Fesenko}}, \bibinfo {author} {\bibfnamefont
  {S.}~\bibnamefont {El~Moussaoui}}, \bibinfo {author} {\bibfnamefont
  {L.}~\bibnamefont {Le~Guyader}}, \bibinfo {author} {\bibfnamefont
  {E.}~\bibnamefont {Mengotti}}, \bibinfo {author} {\bibfnamefont
  {L.}~\bibnamefont {Heyderman}}, \bibinfo {author} {\bibfnamefont
  {F.}~\bibnamefont {Nolting}}, \bibinfo {author} {\bibfnamefont
  {A.}~\bibnamefont {Tsukamoto}}, \bibinfo {author} {\bibfnamefont
  {A.}~\bibnamefont {Itoh}}, \bibinfo {author} {\bibfnamefont {D.}~\bibnamefont
  {Afanasiev}}, \bibinfo {author} {\bibfnamefont {B.}~\bibnamefont {Ivanov}},
  \bibinfo {author} {\bibfnamefont {A.}~\bibnamefont {Kalashnikova}}, \bibinfo
  {author} {\bibfnamefont {K.}~\bibnamefont {Vahaplar}}, \bibinfo {author}
  {\bibfnamefont {J.}~\bibnamefont {Mentink}}, \bibinfo {author} {\bibfnamefont
  {A.}~\bibnamefont {Kirilyuk}}, \bibinfo {author} {\bibfnamefont
  {T.}~\bibnamefont {Rasing}}, \ and\ \bibinfo {author} {\bibfnamefont
  {A.}~\bibnamefont {Kimel}},\ }\href {\doibase 10.1038/ncomms1666} {\bibfield
  {journal} {\bibinfo  {journal} {Nat. Commun.}\ }\textbf {\bibinfo {volume}
  {3}},\ \bibinfo {pages} {666} (\bibinfo {year} {2012})}\BibitemShut {NoStop}%
\bibitem [{\citenamefont {Ma}\ \emph {et~al.}(2012)\citenamefont {Ma},
  \citenamefont {Dudarev},\ and\ \citenamefont {Woo}}]{Ma_2012_PhysRevB_85_18}%
  \BibitemOpen
  \bibfield  {author} {\bibinfo {author} {\bibfnamefont {P.-W.}\ \bibnamefont
  {Ma}}, \bibinfo {author} {\bibfnamefont {S.~L.}\ \bibnamefont {Dudarev}}, \
  and\ \bibinfo {author} {\bibfnamefont {C.~H.}\ \bibnamefont {Woo}},\ }\href
  {\doibase 10.1103/physrevb.85.184301} {\bibfield  {journal} {\bibinfo
  {journal} {Phys. Rev. B}\ }\textbf {\bibinfo {volume} {85}},\ \bibinfo
  {pages} {184301} (\bibinfo {year} {2012})}\BibitemShut {NoStop}%
\bibitem [{\citenamefont {R\"{u}ckriegel}\ \emph {et~al.}(2014)\citenamefont
  {R\"{u}ckriegel}, \citenamefont {Kopietz}, \citenamefont {Bozhko},
  \citenamefont {Serga},\ and\ \citenamefont
  {Hillebrands}}]{Ruckriegel_2014_PhysRevB_89_18}%
  \BibitemOpen
  \bibfield  {author} {\bibinfo {author} {\bibfnamefont {A.}~\bibnamefont
  {R\"{u}ckriegel}}, \bibinfo {author} {\bibfnamefont {P.}~\bibnamefont
  {Kopietz}}, \bibinfo {author} {\bibfnamefont {D.~A.}\ \bibnamefont {Bozhko}},
  \bibinfo {author} {\bibfnamefont {A.~A.}\ \bibnamefont {Serga}}, \ and\
  \bibinfo {author} {\bibfnamefont {B.}~\bibnamefont {Hillebrands}},\ }\href
  {\doibase 10.1103/physrevb.89.184413} {\bibfield  {journal} {\bibinfo
  {journal} {Phys. Rev. B}\ }\textbf {\bibinfo {volume} {89}},\ \bibinfo
  {pages} {184413} (\bibinfo {year} {2014})}\BibitemShut {NoStop}%
\bibitem [{\citenamefont {Gurevich}\ and\ \citenamefont
  {Melkov}(1996)}]{Gurevich_Book}%
  \BibitemOpen
  \bibfield  {author} {\bibinfo {author} {\bibfnamefont {A.~G.}\ \bibnamefont
  {Gurevich}}\ and\ \bibinfo {author} {\bibfnamefont {G.~A.}\ \bibnamefont
  {Melkov}},\ }\href@noop {} {\emph {\bibinfo {title} {{M}agnetization
  {O}scillations and {W}aves}}}\ (\bibinfo  {publisher} {CRC Press},\ \bibinfo
  {year} {1996})\BibitemShut {NoStop}%
\bibitem [{\citenamefont {Wangsness}(1953)}]{Wangsness_1953_PhysRev_91_5}%
  \BibitemOpen
  \bibfield  {author} {\bibinfo {author} {\bibfnamefont {R.~K.}\ \bibnamefont
  {Wangsness}},\ }\href {\doibase 10.1103/physrev.91.1085} {\bibfield
  {journal} {\bibinfo  {journal} {Phys. Rev.}\ }\textbf {\bibinfo {volume}
  {91}},\ \bibinfo {pages} {1085} (\bibinfo {year} {1953})}\BibitemShut
  {NoStop}%
\bibitem [{oom()}]{oommf}%
  \BibitemOpen
  \href@noop {} {\enquote {\bibinfo {title} {Oommf},}\ }\bibinfo {howpublished}
  {\url{https://math.nist.gov/oommf/}}\BibitemShut {NoStop}%
\bibitem [{\citenamefont {Vansteenkiste}\ \emph {et~al.}(2014)\citenamefont
  {Vansteenkiste}, \citenamefont {Leliaert}, \citenamefont {Dvornik},
  \citenamefont {Helsen}, \citenamefont {Garcia-Sanchez},\ and\ \citenamefont
  {Van~Waeyenberge}}]{Vansteenkiste_2014_AIPAdv_4_10}%
  \BibitemOpen
  \bibfield  {author} {\bibinfo {author} {\bibfnamefont {A.}~\bibnamefont
  {Vansteenkiste}}, \bibinfo {author} {\bibfnamefont {J.}~\bibnamefont
  {Leliaert}}, \bibinfo {author} {\bibfnamefont {M.}~\bibnamefont {Dvornik}},
  \bibinfo {author} {\bibfnamefont {M.}~\bibnamefont {Helsen}}, \bibinfo
  {author} {\bibfnamefont {F.}~\bibnamefont {Garcia-Sanchez}}, \ and\ \bibinfo
  {author} {\bibfnamefont {B.}~\bibnamefont {Van~Waeyenberge}},\ }\href
  {\doibase 10.1063/1.4899186} {\bibfield  {journal} {\bibinfo  {journal} {AIP
  Adv.}\ }\textbf {\bibinfo {volume} {4}},\ \bibinfo {pages} {107133} (\bibinfo
  {year} {2014})}\BibitemShut {NoStop}%
\bibitem [{\citenamefont {Bisotti}\ \emph {et~al.}(2018)\citenamefont
  {Bisotti}, \citenamefont {Cort\'{e}s-Ortu\~{n}o}, \citenamefont {Pepper},
  \citenamefont {Wang}, \citenamefont {Beg}, \citenamefont {Kluyver},\ and\
  \citenamefont {Fangohr}}]{Bisotti_2018_JOpenResSoftw_6}%
  \BibitemOpen
  \bibfield  {author} {\bibinfo {author} {\bibfnamefont {M.-A.}\ \bibnamefont
  {Bisotti}}, \bibinfo {author} {\bibfnamefont {D.}~\bibnamefont
  {Cort\'{e}s-Ortu\~{n}o}}, \bibinfo {author} {\bibfnamefont {R.}~\bibnamefont
  {Pepper}}, \bibinfo {author} {\bibfnamefont {W.}~\bibnamefont {Wang}},
  \bibinfo {author} {\bibfnamefont {M.}~\bibnamefont {Beg}}, \bibinfo {author}
  {\bibfnamefont {T.}~\bibnamefont {Kluyver}}, \ and\ \bibinfo {author}
  {\bibfnamefont {H.}~\bibnamefont {Fangohr}},\ }\href {\doibase
  10.5334/jors.223} {\bibfield  {journal} {\bibinfo  {journal} {J. Open Res.
  Softw.}\ }\textbf {\bibinfo {volume} {6}} (\bibinfo {year} {2018}),\
  10.5334/jors.223}\BibitemShut {NoStop}%
\bibitem [{\citenamefont {Lepadatu}(2020)}]{Lepadatu_2020_JApplPhys_128_24}%
  \BibitemOpen
  \bibfield  {author} {\bibinfo {author} {\bibfnamefont {S.}~\bibnamefont
  {Lepadatu}},\ }\href {\doibase 10.1063/5.0024382} {\bibfield  {journal}
  {\bibinfo  {journal} {J. Appl. Phys.}\ }\textbf {\bibinfo {volume} {128}},\
  \bibinfo {pages} {243902} (\bibinfo {year} {2020})}\BibitemShut {NoStop}%
\bibitem [{\citenamefont {Oezelt}\ \emph {et~al.}(2015)\citenamefont {Oezelt},
  \citenamefont {Kovacs}, \citenamefont {Reichel}, \citenamefont {Fischbacher},
  \citenamefont {Bance}, \citenamefont {Gusenbauer}, \citenamefont {Schubert},
  \citenamefont {Albrecht},\ and\ \citenamefont
  {Schrefl}}]{Oezelt_2015_JMagnMagnMater_381}%
  \BibitemOpen
  \bibfield  {author} {\bibinfo {author} {\bibfnamefont {H.}~\bibnamefont
  {Oezelt}}, \bibinfo {author} {\bibfnamefont {A.}~\bibnamefont {Kovacs}},
  \bibinfo {author} {\bibfnamefont {F.}~\bibnamefont {Reichel}}, \bibinfo
  {author} {\bibfnamefont {J.}~\bibnamefont {Fischbacher}}, \bibinfo {author}
  {\bibfnamefont {S.}~\bibnamefont {Bance}}, \bibinfo {author} {\bibfnamefont
  {M.}~\bibnamefont {Gusenbauer}}, \bibinfo {author} {\bibfnamefont
  {C.}~\bibnamefont {Schubert}}, \bibinfo {author} {\bibfnamefont
  {M.}~\bibnamefont {Albrecht}}, \ and\ \bibinfo {author} {\bibfnamefont
  {T.}~\bibnamefont {Schrefl}},\ }\href {\doibase 10.1016/j.jmmm.2014.12.045}
  {\bibfield  {journal} {\bibinfo  {journal} {J. Magn. Magn. Mater.}\ }\textbf
  {\bibinfo {volume} {381}},\ \bibinfo {pages} {28} (\bibinfo {year}
  {2015})}\BibitemShut {NoStop}%
\bibitem [{\citenamefont {Mart\'{i}nez}\ \emph {et~al.}(2019)\citenamefont
  {Mart\'{i}nez}, \citenamefont {Raposo},\ and\ \citenamefont
  {Alejos}}]{Martinez_2019_JMagnMagnMater_491}%
  \BibitemOpen
  \bibfield  {author} {\bibinfo {author} {\bibfnamefont {E.}~\bibnamefont
  {Mart\'{i}nez}}, \bibinfo {author} {\bibfnamefont {V.}~\bibnamefont
  {Raposo}}, \ and\ \bibinfo {author} {\bibfnamefont {O.}~\bibnamefont
  {Alejos}},\ }\href {\doibase 10.1016/j.jmmm.2019.165545} {\bibfield
  {journal} {\bibinfo  {journal} {J. Magn. Magn. Mater.}\ }\textbf {\bibinfo
  {volume} {491}},\ \bibinfo {pages} {165545} (\bibinfo {year}
  {2019})}\BibitemShut {NoStop}%
\bibitem [{\citenamefont {Alejos}\ \emph {et~al.}(2018)\citenamefont {Alejos},
  \citenamefont {Raposo}, \citenamefont {Sanchez-Tejerina}, \citenamefont
  {Tomasello}, \citenamefont {Finocchio},\ and\ \citenamefont
  {Martinez}}]{Alejos_2018_JApplPhys_123_1}%
  \BibitemOpen
  \bibfield  {author} {\bibinfo {author} {\bibfnamefont {O.}~\bibnamefont
  {Alejos}}, \bibinfo {author} {\bibfnamefont {V.}~\bibnamefont {Raposo}},
  \bibinfo {author} {\bibfnamefont {L.}~\bibnamefont {Sanchez-Tejerina}},
  \bibinfo {author} {\bibfnamefont {R.}~\bibnamefont {Tomasello}}, \bibinfo
  {author} {\bibfnamefont {G.}~\bibnamefont {Finocchio}}, \ and\ \bibinfo
  {author} {\bibfnamefont {E.}~\bibnamefont {Martinez}},\ }\href {\doibase
  10.1063/1.5009739} {\bibfield  {journal} {\bibinfo  {journal} {J. Appl.
  Phys.}\ }\textbf {\bibinfo {volume} {123}},\ \bibinfo {pages} {013901}
  (\bibinfo {year} {2018})}\BibitemShut {NoStop}%
\bibitem [{\citenamefont {Mart\'{i}nez}\ \emph {et~al.}(2020)\citenamefont
  {Mart\'{i}nez}, \citenamefont {Raposo},\ and\ \citenamefont
  {Alejos}}]{Martinez_2020_AipAdv_10_1}%
  \BibitemOpen
  \bibfield  {author} {\bibinfo {author} {\bibfnamefont {E.}~\bibnamefont
  {Mart\'{i}nez}}, \bibinfo {author} {\bibfnamefont {V.}~\bibnamefont
  {Raposo}}, \ and\ \bibinfo {author} {\bibfnamefont {O.}~\bibnamefont
  {Alejos}},\ }\href {\doibase 10.1063/1.5130054} {\bibfield  {journal}
  {\bibinfo  {journal} {AIP Adv.}\ }\textbf {\bibinfo {volume} {10}},\ \bibinfo
  {pages} {015202} (\bibinfo {year} {2020})}\BibitemShut {NoStop}%
\bibitem [{\citenamefont {Caretta}\ \emph {et~al.}(2020)\citenamefont
  {Caretta}, \citenamefont {Oh}, \citenamefont {Fakhrul}, \citenamefont {Lee},
  \citenamefont {Lee}, \citenamefont {Kim}, \citenamefont {Ross}, \citenamefont
  {Lee},\ and\ \citenamefont {Beach}}]{Caretta_2020_Science_370_6523}%
  \BibitemOpen
  \bibfield  {author} {\bibinfo {author} {\bibfnamefont {L.}~\bibnamefont
  {Caretta}}, \bibinfo {author} {\bibfnamefont {S.~H.}\ \bibnamefont {Oh}},
  \bibinfo {author} {\bibfnamefont {T.}~\bibnamefont {Fakhrul}}, \bibinfo
  {author} {\bibfnamefont {D.~K.}\ \bibnamefont {Lee}}, \bibinfo {author}
  {\bibfnamefont {B.~H.}\ \bibnamefont {Lee}}, \bibinfo {author} {\bibfnamefont
  {S.~K.}\ \bibnamefont {Kim}}, \bibinfo {author} {\bibfnamefont {C.~A.}\
  \bibnamefont {Ross}}, \bibinfo {author} {\bibfnamefont {K.~J.}\ \bibnamefont
  {Lee}}, \ and\ \bibinfo {author} {\bibfnamefont {G.~S.}\ \bibnamefont
  {Beach}},\ }\href {\doibase 10.1126/science.aba5555} {\bibfield  {journal}
  {\bibinfo  {journal} {Science}\ }\textbf {\bibinfo {volume} {370}},\ \bibinfo
  {pages} {1438} (\bibinfo {year} {2020})}\BibitemShut {NoStop}%
\bibitem [{\citenamefont {Bastardis}\ \emph {et~al.}(2012)\citenamefont
  {Bastardis}, \citenamefont {Atxitia}, \citenamefont {Chubykalo-Fesenko},\
  and\ \citenamefont {Kachkachi}}]{Bastardis_2012_PhysRevB_86_9}%
  \BibitemOpen
  \bibfield  {author} {\bibinfo {author} {\bibfnamefont {R.}~\bibnamefont
  {Bastardis}}, \bibinfo {author} {\bibfnamefont {U.}~\bibnamefont {Atxitia}},
  \bibinfo {author} {\bibfnamefont {O.}~\bibnamefont {Chubykalo-Fesenko}}, \
  and\ \bibinfo {author} {\bibfnamefont {H.}~\bibnamefont {Kachkachi}},\ }\href
  {\doibase 10.1103/physrevb.86.094415} {\bibfield  {journal} {\bibinfo
  {journal} {Phys. Rev. B}\ }\textbf {\bibinfo {volume} {86}},\ \bibinfo
  {pages} {094415} (\bibinfo {year} {2012})}\BibitemShut {NoStop}%
\bibitem [{\citenamefont {Schlickeiser}\ \emph {et~al.}(2012)\citenamefont
  {Schlickeiser}, \citenamefont {Atxitia}, \citenamefont {Wienholdt},
  \citenamefont {Hinzke}, \citenamefont {Chubykalo-Fesenko},\ and\
  \citenamefont {Nowak}}]{Schlickeiser_2012_PhysRevB_86_21}%
  \BibitemOpen
  \bibfield  {author} {\bibinfo {author} {\bibfnamefont {F.}~\bibnamefont
  {Schlickeiser}}, \bibinfo {author} {\bibfnamefont {U.}~\bibnamefont
  {Atxitia}}, \bibinfo {author} {\bibfnamefont {S.}~\bibnamefont {Wienholdt}},
  \bibinfo {author} {\bibfnamefont {D.}~\bibnamefont {Hinzke}}, \bibinfo
  {author} {\bibfnamefont {O.}~\bibnamefont {Chubykalo-Fesenko}}, \ and\
  \bibinfo {author} {\bibfnamefont {U.}~\bibnamefont {Nowak}},\ }\href
  {\doibase 10.1103/physrevb.86.214416} {\bibfield  {journal} {\bibinfo
  {journal} {Phys. Rev. B}\ }\textbf {\bibinfo {volume} {86}},\ \bibinfo
  {pages} {214416} (\bibinfo {year} {2012})}\BibitemShut {NoStop}%
\bibitem [{\citenamefont {Vogler}\ \emph {et~al.}(2019)\citenamefont {Vogler},
  \citenamefont {Abert}, \citenamefont {Bruckner},\ and\ \citenamefont
  {Suess}}]{Vogler_2019_PhysRevB_100_5}%
  \BibitemOpen
  \bibfield  {author} {\bibinfo {author} {\bibfnamefont {C.}~\bibnamefont
  {Vogler}}, \bibinfo {author} {\bibfnamefont {C.}~\bibnamefont {Abert}},
  \bibinfo {author} {\bibfnamefont {F.}~\bibnamefont {Bruckner}}, \ and\
  \bibinfo {author} {\bibfnamefont {D.}~\bibnamefont {Suess}},\ }\href
  {\doibase 10.1103/physrevb.100.054401} {\bibfield  {journal} {\bibinfo
  {journal} {Phys. Rev. B}\ }\textbf {\bibinfo {volume} {100}},\ \bibinfo
  {pages} {054401} (\bibinfo {year} {2019})}\BibitemShut {NoStop}%
\bibitem [{\citenamefont {Barker}\ \emph {et~al.}(2013)\citenamefont {Barker},
  \citenamefont {Atxitia}, \citenamefont {Ostler}, \citenamefont {Hovorka},
  \citenamefont {Chubykalo-Fesenko},\ and\ \citenamefont
  {Chantrell}}]{Barker_2013_SciRep_3_1}%
  \BibitemOpen
  \bibfield  {author} {\bibinfo {author} {\bibfnamefont {J.}~\bibnamefont
  {Barker}}, \bibinfo {author} {\bibfnamefont {U.}~\bibnamefont {Atxitia}},
  \bibinfo {author} {\bibfnamefont {T.~A.}\ \bibnamefont {Ostler}}, \bibinfo
  {author} {\bibfnamefont {O.}~\bibnamefont {Hovorka}}, \bibinfo {author}
  {\bibfnamefont {O.}~\bibnamefont {Chubykalo-Fesenko}}, \ and\ \bibinfo
  {author} {\bibfnamefont {R.~W.}\ \bibnamefont {Chantrell}},\ }\href {\doibase
  10.1038/srep03262} {\bibfield  {journal} {\bibinfo  {journal} {Sci. Rep.}\
  }\textbf {\bibinfo {volume} {3}},\ \bibinfo {pages} {3262} (\bibinfo {year}
  {2013})}\BibitemShut {NoStop}%
\bibitem [{\citenamefont {Baryakhtar}(1988)}]{Baryakhtar_1988}%
  \BibitemOpen
  \bibfield  {author} {\bibinfo {author} {\bibfnamefont {V.~G.}\ \bibnamefont
  {Baryakhtar}},\ }\href@noop {} {\bibfield  {journal} {\bibinfo  {journal}
  {Zh. Eksp. Theor. Fiz.}\ }\textbf {\bibinfo {volume} {94}},\ \bibinfo {pages}
  {196} (\bibinfo {year} {1988})}\BibitemShut {NoStop}%
\bibitem [{\citenamefont {Nieves}\ \emph {et~al.}(2015)\citenamefont {Nieves},
  \citenamefont {Atxitia}, \citenamefont {Chantrell},\ and\ \citenamefont
  {Chubykalo-Fesenko}}]{Nieves_2015_LowTempPhys_41_9}%
  \BibitemOpen
  \bibfield  {author} {\bibinfo {author} {\bibfnamefont {P.}~\bibnamefont
  {Nieves}}, \bibinfo {author} {\bibfnamefont {U.}~\bibnamefont {Atxitia}},
  \bibinfo {author} {\bibfnamefont {R.~W.}\ \bibnamefont {Chantrell}}, \ and\
  \bibinfo {author} {\bibfnamefont {O.}~\bibnamefont {Chubykalo-Fesenko}},\
  }\href {\doibase 10.1063/1.4930973} {\bibfield  {journal} {\bibinfo
  {journal} {Low Temp. Phys.}\ }\textbf {\bibinfo {volume} {41}},\ \bibinfo
  {pages} {739} (\bibinfo {year} {2015})}\BibitemShut {NoStop}%
\bibitem [{\citenamefont {Hennecke}\ \emph {et~al.}(2019)\citenamefont
  {Hennecke}, \citenamefont {Radu}, \citenamefont {Abrudan}, \citenamefont
  {Kachel}, \citenamefont {Holldack}, \citenamefont {Mitzner}, \citenamefont
  {Tsukamoto},\ and\ \citenamefont
  {Eisebitt}}]{Hennecke_2019_PhysRevLett_122_15}%
  \BibitemOpen
  \bibfield  {author} {\bibinfo {author} {\bibfnamefont {M.}~\bibnamefont
  {Hennecke}}, \bibinfo {author} {\bibfnamefont {I.}~\bibnamefont {Radu}},
  \bibinfo {author} {\bibfnamefont {R.}~\bibnamefont {Abrudan}}, \bibinfo
  {author} {\bibfnamefont {T.}~\bibnamefont {Kachel}}, \bibinfo {author}
  {\bibfnamefont {K.}~\bibnamefont {Holldack}}, \bibinfo {author}
  {\bibfnamefont {R.}~\bibnamefont {Mitzner}}, \bibinfo {author} {\bibfnamefont
  {A.}~\bibnamefont {Tsukamoto}}, \ and\ \bibinfo {author} {\bibfnamefont
  {S.}~\bibnamefont {Eisebitt}},\ }\href {\doibase
  10.1103/physrevlett.122.157202} {\bibfield  {journal} {\bibinfo  {journal}
  {Phys. Rev. Lett.}\ }\textbf {\bibinfo {volume} {122}},\ \bibinfo {pages}
  {157202} (\bibinfo {year} {2019})}\BibitemShut {NoStop}%
\bibitem [{\citenamefont {Dionne}(2009)}]{Dionne_Book}%
  \BibitemOpen
  \bibfield  {author} {\bibinfo {author} {\bibfnamefont {G.~F.}\ \bibnamefont
  {Dionne}},\ }\href@noop {} {\emph {\bibinfo {title} {{M}agnetic {O}xides}}}\
  (\bibinfo  {publisher} {Springer},\ \bibinfo {address} {New York},\ \bibinfo
  {year} {2009})\BibitemShut {NoStop}%
\bibitem [{\citenamefont {Yu}\ \emph {et~al.}(2018)\citenamefont {Yu},
  \citenamefont {Bang}, \citenamefont {Mishra}, \citenamefont {Ramaswamy},
  \citenamefont {Oh}, \citenamefont {Park}, \citenamefont {Jeong},
  \citenamefont {Van~Thach}, \citenamefont {Lee}, \citenamefont {Go},
  \citenamefont {Lee}, \citenamefont {Wang}, \citenamefont {Shi}, \citenamefont
  {Qiu}, \citenamefont {Awano}, \citenamefont {Lee},\ and\ \citenamefont
  {Yang}}]{Yu_2018_NatMater_18_1}%
  \BibitemOpen
  \bibfield  {author} {\bibinfo {author} {\bibfnamefont {J.}~\bibnamefont
  {Yu}}, \bibinfo {author} {\bibfnamefont {D.}~\bibnamefont {Bang}}, \bibinfo
  {author} {\bibfnamefont {R.}~\bibnamefont {Mishra}}, \bibinfo {author}
  {\bibfnamefont {R.}~\bibnamefont {Ramaswamy}}, \bibinfo {author}
  {\bibfnamefont {J.~H.}\ \bibnamefont {Oh}}, \bibinfo {author} {\bibfnamefont
  {H.-J.}\ \bibnamefont {Park}}, \bibinfo {author} {\bibfnamefont
  {Y.}~\bibnamefont {Jeong}}, \bibinfo {author} {\bibfnamefont
  {P.}~\bibnamefont {Van~Thach}}, \bibinfo {author} {\bibfnamefont {D.-K.}\
  \bibnamefont {Lee}}, \bibinfo {author} {\bibfnamefont {G.}~\bibnamefont
  {Go}}, \bibinfo {author} {\bibfnamefont {S.-W.}\ \bibnamefont {Lee}},
  \bibinfo {author} {\bibfnamefont {Y.}~\bibnamefont {Wang}}, \bibinfo {author}
  {\bibfnamefont {S.}~\bibnamefont {Shi}}, \bibinfo {author} {\bibfnamefont
  {X.}~\bibnamefont {Qiu}}, \bibinfo {author} {\bibfnamefont {H.}~\bibnamefont
  {Awano}}, \bibinfo {author} {\bibfnamefont {K.-J.}\ \bibnamefont {Lee}}, \
  and\ \bibinfo {author} {\bibfnamefont {H.}~\bibnamefont {Yang}},\ }\href
  {\doibase 10.1038/s41563-018-0236-9} {\bibfield  {journal} {\bibinfo
  {journal} {Nat. Mater.}\ }\textbf {\bibinfo {volume} {18}},\ \bibinfo {pages}
  {29} (\bibinfo {year} {2018})}\BibitemShut {NoStop}%
\bibitem [{\citenamefont {Kim}\ \emph {et~al.}(2019)\citenamefont {Kim},
  \citenamefont {Haruta}, \citenamefont {Ko}, \citenamefont {Go}, \citenamefont
  {Park}, \citenamefont {Nishimura}, \citenamefont {Kim}, \citenamefont
  {Okuno}, \citenamefont {Hirata}, \citenamefont {Futakawa}, \citenamefont
  {Yoshikawa}, \citenamefont {Ham}, \citenamefont {Kim}, \citenamefont
  {Kurata}, \citenamefont {Tsukamoto}, \citenamefont {Shiota}, \citenamefont
  {Moriyama}, \citenamefont {Choe}, \citenamefont {Lee},\ and\ \citenamefont
  {Ono}}]{Kim_2019_NatMater_18_7}%
  \BibitemOpen
  \bibfield  {author} {\bibinfo {author} {\bibfnamefont {D.-H.}\ \bibnamefont
  {Kim}}, \bibinfo {author} {\bibfnamefont {M.}~\bibnamefont {Haruta}},
  \bibinfo {author} {\bibfnamefont {H.-W.}\ \bibnamefont {Ko}}, \bibinfo
  {author} {\bibfnamefont {G.}~\bibnamefont {Go}}, \bibinfo {author}
  {\bibfnamefont {H.-J.}\ \bibnamefont {Park}}, \bibinfo {author}
  {\bibfnamefont {T.}~\bibnamefont {Nishimura}}, \bibinfo {author}
  {\bibfnamefont {D.-Y.}\ \bibnamefont {Kim}}, \bibinfo {author} {\bibfnamefont
  {T.}~\bibnamefont {Okuno}}, \bibinfo {author} {\bibfnamefont
  {Y.}~\bibnamefont {Hirata}}, \bibinfo {author} {\bibfnamefont
  {Y.}~\bibnamefont {Futakawa}}, \bibinfo {author} {\bibfnamefont
  {H.}~\bibnamefont {Yoshikawa}}, \bibinfo {author} {\bibfnamefont
  {W.}~\bibnamefont {Ham}}, \bibinfo {author} {\bibfnamefont {S.}~\bibnamefont
  {Kim}}, \bibinfo {author} {\bibfnamefont {H.}~\bibnamefont {Kurata}},
  \bibinfo {author} {\bibfnamefont {A.}~\bibnamefont {Tsukamoto}}, \bibinfo
  {author} {\bibfnamefont {Y.}~\bibnamefont {Shiota}}, \bibinfo {author}
  {\bibfnamefont {T.}~\bibnamefont {Moriyama}}, \bibinfo {author}
  {\bibfnamefont {S.-B.}\ \bibnamefont {Choe}}, \bibinfo {author}
  {\bibfnamefont {K.-J.}\ \bibnamefont {Lee}}, \ and\ \bibinfo {author}
  {\bibfnamefont {T.}~\bibnamefont {Ono}},\ }\href {\doibase
  10.1038/s41563-019-0380-x} {\bibfield  {journal} {\bibinfo  {journal} {Nat.
  Mater.}\ }\textbf {\bibinfo {volume} {18}},\ \bibinfo {pages} {685} (\bibinfo
  {year} {2019})}\BibitemShut {NoStop}%
\bibitem [{\citenamefont {Sala}\ \emph {et~al.}(2021)\citenamefont {Sala},
  \citenamefont {Krizakova}, \citenamefont {Grimaldi}, \citenamefont {Lambert},
  \citenamefont {Devolder},\ and\ \citenamefont
  {Gambardella}}]{Sala_2021_NatCommun_12_1}%
  \BibitemOpen
  \bibfield  {author} {\bibinfo {author} {\bibfnamefont {G.}~\bibnamefont
  {Sala}}, \bibinfo {author} {\bibfnamefont {V.}~\bibnamefont {Krizakova}},
  \bibinfo {author} {\bibfnamefont {E.}~\bibnamefont {Grimaldi}}, \bibinfo
  {author} {\bibfnamefont {C.-H.}\ \bibnamefont {Lambert}}, \bibinfo {author}
  {\bibfnamefont {T.}~\bibnamefont {Devolder}}, \ and\ \bibinfo {author}
  {\bibfnamefont {P.}~\bibnamefont {Gambardella}},\ }\href {\doibase
  10.1038/s41467-021-20968-0} {\bibfield  {journal} {\bibinfo  {journal} {Nat.
  Commun.}\ }\textbf {\bibinfo {volume} {12}},\ \bibinfo {pages} {656}
  (\bibinfo {year} {2021})}\BibitemShut {NoStop}%
\bibitem [{\citenamefont {Ostler}\ \emph {et~al.}(2011)\citenamefont {Ostler},
  \citenamefont {Evans}, \citenamefont {Chantrell}, \citenamefont {Atxitia},
  \citenamefont {Chubykalo-Fesenko}, \citenamefont {Radu}, \citenamefont
  {Abrudan}, \citenamefont {Radu}, \citenamefont {Tsukamoto}, \citenamefont
  {Itoh}, \citenamefont {Kirilyuk}, \citenamefont {Rasing},\ and\ \citenamefont
  {Kimel}}]{Ostler_2011_PhysRevB_84_2}%
  \BibitemOpen
  \bibfield  {author} {\bibinfo {author} {\bibfnamefont {T.~A.}\ \bibnamefont
  {Ostler}}, \bibinfo {author} {\bibfnamefont {R.~F.~L.}\ \bibnamefont
  {Evans}}, \bibinfo {author} {\bibfnamefont {R.~W.}\ \bibnamefont
  {Chantrell}}, \bibinfo {author} {\bibfnamefont {U.}~\bibnamefont {Atxitia}},
  \bibinfo {author} {\bibfnamefont {O.}~\bibnamefont {Chubykalo-Fesenko}},
  \bibinfo {author} {\bibfnamefont {I.}~\bibnamefont {Radu}}, \bibinfo {author}
  {\bibfnamefont {R.}~\bibnamefont {Abrudan}}, \bibinfo {author} {\bibfnamefont
  {F.}~\bibnamefont {Radu}}, \bibinfo {author} {\bibfnamefont {A.}~\bibnamefont
  {Tsukamoto}}, \bibinfo {author} {\bibfnamefont {A.}~\bibnamefont {Itoh}},
  \bibinfo {author} {\bibfnamefont {A.}~\bibnamefont {Kirilyuk}}, \bibinfo
  {author} {\bibfnamefont {T.}~\bibnamefont {Rasing}}, \ and\ \bibinfo {author}
  {\bibfnamefont {A.}~\bibnamefont {Kimel}},\ }\href {\doibase
  10.1103/physrevb.84.024407} {\bibfield  {journal} {\bibinfo  {journal} {Phys.
  Rev. B}\ }\textbf {\bibinfo {volume} {84}},\ \bibinfo {pages} {024407}
  (\bibinfo {year} {2011})}\BibitemShut {NoStop}%
\bibitem [{\citenamefont {Radu}\ \emph {et~al.}(2011)\citenamefont {Radu},
  \citenamefont {Vahaplar}, \citenamefont {Stamm}, \citenamefont {Kachel},
  \citenamefont {Pontius}, \citenamefont {D\"{u}rr}, \citenamefont {Ostler},
  \citenamefont {Barker}, \citenamefont {Evans}, \citenamefont {Chantrell},
  \citenamefont {Tsukamoto}, \citenamefont {Itoh}, \citenamefont {Kirilyuk},
  \citenamefont {Rasing},\ and\ \citenamefont
  {Kimel}}]{Radu_2011_Nature_472_7342}%
  \BibitemOpen
  \bibfield  {author} {\bibinfo {author} {\bibfnamefont {I.}~\bibnamefont
  {Radu}}, \bibinfo {author} {\bibfnamefont {K.}~\bibnamefont {Vahaplar}},
  \bibinfo {author} {\bibfnamefont {C.}~\bibnamefont {Stamm}}, \bibinfo
  {author} {\bibfnamefont {T.}~\bibnamefont {Kachel}}, \bibinfo {author}
  {\bibfnamefont {N.}~\bibnamefont {Pontius}}, \bibinfo {author} {\bibfnamefont
  {H.~A.}\ \bibnamefont {D\"{u}rr}}, \bibinfo {author} {\bibfnamefont {T.~A.}\
  \bibnamefont {Ostler}}, \bibinfo {author} {\bibfnamefont {J.}~\bibnamefont
  {Barker}}, \bibinfo {author} {\bibfnamefont {R.~F.~L.}\ \bibnamefont
  {Evans}}, \bibinfo {author} {\bibfnamefont {R.~W.}\ \bibnamefont
  {Chantrell}}, \bibinfo {author} {\bibfnamefont {A.}~\bibnamefont
  {Tsukamoto}}, \bibinfo {author} {\bibfnamefont {A.}~\bibnamefont {Itoh}},
  \bibinfo {author} {\bibfnamefont {A.}~\bibnamefont {Kirilyuk}}, \bibinfo
  {author} {\bibfnamefont {T.}~\bibnamefont {Rasing}}, \ and\ \bibinfo {author}
  {\bibfnamefont {A.~V.}\ \bibnamefont {Kimel}},\ }\href {\doibase
  10.1038/nature09901} {\bibfield  {journal} {\bibinfo  {journal} {Nature}\
  }\textbf {\bibinfo {volume} {472}},\ \bibinfo {pages} {205} (\bibinfo {year}
  {2011})}\BibitemShut {NoStop}%
\bibitem [{\citenamefont {Cherepanov}\ \emph {et~al.}(1993)\citenamefont
  {Cherepanov}, \citenamefont {Kolokolov},\ and\ \citenamefont
  {L'vov}}]{Cherepanov_1993_PhysRep_229_3}%
  \BibitemOpen
  \bibfield  {author} {\bibinfo {author} {\bibfnamefont {V.}~\bibnamefont
  {Cherepanov}}, \bibinfo {author} {\bibfnamefont {I.}~\bibnamefont
  {Kolokolov}}, \ and\ \bibinfo {author} {\bibfnamefont {V.}~\bibnamefont
  {L'vov}},\ }\href {\doibase 10.1016/0370-1573(93)90107-o} {\bibfield
  {journal} {\bibinfo  {journal} {Phys. Rep.}\ }\textbf {\bibinfo {volume}
  {229}},\ \bibinfo {pages} {81} (\bibinfo {year} {1993})}\BibitemShut
  {NoStop}%
\bibitem [{\citenamefont {Bauer}\ \emph {et~al.}(2012)\citenamefont {Bauer},
  \citenamefont {Saitoh},\ and\ \citenamefont {van
  Wees}}]{Bauer_2012_NatMater_11_5}%
  \BibitemOpen
  \bibfield  {author} {\bibinfo {author} {\bibfnamefont {G.~E.~W.}\
  \bibnamefont {Bauer}}, \bibinfo {author} {\bibfnamefont {E.}~\bibnamefont
  {Saitoh}}, \ and\ \bibinfo {author} {\bibfnamefont {B.~J.}\ \bibnamefont {van
  Wees}},\ }\href {\doibase 10.1038/nmat3301} {\bibfield  {journal} {\bibinfo
  {journal} {Nat. Mater.}\ }\textbf {\bibinfo {volume} {11}},\ \bibinfo {pages}
  {391} (\bibinfo {year} {2012})}\BibitemShut {NoStop}%
\bibitem [{\citenamefont {Toth}\ and\ \citenamefont
  {Lake}(2015)}]{Toth_2015_JPhysCondensMatter_27_16}%
  \BibitemOpen
  \bibfield  {author} {\bibinfo {author} {\bibfnamefont {S.}~\bibnamefont
  {Toth}}\ and\ \bibinfo {author} {\bibfnamefont {B.}~\bibnamefont {Lake}},\
  }\href {\doibase 10.1088/0953-8984/27/16/166002} {\bibfield  {journal}
  {\bibinfo  {journal} {J. Phys.: Condens. Matter}\ }\textbf {\bibinfo {volume}
  {27}},\ \bibinfo {pages} {166002} (\bibinfo {year} {2015})}\BibitemShut
  {NoStop}%
\bibitem [{\citenamefont {Princep}\ \emph {et~al.}(2017)\citenamefont
  {Princep}, \citenamefont {Ewings}, \citenamefont {Ward}, \citenamefont
  {T\'{o}th}, \citenamefont {Dubs}, \citenamefont {Prabhakaran},\ and\
  \citenamefont {Boothroyd}}]{Princep_2017_npjQuantumMater_2_1}%
  \BibitemOpen
  \bibfield  {author} {\bibinfo {author} {\bibfnamefont {A.~J.}\ \bibnamefont
  {Princep}}, \bibinfo {author} {\bibfnamefont {R.~A.}\ \bibnamefont {Ewings}},
  \bibinfo {author} {\bibfnamefont {S.}~\bibnamefont {Ward}}, \bibinfo {author}
  {\bibfnamefont {S.}~\bibnamefont {T\'{o}th}}, \bibinfo {author}
  {\bibfnamefont {C.}~\bibnamefont {Dubs}}, \bibinfo {author} {\bibfnamefont
  {D.}~\bibnamefont {Prabhakaran}}, \ and\ \bibinfo {author} {\bibfnamefont
  {A.~T.}\ \bibnamefont {Boothroyd}},\ }\href {\doibase
  10.1038/s41535-017-0067-y} {\bibfield  {journal} {\bibinfo  {journal} {npj
  Quantum Mater.}\ }\textbf {\bibinfo {volume} {2}},\ \bibinfo {pages} {63}
  (\bibinfo {year} {2017})}\BibitemShut {NoStop}%
\bibitem [{\citenamefont {Xie}\ \emph {et~al.}(2017)\citenamefont {Xie},
  \citenamefont {Jin}, \citenamefont {He}, \citenamefont {Bauer}, \citenamefont
  {Barker},\ and\ \citenamefont {Xia}}]{Xie_2017_PhysRevB_95_1}%
  \BibitemOpen
  \bibfield  {author} {\bibinfo {author} {\bibfnamefont {L.-S.}\ \bibnamefont
  {Xie}}, \bibinfo {author} {\bibfnamefont {G.-X.}\ \bibnamefont {Jin}},
  \bibinfo {author} {\bibfnamefont {L.}~\bibnamefont {He}}, \bibinfo {author}
  {\bibfnamefont {G.~E.~W.}\ \bibnamefont {Bauer}}, \bibinfo {author}
  {\bibfnamefont {J.}~\bibnamefont {Barker}}, \ and\ \bibinfo {author}
  {\bibfnamefont {K.}~\bibnamefont {Xia}},\ }\href {\doibase
  10.1103/physrevb.95.014423} {\bibfield  {journal} {\bibinfo  {journal} {Phys.
  Rev. B}\ }\textbf {\bibinfo {volume} {95}},\ \bibinfo {pages} {014423}
  (\bibinfo {year} {2017})}\BibitemShut {NoStop}%
\bibitem [{\citenamefont {Nakamoto}\ \emph {et~al.}(2017)\citenamefont
  {Nakamoto}, \citenamefont {Xu}, \citenamefont {Xu}, \citenamefont {Xu},\ and\
  \citenamefont {Bellaiche}}]{Nakamoto_2017_PhysRevB_95_2}%
  \BibitemOpen
  \bibfield  {author} {\bibinfo {author} {\bibfnamefont {R.}~\bibnamefont
  {Nakamoto}}, \bibinfo {author} {\bibfnamefont {B.}~\bibnamefont {Xu}},
  \bibinfo {author} {\bibfnamefont {C.}~\bibnamefont {Xu}}, \bibinfo {author}
  {\bibfnamefont {H.}~\bibnamefont {Xu}}, \ and\ \bibinfo {author}
  {\bibfnamefont {L.}~\bibnamefont {Bellaiche}},\ }\href {\doibase
  10.1103/physrevb.95.024434} {\bibfield  {journal} {\bibinfo  {journal} {Phys.
  Rev. B}\ }\textbf {\bibinfo {volume} {95}},\ \bibinfo {pages} {024434}
  (\bibinfo {year} {2017})}\BibitemShut {NoStop}%
\bibitem [{\citenamefont {Iori}\ \emph {et~al.}(2019)\citenamefont {Iori},
  \citenamefont {Teurtrie}, \citenamefont {Bocher}, \citenamefont {Popova},
  \citenamefont {Keller}, \citenamefont {St\'{e}phan},\ and\ \citenamefont
  {Gloter}}]{Iori_2019_PhysRevB_100_24}%
  \BibitemOpen
  \bibfield  {author} {\bibinfo {author} {\bibfnamefont {F.}~\bibnamefont
  {Iori}}, \bibinfo {author} {\bibfnamefont {A.}~\bibnamefont {Teurtrie}},
  \bibinfo {author} {\bibfnamefont {L.}~\bibnamefont {Bocher}}, \bibinfo
  {author} {\bibfnamefont {E.}~\bibnamefont {Popova}}, \bibinfo {author}
  {\bibfnamefont {N.}~\bibnamefont {Keller}}, \bibinfo {author} {\bibfnamefont
  {O.}~\bibnamefont {St\'{e}phan}}, \ and\ \bibinfo {author} {\bibfnamefont
  {A.}~\bibnamefont {Gloter}},\ }\href {\doibase 10.1103/physrevb.100.245150}
  {\bibfield  {journal} {\bibinfo  {journal} {Phys. Rev. B}\ }\textbf {\bibinfo
  {volume} {100}},\ \bibinfo {pages} {245150} (\bibinfo {year}
  {2019})}\BibitemShut {NoStop}%
\bibitem [{\citenamefont {Campbell}\ \emph {et~al.}(2020)\citenamefont
  {Campbell}, \citenamefont {Xu}, \citenamefont {Bayaraa},\ and\ \citenamefont
  {Bellaiche}}]{Campbell_2020_PhysRevB_102_14}%
  \BibitemOpen
  \bibfield  {author} {\bibinfo {author} {\bibfnamefont {D.}~\bibnamefont
  {Campbell}}, \bibinfo {author} {\bibfnamefont {C.}~\bibnamefont {Xu}},
  \bibinfo {author} {\bibfnamefont {T.}~\bibnamefont {Bayaraa}}, \ and\
  \bibinfo {author} {\bibfnamefont {L.}~\bibnamefont {Bellaiche}},\ }\href
  {\doibase 10.1103/physrevb.102.144406} {\bibfield  {journal} {\bibinfo
  {journal} {Phys. Rev. B}\ }\textbf {\bibinfo {volume} {102}},\ \bibinfo
  {pages} {144406} (\bibinfo {year} {2020})}\BibitemShut {NoStop}%
\bibitem [{\citenamefont {Faleev}\ \emph {et~al.}(2004)\citenamefont {Faleev},
  \citenamefont {van Schilfgaarde},\ and\ \citenamefont
  {Kotani}}]{Faleev_2004_PhysRevLett_93_12}%
  \BibitemOpen
  \bibfield  {author} {\bibinfo {author} {\bibfnamefont {S.~V.}\ \bibnamefont
  {Faleev}}, \bibinfo {author} {\bibfnamefont {M.}~\bibnamefont {van
  Schilfgaarde}}, \ and\ \bibinfo {author} {\bibfnamefont {T.}~\bibnamefont
  {Kotani}},\ }\href {\doibase 10.1103/physrevlett.93.126406} {\bibfield
  {journal} {\bibinfo  {journal} {Phys. Rev. Lett.}\ }\textbf {\bibinfo
  {volume} {93}},\ \bibinfo {pages} {126406} (\bibinfo {year}
  {2004})}\BibitemShut {NoStop}%
\bibitem [{\citenamefont {Barker}\ \emph {et~al.}(2020)\citenamefont {Barker},
  \citenamefont {Pashov},\ and\ \citenamefont
  {Jackson}}]{Barker_2020_ElectronStruct_2_4}%
  \BibitemOpen
  \bibfield  {author} {\bibinfo {author} {\bibfnamefont {J.}~\bibnamefont
  {Barker}}, \bibinfo {author} {\bibfnamefont {D.}~\bibnamefont {Pashov}}, \
  and\ \bibinfo {author} {\bibfnamefont {J.}~\bibnamefont {Jackson}},\ }\href
  {\doibase 10.1088/2516-1075/abd097} {\bibfield  {journal} {\bibinfo
  {journal} {Electron. Struct.}\ }\textbf {\bibinfo {volume} {2}},\ \bibinfo
  {pages} {044002} (\bibinfo {year} {2020})}\BibitemShut {NoStop}%
\bibitem [{\citenamefont {Nambu}\ \emph {et~al.}(2020)\citenamefont {Nambu},
  \citenamefont {Barker}, \citenamefont {Okino}, \citenamefont {Kikkawa},
  \citenamefont {Shiomi}, \citenamefont {Enderle}, \citenamefont {Weber},
  \citenamefont {Winn}, \citenamefont {Graves-Brook}, \citenamefont
  {Tranquada}, \citenamefont {Ziman}, \citenamefont {Fujita}, \citenamefont
  {Bauer}, \citenamefont {Saitoh},\ and\ \citenamefont
  {Kakurai}}]{Nambu_2020_PhysRevLett_125_2}%
  \BibitemOpen
  \bibfield  {author} {\bibinfo {author} {\bibfnamefont {Y.}~\bibnamefont
  {Nambu}}, \bibinfo {author} {\bibfnamefont {J.}~\bibnamefont {Barker}},
  \bibinfo {author} {\bibfnamefont {Y.}~\bibnamefont {Okino}}, \bibinfo
  {author} {\bibfnamefont {T.}~\bibnamefont {Kikkawa}}, \bibinfo {author}
  {\bibfnamefont {Y.}~\bibnamefont {Shiomi}}, \bibinfo {author} {\bibfnamefont
  {M.}~\bibnamefont {Enderle}}, \bibinfo {author} {\bibfnamefont
  {T.}~\bibnamefont {Weber}}, \bibinfo {author} {\bibfnamefont
  {B.}~\bibnamefont {Winn}}, \bibinfo {author} {\bibfnamefont {M.}~\bibnamefont
  {Graves-Brook}}, \bibinfo {author} {\bibfnamefont {J.~M.}\ \bibnamefont
  {Tranquada}}, \bibinfo {author} {\bibfnamefont {T.}~\bibnamefont {Ziman}},
  \bibinfo {author} {\bibfnamefont {M.}~\bibnamefont {Fujita}}, \bibinfo
  {author} {\bibfnamefont {G.~E.~W.}\ \bibnamefont {Bauer}}, \bibinfo {author}
  {\bibfnamefont {E.}~\bibnamefont {Saitoh}}, \ and\ \bibinfo {author}
  {\bibfnamefont {K.}~\bibnamefont {Kakurai}},\ }\href {\doibase
  10.1103/physrevlett.125.027201} {\bibfield  {journal} {\bibinfo  {journal}
  {Phys. Rev. Lett.}\ }\textbf {\bibinfo {volume} {125}},\ \bibinfo {pages}
  {027201} (\bibinfo {year} {2020})}\BibitemShut {NoStop}%
\bibitem [{\citenamefont {Cornelissen}\ \emph {et~al.}(2016)\citenamefont
  {Cornelissen}, \citenamefont {Peters}, \citenamefont {Bauer}, \citenamefont
  {Duine},\ and\ \citenamefont {van Wees}}]{Cornelissen_2016_PhysRevB_94_1}%
  \BibitemOpen
  \bibfield  {author} {\bibinfo {author} {\bibfnamefont {L.~J.}\ \bibnamefont
  {Cornelissen}}, \bibinfo {author} {\bibfnamefont {K.~J.~H.}\ \bibnamefont
  {Peters}}, \bibinfo {author} {\bibfnamefont {G.~E.~W.}\ \bibnamefont
  {Bauer}}, \bibinfo {author} {\bibfnamefont {R.~A.}\ \bibnamefont {Duine}}, \
  and\ \bibinfo {author} {\bibfnamefont {B.~J.}\ \bibnamefont {van Wees}},\
  }\href {\doibase 10.1103/physrevb.94.014412} {\bibfield  {journal} {\bibinfo
  {journal} {Phys. Rev. B}\ }\textbf {\bibinfo {volume} {94}},\ \bibinfo
  {pages} {014412} (\bibinfo {year} {2016})}\BibitemShut {NoStop}%
\bibitem [{\citenamefont {Boona}\ and\ \citenamefont
  {Heremans}(2014)}]{Boona_2014_PhysRevB_90_6}%
  \BibitemOpen
  \bibfield  {author} {\bibinfo {author} {\bibfnamefont {S.~R.}\ \bibnamefont
  {Boona}}\ and\ \bibinfo {author} {\bibfnamefont {J.~P.}\ \bibnamefont
  {Heremans}},\ }\href {\doibase 10.1103/physrevb.90.064421} {\bibfield
  {journal} {\bibinfo  {journal} {Phys. Rev. B}\ }\textbf {\bibinfo {volume}
  {90}},\ \bibinfo {pages} {064421} (\bibinfo {year} {2014})}\BibitemShut
  {NoStop}%
\bibitem [{\citenamefont {Rezende}\ and\ \citenamefont
  {L\'{o}pez~Ortiz}(2015)}]{Rezende_2015_PhysRevB_91_10}%
  \BibitemOpen
  \bibfield  {author} {\bibinfo {author} {\bibfnamefont {S.~M.}\ \bibnamefont
  {Rezende}}\ and\ \bibinfo {author} {\bibfnamefont {J.~C.}\ \bibnamefont
  {L\'{o}pez~Ortiz}},\ }\href {\doibase 10.1103/physrevb.91.104416} {\bibfield
  {journal} {\bibinfo  {journal} {Phys. Rev. B}\ }\textbf {\bibinfo {volume}
  {91}},\ \bibinfo {pages} {104416} (\bibinfo {year} {2015})}\BibitemShut
  {NoStop}%
\bibitem [{\citenamefont {Beaurepaire}\ \emph {et~al.}(1996)\citenamefont
  {Beaurepaire}, \citenamefont {Merle}, \citenamefont {Daunois},\ and\
  \citenamefont {Bigot}}]{Beaurepaire_1996_PhysRevLett_76_22}%
  \BibitemOpen
  \bibfield  {author} {\bibinfo {author} {\bibfnamefont {E.}~\bibnamefont
  {Beaurepaire}}, \bibinfo {author} {\bibfnamefont {J.-C.}\ \bibnamefont
  {Merle}}, \bibinfo {author} {\bibfnamefont {A.}~\bibnamefont {Daunois}}, \
  and\ \bibinfo {author} {\bibfnamefont {J.-Y.}\ \bibnamefont {Bigot}},\ }\href
  {\doibase 10.1103/physrevlett.76.4250} {\bibfield  {journal} {\bibinfo
  {journal} {Phys. Rev. Lett.}\ }\textbf {\bibinfo {volume} {76}},\ \bibinfo
  {pages} {4250} (\bibinfo {year} {1996})}\BibitemShut {NoStop}%
\bibitem [{\citenamefont {Koopmans}\ \emph {et~al.}(2009)\citenamefont
  {Koopmans}, \citenamefont {Malinowski}, \citenamefont {Dalla~Longa},
  \citenamefont {Steiauf}, \citenamefont {F\"{a}hnle}, \citenamefont {Roth},
  \citenamefont {Cinchetti},\ and\ \citenamefont
  {Aeschlimann}}]{Koopmans_2009_NatMater_9_3}%
  \BibitemOpen
  \bibfield  {author} {\bibinfo {author} {\bibfnamefont {B.}~\bibnamefont
  {Koopmans}}, \bibinfo {author} {\bibfnamefont {G.}~\bibnamefont
  {Malinowski}}, \bibinfo {author} {\bibfnamefont {F.}~\bibnamefont
  {Dalla~Longa}}, \bibinfo {author} {\bibfnamefont {D.}~\bibnamefont
  {Steiauf}}, \bibinfo {author} {\bibfnamefont {M.}~\bibnamefont {F\"{a}hnle}},
  \bibinfo {author} {\bibfnamefont {T.}~\bibnamefont {Roth}}, \bibinfo {author}
  {\bibfnamefont {M.}~\bibnamefont {Cinchetti}}, \ and\ \bibinfo {author}
  {\bibfnamefont {M.}~\bibnamefont {Aeschlimann}},\ }\href {\doibase
  10.1038/nmat2593} {\bibfield  {journal} {\bibinfo  {journal} {Nat. Mater.}\
  }\textbf {\bibinfo {volume} {9}},\ \bibinfo {pages} {259} (\bibinfo {year}
  {2009})}\BibitemShut {NoStop}%
\bibitem [{\citenamefont {Dornes}\ \emph {et~al.}(2019)\citenamefont {Dornes},
  \citenamefont {Acremann}, \citenamefont {Savoini}, \citenamefont {Kubli},
  \citenamefont {Neugebauer}, \citenamefont {Abreu}, \citenamefont {Huber},
  \citenamefont {Lantz}, \citenamefont {Vaz}, \citenamefont {Lemke},
  \citenamefont {Bothschafter}, \citenamefont {Porer}, \citenamefont
  {Esposito}, \citenamefont {Rettig}, \citenamefont {Buzzi}, \citenamefont
  {Alberca}, \citenamefont {Windsor}, \citenamefont {Beaud}, \citenamefont
  {Staub}, \citenamefont {Zhu}, \citenamefont {Song}, \citenamefont {Glownia},\
  and\ \citenamefont {Johnson}}]{Dornes_2019_Nature_565_7738}%
  \BibitemOpen
  \bibfield  {author} {\bibinfo {author} {\bibfnamefont {C.}~\bibnamefont
  {Dornes}}, \bibinfo {author} {\bibfnamefont {Y.}~\bibnamefont {Acremann}},
  \bibinfo {author} {\bibfnamefont {M.}~\bibnamefont {Savoini}}, \bibinfo
  {author} {\bibfnamefont {M.}~\bibnamefont {Kubli}}, \bibinfo {author}
  {\bibfnamefont {M.~J.}\ \bibnamefont {Neugebauer}}, \bibinfo {author}
  {\bibfnamefont {E.}~\bibnamefont {Abreu}}, \bibinfo {author} {\bibfnamefont
  {L.}~\bibnamefont {Huber}}, \bibinfo {author} {\bibfnamefont
  {G.}~\bibnamefont {Lantz}}, \bibinfo {author} {\bibfnamefont {C.~A.~F.}\
  \bibnamefont {Vaz}}, \bibinfo {author} {\bibfnamefont {H.}~\bibnamefont
  {Lemke}}, \bibinfo {author} {\bibfnamefont {E.~M.}\ \bibnamefont
  {Bothschafter}}, \bibinfo {author} {\bibfnamefont {M.}~\bibnamefont {Porer}},
  \bibinfo {author} {\bibfnamefont {V.}~\bibnamefont {Esposito}}, \bibinfo
  {author} {\bibfnamefont {L.}~\bibnamefont {Rettig}}, \bibinfo {author}
  {\bibfnamefont {M.}~\bibnamefont {Buzzi}}, \bibinfo {author} {\bibfnamefont
  {A.}~\bibnamefont {Alberca}}, \bibinfo {author} {\bibfnamefont {Y.~W.}\
  \bibnamefont {Windsor}}, \bibinfo {author} {\bibfnamefont {P.}~\bibnamefont
  {Beaud}}, \bibinfo {author} {\bibfnamefont {U.}~\bibnamefont {Staub}},
  \bibinfo {author} {\bibfnamefont {D.}~\bibnamefont {Zhu}}, \bibinfo {author}
  {\bibfnamefont {S.}~\bibnamefont {Song}}, \bibinfo {author} {\bibfnamefont
  {J.~M.}\ \bibnamefont {Glownia}}, \ and\ \bibinfo {author} {\bibfnamefont
  {S.~L.}\ \bibnamefont {Johnson}},\ }\href {\doibase
  10.1038/s41586-018-0822-7} {\bibfield  {journal} {\bibinfo  {journal}
  {Nature}\ }\textbf {\bibinfo {volume} {565}},\ \bibinfo {pages} {209}
  (\bibinfo {year} {2019})}\BibitemShut {NoStop}%
\bibitem [{\citenamefont {Battiato}\ \emph {et~al.}(2010)\citenamefont
  {Battiato}, \citenamefont {Carva},\ and\ \citenamefont
  {Oppeneer}}]{Battiato_2010_PhysRevLett_105_2}%
  \BibitemOpen
  \bibfield  {author} {\bibinfo {author} {\bibfnamefont {M.}~\bibnamefont
  {Battiato}}, \bibinfo {author} {\bibfnamefont {K.}~\bibnamefont {Carva}}, \
  and\ \bibinfo {author} {\bibfnamefont {P.~M.}\ \bibnamefont {Oppeneer}},\
  }\href {\doibase 10.1103/physrevlett.105.027203} {\bibfield  {journal}
  {\bibinfo  {journal} {Phys. Rev. Lett.}\ }\textbf {\bibinfo {volume} {105}},\
  \bibinfo {pages} {027203} (\bibinfo {year} {2010})}\BibitemShut {NoStop}%
\bibitem [{\citenamefont {Rudolf}\ \emph {et~al.}(2012)\citenamefont {Rudolf},
  \citenamefont {La-O-Vorakiat}, \citenamefont {Battiato}, \citenamefont
  {Adam}, \citenamefont {Shaw}, \citenamefont {Turgut}, \citenamefont
  {Maldonado}, \citenamefont {Mathias}, \citenamefont {Grychtol}, \citenamefont
  {Nembach}, \citenamefont {Silva}, \citenamefont {Aeschlimann}, \citenamefont
  {Kapteyn}, \citenamefont {Murnane}, \citenamefont {Schneider},\ and\
  \citenamefont {Oppeneer}}]{Rudolf_2012_NatCommun_3_1}%
  \BibitemOpen
  \bibfield  {author} {\bibinfo {author} {\bibfnamefont {D.}~\bibnamefont
  {Rudolf}}, \bibinfo {author} {\bibfnamefont {C.}~\bibnamefont
  {La-O-Vorakiat}}, \bibinfo {author} {\bibfnamefont {M.}~\bibnamefont
  {Battiato}}, \bibinfo {author} {\bibfnamefont {R.}~\bibnamefont {Adam}},
  \bibinfo {author} {\bibfnamefont {J.~M.}\ \bibnamefont {Shaw}}, \bibinfo
  {author} {\bibfnamefont {E.}~\bibnamefont {Turgut}}, \bibinfo {author}
  {\bibfnamefont {P.}~\bibnamefont {Maldonado}}, \bibinfo {author}
  {\bibfnamefont {S.}~\bibnamefont {Mathias}}, \bibinfo {author} {\bibfnamefont
  {P.}~\bibnamefont {Grychtol}}, \bibinfo {author} {\bibfnamefont {H.~T.}\
  \bibnamefont {Nembach}}, \bibinfo {author} {\bibfnamefont {T.~J.}\
  \bibnamefont {Silva}}, \bibinfo {author} {\bibfnamefont {M.}~\bibnamefont
  {Aeschlimann}}, \bibinfo {author} {\bibfnamefont {H.~C.}\ \bibnamefont
  {Kapteyn}}, \bibinfo {author} {\bibfnamefont {M.~M.}\ \bibnamefont
  {Murnane}}, \bibinfo {author} {\bibfnamefont {C.~M.}\ \bibnamefont
  {Schneider}}, \ and\ \bibinfo {author} {\bibfnamefont {P.~M.}\ \bibnamefont
  {Oppeneer}},\ }\href {\doibase 10.1038/ncomms2029} {\bibfield  {journal}
  {\bibinfo  {journal} {Nat. Commun.}\ }\textbf {\bibinfo {volume} {3}},\
  \bibinfo {pages} {1037} (\bibinfo {year} {2012})}\BibitemShut {NoStop}%
\bibitem [{\citenamefont {Stanciu}\ \emph {et~al.}(2007)\citenamefont
  {Stanciu}, \citenamefont {Hansteen}, \citenamefont {Kimel}, \citenamefont
  {Kirilyuk}, \citenamefont {Tsukamoto}, \citenamefont {Itoh},\ and\
  \citenamefont {Rasing}}]{Stanciu_2007_PhysRevLett_99_4}%
  \BibitemOpen
  \bibfield  {author} {\bibinfo {author} {\bibfnamefont {C.~D.}\ \bibnamefont
  {Stanciu}}, \bibinfo {author} {\bibfnamefont {F.}~\bibnamefont {Hansteen}},
  \bibinfo {author} {\bibfnamefont {A.~V.}\ \bibnamefont {Kimel}}, \bibinfo
  {author} {\bibfnamefont {A.}~\bibnamefont {Kirilyuk}}, \bibinfo {author}
  {\bibfnamefont {A.}~\bibnamefont {Tsukamoto}}, \bibinfo {author}
  {\bibfnamefont {A.}~\bibnamefont {Itoh}}, \ and\ \bibinfo {author}
  {\bibfnamefont {T.}~\bibnamefont {Rasing}},\ }\href {\doibase
  10.1103/physrevlett.99.047601} {\bibfield  {journal} {\bibinfo  {journal}
  {Phys. Rev. Lett.}\ }\textbf {\bibinfo {volume} {99}},\ \bibinfo {pages}
  {047601} (\bibinfo {year} {2007})}\BibitemShut {NoStop}%
\bibitem [{\citenamefont {Stupakiewicz}\ \emph {et~al.}(2017)\citenamefont
  {Stupakiewicz}, \citenamefont {Szerenos}, \citenamefont {Afanasiev},
  \citenamefont {Kirilyuk},\ and\ \citenamefont
  {Kimel}}]{Stupakiewicz_2017_Nature_542_7639}%
  \BibitemOpen
  \bibfield  {author} {\bibinfo {author} {\bibfnamefont {A.}~\bibnamefont
  {Stupakiewicz}}, \bibinfo {author} {\bibfnamefont {K.}~\bibnamefont
  {Szerenos}}, \bibinfo {author} {\bibfnamefont {D.}~\bibnamefont {Afanasiev}},
  \bibinfo {author} {\bibfnamefont {A.}~\bibnamefont {Kirilyuk}}, \ and\
  \bibinfo {author} {\bibfnamefont {A.~V.}\ \bibnamefont {Kimel}},\ }\href
  {\doibase 10.1038/nature20807} {\bibfield  {journal} {\bibinfo  {journal}
  {Nature}\ }\textbf {\bibinfo {volume} {542}},\ \bibinfo {pages} {71}
  (\bibinfo {year} {2017})}\BibitemShut {NoStop}%
\bibitem [{\citenamefont {Schlauderer}\ \emph {et~al.}(2019)\citenamefont
  {Schlauderer}, \citenamefont {Lange}, \citenamefont {Baierl}, \citenamefont
  {Ebnet}, \citenamefont {Schmid}, \citenamefont {Valovcin}, \citenamefont
  {Zvezdin}, \citenamefont {Kimel}, \citenamefont {Mikhaylovskiy},\ and\
  \citenamefont {Huber}}]{Schlauderer_2019_Nature_569_7756}%
  \BibitemOpen
  \bibfield  {author} {\bibinfo {author} {\bibfnamefont {S.}~\bibnamefont
  {Schlauderer}}, \bibinfo {author} {\bibfnamefont {C.}~\bibnamefont {Lange}},
  \bibinfo {author} {\bibfnamefont {S.}~\bibnamefont {Baierl}}, \bibinfo
  {author} {\bibfnamefont {T.}~\bibnamefont {Ebnet}}, \bibinfo {author}
  {\bibfnamefont {C.~P.}\ \bibnamefont {Schmid}}, \bibinfo {author}
  {\bibfnamefont {D.~C.}\ \bibnamefont {Valovcin}}, \bibinfo {author}
  {\bibfnamefont {A.~K.}\ \bibnamefont {Zvezdin}}, \bibinfo {author}
  {\bibfnamefont {A.~V.}\ \bibnamefont {Kimel}}, \bibinfo {author}
  {\bibfnamefont {R.~V.}\ \bibnamefont {Mikhaylovskiy}}, \ and\ \bibinfo
  {author} {\bibfnamefont {R.}~\bibnamefont {Huber}},\ }\href {\doibase
  10.1038/s41586-019-1174-7} {\bibfield  {journal} {\bibinfo  {journal}
  {Nature}\ }\textbf {\bibinfo {volume} {569}},\ \bibinfo {pages} {383}
  (\bibinfo {year} {2019})}\BibitemShut {NoStop}%
\bibitem [{\citenamefont {Yang}\ \emph {et~al.}(2017)\citenamefont {Yang},
  \citenamefont {Wilson}, \citenamefont {Gorchon}, \citenamefont {Lambert},
  \citenamefont {Salahuddin},\ and\ \citenamefont
  {Bokor}}]{Yang_2017_SciAdv_3_11}%
  \BibitemOpen
  \bibfield  {author} {\bibinfo {author} {\bibfnamefont {Y.}~\bibnamefont
  {Yang}}, \bibinfo {author} {\bibfnamefont {R.~B.}\ \bibnamefont {Wilson}},
  \bibinfo {author} {\bibfnamefont {J.}~\bibnamefont {Gorchon}}, \bibinfo
  {author} {\bibfnamefont {C.-H.}\ \bibnamefont {Lambert}}, \bibinfo {author}
  {\bibfnamefont {S.}~\bibnamefont {Salahuddin}}, \ and\ \bibinfo {author}
  {\bibfnamefont {J.}~\bibnamefont {Bokor}},\ }\href {\doibase
  10.1126/sciadv.1603117} {\bibfield  {journal} {\bibinfo  {journal} {Sci.
  Adv.}\ }\textbf {\bibinfo {volume} {3}},\ \bibinfo {pages} {e1603117}
  (\bibinfo {year} {2017})}\BibitemShut {NoStop}%
\bibitem [{\citenamefont {van Hees}\ \emph {et~al.}(2020)\citenamefont {van
  Hees}, \citenamefont {van~de Meugheuvel}, \citenamefont {Koopmans},\ and\
  \citenamefont {Lavrijsen}}]{van_Hees_2020_NatCommun_11_1}%
  \BibitemOpen
  \bibfield  {author} {\bibinfo {author} {\bibfnamefont {Y.~L.~W.}\
  \bibnamefont {van Hees}}, \bibinfo {author} {\bibfnamefont {P.}~\bibnamefont
  {van~de Meugheuvel}}, \bibinfo {author} {\bibfnamefont {B.}~\bibnamefont
  {Koopmans}}, \ and\ \bibinfo {author} {\bibfnamefont {R.}~\bibnamefont
  {Lavrijsen}},\ }\href {\doibase 10.1038/s41467-020-17676-6} {\bibfield
  {journal} {\bibinfo  {journal} {Nat. Commun.}\ }\textbf {\bibinfo {volume}
  {11}},\ \bibinfo {pages} {3835} (\bibinfo {year} {2020})}\BibitemShut
  {NoStop}%
\bibitem [{\citenamefont {Banerjee}\ \emph {et~al.}(2020)\citenamefont
  {Banerjee}, \citenamefont {Teichert}, \citenamefont {Siewierska},
  \citenamefont {Gercsi}, \citenamefont {Atcheson}, \citenamefont {Stamenov},
  \citenamefont {Rode}, \citenamefont {Coey},\ and\ \citenamefont
  {Besbas}}]{Banerjee_2020_NatCommun_11_1}%
  \BibitemOpen
  \bibfield  {author} {\bibinfo {author} {\bibfnamefont {C.}~\bibnamefont
  {Banerjee}}, \bibinfo {author} {\bibfnamefont {N.}~\bibnamefont {Teichert}},
  \bibinfo {author} {\bibfnamefont {K.~E.}\ \bibnamefont {Siewierska}},
  \bibinfo {author} {\bibfnamefont {Z.}~\bibnamefont {Gercsi}}, \bibinfo
  {author} {\bibfnamefont {G.~Y.~P.}\ \bibnamefont {Atcheson}}, \bibinfo
  {author} {\bibfnamefont {P.}~\bibnamefont {Stamenov}}, \bibinfo {author}
  {\bibfnamefont {K.}~\bibnamefont {Rode}}, \bibinfo {author} {\bibfnamefont
  {J.~M.~D.}\ \bibnamefont {Coey}}, \ and\ \bibinfo {author} {\bibfnamefont
  {J.}~\bibnamefont {Besbas}},\ }\href {\doibase 10.1038/s41467-020-18340-9}
  {\bibfield  {journal} {\bibinfo  {journal} {Nat. Commun.}\ }\textbf {\bibinfo
  {volume} {11}},\ \bibinfo {pages} {4444} (\bibinfo {year}
  {2020})}\BibitemShut {NoStop}%
\bibitem [{\citenamefont {Le~Guyader}\ \emph {et~al.}(2012)\citenamefont
  {Le~Guyader}, \citenamefont {El~Moussaoui}, \citenamefont {Buzzi},
  \citenamefont {Chopdekar}, \citenamefont {Heyderman}, \citenamefont
  {Tsukamoto}, \citenamefont {Itoh}, \citenamefont {Kirilyuk}, \citenamefont
  {Rasing}, \citenamefont {Kimel},\ and\ \citenamefont
  {Nolting}}]{Le_Guyader_2012_ApplPhysLett_101_2}%
  \BibitemOpen
  \bibfield  {author} {\bibinfo {author} {\bibfnamefont {L.}~\bibnamefont
  {Le~Guyader}}, \bibinfo {author} {\bibfnamefont {S.}~\bibnamefont
  {El~Moussaoui}}, \bibinfo {author} {\bibfnamefont {M.}~\bibnamefont {Buzzi}},
  \bibinfo {author} {\bibfnamefont {R.~V.}\ \bibnamefont {Chopdekar}}, \bibinfo
  {author} {\bibfnamefont {L.~J.}\ \bibnamefont {Heyderman}}, \bibinfo {author}
  {\bibfnamefont {A.}~\bibnamefont {Tsukamoto}}, \bibinfo {author}
  {\bibfnamefont {A.}~\bibnamefont {Itoh}}, \bibinfo {author} {\bibfnamefont
  {A.}~\bibnamefont {Kirilyuk}}, \bibinfo {author} {\bibfnamefont
  {T.}~\bibnamefont {Rasing}}, \bibinfo {author} {\bibfnamefont {A.~V.}\
  \bibnamefont {Kimel}}, \ and\ \bibinfo {author} {\bibfnamefont
  {F.}~\bibnamefont {Nolting}},\ }\href {\doibase 10.1063/1.4733965} {\bibfield
   {journal} {\bibinfo  {journal} {Appl. Phys. Lett.}\ }\textbf {\bibinfo
  {volume} {101}},\ \bibinfo {pages} {022410} (\bibinfo {year}
  {2012})}\BibitemShut {NoStop}%
\bibitem [{\citenamefont {Graves}\ \emph {et~al.}(2013)\citenamefont {Graves},
  \citenamefont {Reid}, \citenamefont {Wang}, \citenamefont {Wu}, \citenamefont
  {de~Jong}, \citenamefont {Vahaplar}, \citenamefont {Radu}, \citenamefont
  {Bernstein}, \citenamefont {Messerschmidt}, \citenamefont {M\"{u}ller},
  \citenamefont {Coffee}, \citenamefont {Bionta}, \citenamefont {Epp},
  \citenamefont {Hartmann}, \citenamefont {Kimmel}, \citenamefont {Hauser},
  \citenamefont {Hartmann}, \citenamefont {Holl}, \citenamefont {Gorke},
  \citenamefont {Mentink}, \citenamefont {Tsukamoto}, \citenamefont {Fognini},
  \citenamefont {Turner}, \citenamefont {Schlotter}, \citenamefont {Rolles},
  \citenamefont {Soltau}, \citenamefont {Str\"{u}der}, \citenamefont
  {Acremann}, \citenamefont {Kimel}, \citenamefont {Kirilyuk}, \citenamefont
  {Rasing}, \citenamefont {St\"{o}hr}, \citenamefont {Scherz},\ and\
  \citenamefont {D\"{u}rr}}]{Graves_2013_NatMater_12_4}%
  \BibitemOpen
  \bibfield  {author} {\bibinfo {author} {\bibfnamefont {C.~E.}\ \bibnamefont
  {Graves}}, \bibinfo {author} {\bibfnamefont {A.~H.}\ \bibnamefont {Reid}},
  \bibinfo {author} {\bibfnamefont {T.}~\bibnamefont {Wang}}, \bibinfo {author}
  {\bibfnamefont {B.}~\bibnamefont {Wu}}, \bibinfo {author} {\bibfnamefont
  {S.}~\bibnamefont {de~Jong}}, \bibinfo {author} {\bibfnamefont
  {K.}~\bibnamefont {Vahaplar}}, \bibinfo {author} {\bibfnamefont
  {I.}~\bibnamefont {Radu}}, \bibinfo {author} {\bibfnamefont {D.~P.}\
  \bibnamefont {Bernstein}}, \bibinfo {author} {\bibfnamefont {M.}~\bibnamefont
  {Messerschmidt}}, \bibinfo {author} {\bibfnamefont {L.}~\bibnamefont
  {M\"{u}ller}}, \bibinfo {author} {\bibfnamefont {R.}~\bibnamefont {Coffee}},
  \bibinfo {author} {\bibfnamefont {M.}~\bibnamefont {Bionta}}, \bibinfo
  {author} {\bibfnamefont {S.~W.}\ \bibnamefont {Epp}}, \bibinfo {author}
  {\bibfnamefont {R.}~\bibnamefont {Hartmann}}, \bibinfo {author}
  {\bibfnamefont {N.}~\bibnamefont {Kimmel}}, \bibinfo {author} {\bibfnamefont
  {G.}~\bibnamefont {Hauser}}, \bibinfo {author} {\bibfnamefont
  {A.}~\bibnamefont {Hartmann}}, \bibinfo {author} {\bibfnamefont
  {P.}~\bibnamefont {Holl}}, \bibinfo {author} {\bibfnamefont {H.}~\bibnamefont
  {Gorke}}, \bibinfo {author} {\bibfnamefont {J.~H.}\ \bibnamefont {Mentink}},
  \bibinfo {author} {\bibfnamefont {A.}~\bibnamefont {Tsukamoto}}, \bibinfo
  {author} {\bibfnamefont {A.}~\bibnamefont {Fognini}}, \bibinfo {author}
  {\bibfnamefont {J.~J.}\ \bibnamefont {Turner}}, \bibinfo {author}
  {\bibfnamefont {W.~F.}\ \bibnamefont {Schlotter}}, \bibinfo {author}
  {\bibfnamefont {D.}~\bibnamefont {Rolles}}, \bibinfo {author} {\bibfnamefont
  {H.}~\bibnamefont {Soltau}}, \bibinfo {author} {\bibfnamefont
  {L.}~\bibnamefont {Str\"{u}der}}, \bibinfo {author} {\bibfnamefont
  {Y.}~\bibnamefont {Acremann}}, \bibinfo {author} {\bibfnamefont {A.~V.}\
  \bibnamefont {Kimel}}, \bibinfo {author} {\bibfnamefont {A.}~\bibnamefont
  {Kirilyuk}}, \bibinfo {author} {\bibfnamefont {T.}~\bibnamefont {Rasing}},
  \bibinfo {author} {\bibfnamefont {J.}~\bibnamefont {St\"{o}hr}}, \bibinfo
  {author} {\bibfnamefont {A.~O.}\ \bibnamefont {Scherz}}, \ and\ \bibinfo
  {author} {\bibfnamefont {H.~A.}\ \bibnamefont {D\"{u}rr}},\ }\href {\doibase
  10.1038/nmat3597} {\bibfield  {journal} {\bibinfo  {journal} {Nat. Mater.}\
  }\textbf {\bibinfo {volume} {12}},\ \bibinfo {pages} {293} (\bibinfo {year}
  {2013})}\BibitemShut {NoStop}%
\bibitem [{\citenamefont {Baryakhtar}\ and\ \citenamefont
  {Danilevich}(2013)}]{Baryakhtar_2013_LowTempPhys_39_12}%
  \BibitemOpen
  \bibfield  {author} {\bibinfo {author} {\bibfnamefont {V.~G.}\ \bibnamefont
  {Baryakhtar}}\ and\ \bibinfo {author} {\bibfnamefont {A.~G.}\ \bibnamefont
  {Danilevich}},\ }\href {\doibase 10.1063/1.4843275} {\bibfield  {journal}
  {\bibinfo  {journal} {Low Temp. Phys.}\ }\textbf {\bibinfo {volume} {39}},\
  \bibinfo {pages} {993} (\bibinfo {year} {2013})}\BibitemShut {NoStop}%
\bibitem [{\citenamefont {Atxitia}\ \emph {et~al.}(2014)\citenamefont
  {Atxitia}, \citenamefont {Barker}, \citenamefont {Chantrell},\ and\
  \citenamefont {Chubykalo-Fesenko}}]{Atxitia_2014_PhysRevB_89_22}%
  \BibitemOpen
  \bibfield  {author} {\bibinfo {author} {\bibfnamefont {U.}~\bibnamefont
  {Atxitia}}, \bibinfo {author} {\bibfnamefont {J.}~\bibnamefont {Barker}},
  \bibinfo {author} {\bibfnamefont {R.~W.}\ \bibnamefont {Chantrell}}, \ and\
  \bibinfo {author} {\bibfnamefont {O.}~\bibnamefont {Chubykalo-Fesenko}},\
  }\href {\doibase 10.1103/physrevb.89.224421} {\bibfield  {journal} {\bibinfo
  {journal} {Phys. Rev. B}\ }\textbf {\bibinfo {volume} {89}},\ \bibinfo
  {pages} {224421} (\bibinfo {year} {2014})}\BibitemShut {NoStop}%
\bibitem [{\citenamefont
  {Gridnev}(2016)}]{Gridnev_2016_JPhysCondensMatter_28_47}%
  \BibitemOpen
  \bibfield  {author} {\bibinfo {author} {\bibfnamefont {V.~N.}\ \bibnamefont
  {Gridnev}},\ }\href {\doibase 10.1088/0953-8984/28/47/476007} {\bibfield
  {journal} {\bibinfo  {journal} {J. Phys.: Condens. Matter}\ }\textbf
  {\bibinfo {volume} {28}},\ \bibinfo {pages} {476007} (\bibinfo {year}
  {2016})}\BibitemShut {NoStop}%
\bibitem [{\citenamefont {Schellekens}\ and\ \citenamefont
  {Koopmans}(2013)}]{Schellekens_2013_PhysRevB_87_2}%
  \BibitemOpen
  \bibfield  {author} {\bibinfo {author} {\bibfnamefont {A.~J.}\ \bibnamefont
  {Schellekens}}\ and\ \bibinfo {author} {\bibfnamefont {B.}~\bibnamefont
  {Koopmans}},\ }\href {\doibase 10.1103/physrevb.87.020407} {\bibfield
  {journal} {\bibinfo  {journal} {Phys. Rev. B}\ }\textbf {\bibinfo {volume}
  {87}},\ \bibinfo {pages} {020407} (\bibinfo {year} {2013})}\BibitemShut
  {NoStop}%
\bibitem [{\citenamefont {Mangin}\ \emph {et~al.}(2014)\citenamefont {Mangin},
  \citenamefont {Gottwald}, \citenamefont {Lambert}, \citenamefont {Steil},
  \citenamefont {Uhl\'{i}\v{r}}, \citenamefont {Pang}, \citenamefont {Hehn},
  \citenamefont {Alebrand}, \citenamefont {Cinchetti}, \citenamefont
  {Malinowski}, \citenamefont {Fainman}, \citenamefont {Aeschlimann},\ and\
  \citenamefont {Fullerton}}]{Mangin_2014_NatMater_13_3}%
  \BibitemOpen
  \bibfield  {author} {\bibinfo {author} {\bibfnamefont {S.}~\bibnamefont
  {Mangin}}, \bibinfo {author} {\bibfnamefont {M.}~\bibnamefont {Gottwald}},
  \bibinfo {author} {\bibfnamefont {C.-H.}\ \bibnamefont {Lambert}}, \bibinfo
  {author} {\bibfnamefont {D.}~\bibnamefont {Steil}}, \bibinfo {author}
  {\bibfnamefont {V.}~\bibnamefont {Uhl\'{i}\v{r}}}, \bibinfo {author}
  {\bibfnamefont {L.}~\bibnamefont {Pang}}, \bibinfo {author} {\bibfnamefont
  {M.}~\bibnamefont {Hehn}}, \bibinfo {author} {\bibfnamefont {S.}~\bibnamefont
  {Alebrand}}, \bibinfo {author} {\bibfnamefont {M.}~\bibnamefont {Cinchetti}},
  \bibinfo {author} {\bibfnamefont {G.}~\bibnamefont {Malinowski}}, \bibinfo
  {author} {\bibfnamefont {Y.}~\bibnamefont {Fainman}}, \bibinfo {author}
  {\bibfnamefont {M.}~\bibnamefont {Aeschlimann}}, \ and\ \bibinfo {author}
  {\bibfnamefont {E.~E.}\ \bibnamefont {Fullerton}},\ }\href {\doibase
  10.1038/nmat3864} {\bibfield  {journal} {\bibinfo  {journal} {Nat. Mater.}\
  }\textbf {\bibinfo {volume} {13}},\ \bibinfo {pages} {286} (\bibinfo {year}
  {2014})}\BibitemShut {NoStop}%
\bibitem [{\citenamefont
  {Mentink}(2017)}]{Mentink_2017_JPhysCondensMatter_29_45}%
  \BibitemOpen
  \bibfield  {author} {\bibinfo {author} {\bibfnamefont {J.~H.}\ \bibnamefont
  {Mentink}},\ }\href {\doibase 10.1088/1361-648x/aa8abf} {\bibfield  {journal}
  {\bibinfo  {journal} {J. Phys.: Condens. Matter}\ }\textbf {\bibinfo {volume}
  {29}},\ \bibinfo {pages} {453001} (\bibinfo {year} {2017})}\BibitemShut
  {NoStop}%
\bibitem [{\citenamefont {Wienholdt}\ \emph {et~al.}(2013)\citenamefont
  {Wienholdt}, \citenamefont {Hinzke}, \citenamefont {Carva}, \citenamefont
  {Oppeneer},\ and\ \citenamefont {Nowak}}]{Wienholdt_2013_PhysRevB_88_2}%
  \BibitemOpen
  \bibfield  {author} {\bibinfo {author} {\bibfnamefont {S.}~\bibnamefont
  {Wienholdt}}, \bibinfo {author} {\bibfnamefont {D.}~\bibnamefont {Hinzke}},
  \bibinfo {author} {\bibfnamefont {K.}~\bibnamefont {Carva}}, \bibinfo
  {author} {\bibfnamefont {P.~M.}\ \bibnamefont {Oppeneer}}, \ and\ \bibinfo
  {author} {\bibfnamefont {U.}~\bibnamefont {Nowak}},\ }\href {\doibase
  10.1103/physrevb.88.020406} {\bibfield  {journal} {\bibinfo  {journal} {Phys.
  Rev. B}\ }\textbf {\bibinfo {volume} {88}},\ \bibinfo {pages} {020406}
  (\bibinfo {year} {2013})}\BibitemShut {NoStop}%
\bibitem [{\citenamefont {Gerlach}\ \emph {et~al.}(2017)\citenamefont
  {Gerlach}, \citenamefont {Oroszlany}, \citenamefont {Hinzke}, \citenamefont
  {Sievering}, \citenamefont {Wienholdt}, \citenamefont {Szunyogh},\ and\
  \citenamefont {Nowak}}]{Gerlach_2017_PhysRevB_95_22}%
  \BibitemOpen
  \bibfield  {author} {\bibinfo {author} {\bibfnamefont {S.}~\bibnamefont
  {Gerlach}}, \bibinfo {author} {\bibfnamefont {L.}~\bibnamefont {Oroszlany}},
  \bibinfo {author} {\bibfnamefont {D.}~\bibnamefont {Hinzke}}, \bibinfo
  {author} {\bibfnamefont {S.}~\bibnamefont {Sievering}}, \bibinfo {author}
  {\bibfnamefont {S.}~\bibnamefont {Wienholdt}}, \bibinfo {author}
  {\bibfnamefont {L.}~\bibnamefont {Szunyogh}}, \ and\ \bibinfo {author}
  {\bibfnamefont {U.}~\bibnamefont {Nowak}},\ }\href {\doibase
  10.1103/physrevb.95.224435} {\bibfield  {journal} {\bibinfo  {journal} {Phys.
  Rev. B}\ }\textbf {\bibinfo {volume} {95}},\ \bibinfo {pages} {224435}
  (\bibinfo {year} {2017})}\BibitemShut {NoStop}%
\bibitem [{\citenamefont {Atxitia}\ \emph {et~al.}(2013)\citenamefont
  {Atxitia}, \citenamefont {Ostler}, \citenamefont {Barker}, \citenamefont
  {Evans}, \citenamefont {Chantrell},\ and\ \citenamefont
  {Chubykalo-Fesenko}}]{Atxitia_2013_PhysRevB_87_22}%
  \BibitemOpen
  \bibfield  {author} {\bibinfo {author} {\bibfnamefont {U.}~\bibnamefont
  {Atxitia}}, \bibinfo {author} {\bibfnamefont {T.}~\bibnamefont {Ostler}},
  \bibinfo {author} {\bibfnamefont {J.}~\bibnamefont {Barker}}, \bibinfo
  {author} {\bibfnamefont {R.~F.~L.}\ \bibnamefont {Evans}}, \bibinfo {author}
  {\bibfnamefont {R.~W.}\ \bibnamefont {Chantrell}}, \ and\ \bibinfo {author}
  {\bibfnamefont {O.}~\bibnamefont {Chubykalo-Fesenko}},\ }\href {\doibase
  10.1103/physrevb.87.224417} {\bibfield  {journal} {\bibinfo  {journal} {Phys.
  Rev. B}\ }\textbf {\bibinfo {volume} {87}},\ \bibinfo {pages} {224417}
  (\bibinfo {year} {2013})}\BibitemShut {NoStop}%
\bibitem [{\citenamefont {Kaganov}\ \emph {et~al.}(1957)\citenamefont
  {Kaganov}, \citenamefont {Lifshitz},\ and\ \citenamefont
  {Tanatarov}}]{Kaganov1957}%
  \BibitemOpen
  \bibfield  {author} {\bibinfo {author} {\bibfnamefont {M.~I.}\ \bibnamefont
  {Kaganov}}, \bibinfo {author} {\bibfnamefont {I.~M.}\ \bibnamefont
  {Lifshitz}}, \ and\ \bibinfo {author} {\bibfnamefont {L.~V.}\ \bibnamefont
  {Tanatarov}},\ }\href@noop {} {\bibfield  {journal} {\bibinfo  {journal}
  {JETP}\ }\textbf {\bibinfo {volume} {173}} (\bibinfo {year}
  {1957})}\BibitemShut {NoStop}%
\bibitem [{\citenamefont {Chen}\ \emph {et~al.}(2006)\citenamefont {Chen},
  \citenamefont {Tzou},\ and\ \citenamefont
  {Beraun}}]{Chen_2006_IntJHeatMassTransf_49_1-2}%
  \BibitemOpen
  \bibfield  {author} {\bibinfo {author} {\bibfnamefont {J.}~\bibnamefont
  {Chen}}, \bibinfo {author} {\bibfnamefont {D.}~\bibnamefont {Tzou}}, \ and\
  \bibinfo {author} {\bibfnamefont {J.}~\bibnamefont {Beraun}},\ }\href
  {\doibase 10.1016/j.ijheatmasstransfer.2005.06.022} {\bibfield  {journal}
  {\bibinfo  {journal} {Int. J. Heat Mass Transf.}\ }\textbf {\bibinfo {volume}
  {49}},\ \bibinfo {pages} {307} (\bibinfo {year} {2006})}\BibitemShut
  {NoStop}%
\bibitem [{\citenamefont {El-Ghazaly}\ \emph {et~al.}(2019)\citenamefont
  {El-Ghazaly}, \citenamefont {Tran}, \citenamefont {Ceballos}, \citenamefont
  {Lambert}, \citenamefont {Pattabi}, \citenamefont {Salahuddin}, \citenamefont
  {Hellman},\ and\ \citenamefont
  {Bokor}}]{El-Ghazaly_2019_ApplPhysLett_114_23}%
  \BibitemOpen
  \bibfield  {author} {\bibinfo {author} {\bibfnamefont {A.}~\bibnamefont
  {El-Ghazaly}}, \bibinfo {author} {\bibfnamefont {B.}~\bibnamefont {Tran}},
  \bibinfo {author} {\bibfnamefont {A.}~\bibnamefont {Ceballos}}, \bibinfo
  {author} {\bibfnamefont {C.-H.}\ \bibnamefont {Lambert}}, \bibinfo {author}
  {\bibfnamefont {A.}~\bibnamefont {Pattabi}}, \bibinfo {author} {\bibfnamefont
  {S.}~\bibnamefont {Salahuddin}}, \bibinfo {author} {\bibfnamefont
  {F.}~\bibnamefont {Hellman}}, \ and\ \bibinfo {author} {\bibfnamefont
  {J.}~\bibnamefont {Bokor}},\ }\href {\doibase 10.1063/1.5098453} {\bibfield
  {journal} {\bibinfo  {journal} {Appl. Phys. Lett.}\ }\textbf {\bibinfo
  {volume} {114}},\ \bibinfo {pages} {232407} (\bibinfo {year}
  {2019})}\BibitemShut {NoStop}%
\bibitem [{\citenamefont {Gorchon}\ \emph {et~al.}(2016)\citenamefont
  {Gorchon}, \citenamefont {Wilson}, \citenamefont {Yang}, \citenamefont
  {Pattabi}, \citenamefont {Chen}, \citenamefont {He}, \citenamefont {Wang},
  \citenamefont {Li},\ and\ \citenamefont
  {Bokor}}]{Gorchon_2016_PhysRevB_94_18}%
  \BibitemOpen
  \bibfield  {author} {\bibinfo {author} {\bibfnamefont {J.}~\bibnamefont
  {Gorchon}}, \bibinfo {author} {\bibfnamefont {R.~B.}\ \bibnamefont {Wilson}},
  \bibinfo {author} {\bibfnamefont {Y.}~\bibnamefont {Yang}}, \bibinfo {author}
  {\bibfnamefont {A.}~\bibnamefont {Pattabi}}, \bibinfo {author} {\bibfnamefont
  {J.~Y.}\ \bibnamefont {Chen}}, \bibinfo {author} {\bibfnamefont
  {L.}~\bibnamefont {He}}, \bibinfo {author} {\bibfnamefont {J.~P.}\
  \bibnamefont {Wang}}, \bibinfo {author} {\bibfnamefont {M.}~\bibnamefont
  {Li}}, \ and\ \bibinfo {author} {\bibfnamefont {J.}~\bibnamefont {Bokor}},\
  }\href {\doibase 10.1103/physrevb.94.184406} {\bibfield  {journal} {\bibinfo
  {journal} {Phys. Rev. B}\ }\textbf {\bibinfo {volume} {94}},\ \bibinfo
  {pages} {184406} (\bibinfo {year} {2016})}\BibitemShut {NoStop}%
\bibitem [{\citenamefont {Jakobs}\ \emph {et~al.}(2020)\citenamefont {Jakobs},
  \citenamefont {Ostler}, \citenamefont {Lambert}, \citenamefont {Yang},
  \citenamefont {Wilson}, \citenamefont {Gorchon}, \citenamefont {Bokor},\ and\
  \citenamefont {Atxitia}}]{Jakobs_arxiv_2020}%
  \BibitemOpen
  \bibfield  {author} {\bibinfo {author} {\bibfnamefont {F.}~\bibnamefont
  {Jakobs}}, \bibinfo {author} {\bibfnamefont {T.}~\bibnamefont {Ostler}},
  \bibinfo {author} {\bibfnamefont {C.-H.}\ \bibnamefont {Lambert}}, \bibinfo
  {author} {\bibfnamefont {S.}~\bibnamefont {Yang}, \bibfnamefont
  {Yang~Salahuddin}}, \bibinfo {author} {\bibfnamefont {R.~B.}\ \bibnamefont
  {Wilson}}, \bibinfo {author} {\bibfnamefont {J.}~\bibnamefont {Gorchon}},
  \bibinfo {author} {\bibfnamefont {J.}~\bibnamefont {Bokor}}, \ and\ \bibinfo
  {author} {\bibfnamefont {U.}~\bibnamefont {Atxitia}},\ }\href
  {https://arxiv.org/abs/2004.14844} {\bibfield  {journal} {\bibinfo  {journal}
  {arXiv:2004.14844}\ } (\bibinfo {year} {2020})}\BibitemShut {NoStop}%
\bibitem [{\citenamefont {Liao}\ \emph {et~al.}(2019)\citenamefont {Liao},
  \citenamefont {Vallobra}, \citenamefont {O'Brien}, \citenamefont {Atxitia},
  \citenamefont {Raposo}, \citenamefont {Petit}, \citenamefont {Vemulkar},
  \citenamefont {Malinowski}, \citenamefont {Hehn}, \citenamefont
  {Mart\'{i}nez}, \citenamefont {Mangin},\ and\ \citenamefont
  {Cowburn}}]{Liao_2019_AdvSci_6_24}%
  \BibitemOpen
  \bibfield  {author} {\bibinfo {author} {\bibfnamefont {J.-W.}\ \bibnamefont
  {Liao}}, \bibinfo {author} {\bibfnamefont {P.}~\bibnamefont {Vallobra}},
  \bibinfo {author} {\bibfnamefont {L.}~\bibnamefont {O'Brien}}, \bibinfo
  {author} {\bibfnamefont {U.}~\bibnamefont {Atxitia}}, \bibinfo {author}
  {\bibfnamefont {V.}~\bibnamefont {Raposo}}, \bibinfo {author} {\bibfnamefont
  {D.}~\bibnamefont {Petit}}, \bibinfo {author} {\bibfnamefont
  {T.}~\bibnamefont {Vemulkar}}, \bibinfo {author} {\bibfnamefont
  {G.}~\bibnamefont {Malinowski}}, \bibinfo {author} {\bibfnamefont
  {M.}~\bibnamefont {Hehn}}, \bibinfo {author} {\bibfnamefont {E.}~\bibnamefont
  {Mart\'{i}nez}}, \bibinfo {author} {\bibfnamefont {S.}~\bibnamefont
  {Mangin}}, \ and\ \bibinfo {author} {\bibfnamefont {R.~P.}\ \bibnamefont
  {Cowburn}},\ }\href {\doibase 10.1002/advs.201901876} {\bibfield  {journal}
  {\bibinfo  {journal} {Adv. Sci.}\ }\textbf {\bibinfo {volume} {6}},\ \bibinfo
  {pages} {1901876} (\bibinfo {year} {2019})}\BibitemShut {NoStop}%
\bibitem [{\citenamefont {Uchida}\ \emph {et~al.}(2008)\citenamefont {Uchida},
  \citenamefont {Takahashi}, \citenamefont {Harii}, \citenamefont {Ieda},
  \citenamefont {Koshibae}, \citenamefont {Ando}, \citenamefont {Maekawa},\
  and\ \citenamefont {Saitoh}}]{Uchida_2008_Nature_455_7214}%
  \BibitemOpen
  \bibfield  {author} {\bibinfo {author} {\bibfnamefont {K.}~\bibnamefont
  {Uchida}}, \bibinfo {author} {\bibfnamefont {S.}~\bibnamefont {Takahashi}},
  \bibinfo {author} {\bibfnamefont {K.}~\bibnamefont {Harii}}, \bibinfo
  {author} {\bibfnamefont {J.}~\bibnamefont {Ieda}}, \bibinfo {author}
  {\bibfnamefont {W.}~\bibnamefont {Koshibae}}, \bibinfo {author}
  {\bibfnamefont {K.}~\bibnamefont {Ando}}, \bibinfo {author} {\bibfnamefont
  {S.}~\bibnamefont {Maekawa}}, \ and\ \bibinfo {author} {\bibfnamefont
  {E.}~\bibnamefont {Saitoh}},\ }\href {\doibase 10.1038/nature07321}
  {\bibfield  {journal} {\bibinfo  {journal} {Nature}\ }\textbf {\bibinfo
  {volume} {455}},\ \bibinfo {pages} {778} (\bibinfo {year}
  {2008})}\BibitemShut {NoStop}%
\bibitem [{\citenamefont {Uchida}\ \emph {et~al.}(2010)\citenamefont {Uchida},
  \citenamefont {Xiao}, \citenamefont {Adachi}, \citenamefont {Ohe},
  \citenamefont {Takahashi}, \citenamefont {Ieda}, \citenamefont {Ota},
  \citenamefont {Kajiwara}, \citenamefont {Umezawa}, \citenamefont {Kawai},
  \citenamefont {Bauer}, \citenamefont {Maekawa},\ and\ \citenamefont
  {Saitoh}}]{Uchida_2010_NatMater_9_11}%
  \BibitemOpen
  \bibfield  {author} {\bibinfo {author} {\bibfnamefont {K.}~\bibnamefont
  {Uchida}}, \bibinfo {author} {\bibfnamefont {J.}~\bibnamefont {Xiao}},
  \bibinfo {author} {\bibfnamefont {H.}~\bibnamefont {Adachi}}, \bibinfo
  {author} {\bibfnamefont {J.}~\bibnamefont {Ohe}}, \bibinfo {author}
  {\bibfnamefont {S.}~\bibnamefont {Takahashi}}, \bibinfo {author}
  {\bibfnamefont {J.}~\bibnamefont {Ieda}}, \bibinfo {author} {\bibfnamefont
  {T.}~\bibnamefont {Ota}}, \bibinfo {author} {\bibfnamefont {Y.}~\bibnamefont
  {Kajiwara}}, \bibinfo {author} {\bibfnamefont {H.}~\bibnamefont {Umezawa}},
  \bibinfo {author} {\bibfnamefont {H.}~\bibnamefont {Kawai}}, \bibinfo
  {author} {\bibfnamefont {G.~E.~W.}\ \bibnamefont {Bauer}}, \bibinfo {author}
  {\bibfnamefont {S.}~\bibnamefont {Maekawa}}, \ and\ \bibinfo {author}
  {\bibfnamefont {E.}~\bibnamefont {Saitoh}},\ }\href {\doibase
  10.1038/nmat2856} {\bibfield  {journal} {\bibinfo  {journal} {Nat. Mater.}\
  }\textbf {\bibinfo {volume} {9}},\ \bibinfo {pages} {894} (\bibinfo {year}
  {2010})}\BibitemShut {NoStop}%
\bibitem [{\citenamefont {Xiao}\ \emph {et~al.}(2010)\citenamefont {Xiao},
  \citenamefont {Bauer}, \citenamefont {Uchida}, \citenamefont {Saitoh},\ and\
  \citenamefont {Maekawa}}]{Xiao_2010_PhysRevB_81_21}%
  \BibitemOpen
  \bibfield  {author} {\bibinfo {author} {\bibfnamefont {J.}~\bibnamefont
  {Xiao}}, \bibinfo {author} {\bibfnamefont {G.~E.~W.}\ \bibnamefont {Bauer}},
  \bibinfo {author} {\bibfnamefont {K.-c.}\ \bibnamefont {Uchida}}, \bibinfo
  {author} {\bibfnamefont {E.}~\bibnamefont {Saitoh}}, \ and\ \bibinfo {author}
  {\bibfnamefont {S.}~\bibnamefont {Maekawa}},\ }\href {\doibase
  10.1103/physrevb.81.214418} {\bibfield  {journal} {\bibinfo  {journal} {Phys.
  Rev. B}\ }\textbf {\bibinfo {volume} {81}},\ \bibinfo {pages} {214418}
  (\bibinfo {year} {2010})}\BibitemShut {NoStop}%
\bibitem [{\citenamefont {Flipse}\ \emph {et~al.}(2014)\citenamefont {Flipse},
  \citenamefont {Dejene}, \citenamefont {Wagenaar}, \citenamefont {Bauer},
  \citenamefont {Youssef},\ and\ \citenamefont {van
  Wees}}]{Flipse_2014_PhysRevLett_113_2}%
  \BibitemOpen
  \bibfield  {author} {\bibinfo {author} {\bibfnamefont {J.}~\bibnamefont
  {Flipse}}, \bibinfo {author} {\bibfnamefont {F.~K.}\ \bibnamefont {Dejene}},
  \bibinfo {author} {\bibfnamefont {D.}~\bibnamefont {Wagenaar}}, \bibinfo
  {author} {\bibfnamefont {G.~E.~W.}\ \bibnamefont {Bauer}}, \bibinfo {author}
  {\bibfnamefont {J.~B.}\ \bibnamefont {Youssef}}, \ and\ \bibinfo {author}
  {\bibfnamefont {B.~J.}\ \bibnamefont {van Wees}},\ }\href {\doibase
  10.1103/physrevlett.113.027601} {\bibfield  {journal} {\bibinfo  {journal}
  {Phys. Rev. Lett.}\ }\textbf {\bibinfo {volume} {113}},\ \bibinfo {pages}
  {027601} (\bibinfo {year} {2014})}\BibitemShut {NoStop}%
\bibitem [{\citenamefont {Daimon}\ \emph {et~al.}(2016)\citenamefont {Daimon},
  \citenamefont {Iguchi}, \citenamefont {Hioki}, \citenamefont {Saitoh},\ and\
  \citenamefont {Uchida}}]{Daimon_2016_NatCommun_7_1}%
  \BibitemOpen
  \bibfield  {author} {\bibinfo {author} {\bibfnamefont {S.}~\bibnamefont
  {Daimon}}, \bibinfo {author} {\bibfnamefont {R.}~\bibnamefont {Iguchi}},
  \bibinfo {author} {\bibfnamefont {T.}~\bibnamefont {Hioki}}, \bibinfo
  {author} {\bibfnamefont {E.}~\bibnamefont {Saitoh}}, \ and\ \bibinfo {author}
  {\bibfnamefont {K.-i.}\ \bibnamefont {Uchida}},\ }\href {\doibase
  10.1038/ncomms13754} {\bibfield  {journal} {\bibinfo  {journal} {Nat.
  Commun.}\ }\textbf {\bibinfo {volume} {7}},\ \bibinfo {pages} {13754}
  (\bibinfo {year} {2016})}\BibitemShut {NoStop}%
\bibitem [{\citenamefont {Meyer}\ \emph {et~al.}(2017)\citenamefont {Meyer},
  \citenamefont {Chen}, \citenamefont {Wimmer}, \citenamefont {Althammer},
  \citenamefont {Wimmer}, \citenamefont {Schlitz}, \citenamefont {Gepr\"{a}gs},
  \citenamefont {Huebl}, \citenamefont {K\"{o}dderitzsch}, \citenamefont
  {Ebert}, \citenamefont {Bauer}, \citenamefont {Gross},\ and\ \citenamefont
  {Goennenwein}}]{Meyer_2017_NatMater_16_10}%
  \BibitemOpen
  \bibfield  {author} {\bibinfo {author} {\bibfnamefont {S.}~\bibnamefont
  {Meyer}}, \bibinfo {author} {\bibfnamefont {Y.-T.}\ \bibnamefont {Chen}},
  \bibinfo {author} {\bibfnamefont {S.}~\bibnamefont {Wimmer}}, \bibinfo
  {author} {\bibfnamefont {M.}~\bibnamefont {Althammer}}, \bibinfo {author}
  {\bibfnamefont {T.}~\bibnamefont {Wimmer}}, \bibinfo {author} {\bibfnamefont
  {R.}~\bibnamefont {Schlitz}}, \bibinfo {author} {\bibfnamefont
  {S.}~\bibnamefont {Gepr\"{a}gs}}, \bibinfo {author} {\bibfnamefont
  {H.}~\bibnamefont {Huebl}}, \bibinfo {author} {\bibfnamefont
  {D.}~\bibnamefont {K\"{o}dderitzsch}}, \bibinfo {author} {\bibfnamefont
  {H.}~\bibnamefont {Ebert}}, \bibinfo {author} {\bibfnamefont {G.~E.~W.}\
  \bibnamefont {Bauer}}, \bibinfo {author} {\bibfnamefont {R.}~\bibnamefont
  {Gross}}, \ and\ \bibinfo {author} {\bibfnamefont {S.~T.~B.}\ \bibnamefont
  {Goennenwein}},\ }\href {\doibase 10.1038/nmat4964} {\bibfield  {journal}
  {\bibinfo  {journal} {Nat. Mater.}\ }\textbf {\bibinfo {volume} {16}},\
  \bibinfo {pages} {977} (\bibinfo {year} {2017})}\BibitemShut {NoStop}%
\bibitem [{\citenamefont {Ritzmann}\ \emph {et~al.}(2017)\citenamefont
  {Ritzmann}, \citenamefont {Hinzke},\ and\ \citenamefont
  {Nowak}}]{Ritzmann_2017_PhysRevB_95_5}%
  \BibitemOpen
  \bibfield  {author} {\bibinfo {author} {\bibfnamefont {U.}~\bibnamefont
  {Ritzmann}}, \bibinfo {author} {\bibfnamefont {D.}~\bibnamefont {Hinzke}}, \
  and\ \bibinfo {author} {\bibfnamefont {U.}~\bibnamefont {Nowak}},\ }\href
  {\doibase 10.1103/physrevb.95.054411} {\bibfield  {journal} {\bibinfo
  {journal} {Phys. Rev. B}\ }\textbf {\bibinfo {volume} {95}},\ \bibinfo
  {pages} {054411} (\bibinfo {year} {2017})}\BibitemShut {NoStop}%
\bibitem [{\citenamefont {Barker}\ and\ \citenamefont
  {Tretiakov}(2016)}]{Barker_2016_PhysRevLett_116_14}%
  \BibitemOpen
  \bibfield  {author} {\bibinfo {author} {\bibfnamefont {J.}~\bibnamefont
  {Barker}}\ and\ \bibinfo {author} {\bibfnamefont {O.~A.}\ \bibnamefont
  {Tretiakov}},\ }\href {\doibase 10.1103/physrevlett.116.147203} {\bibfield
  {journal} {\bibinfo  {journal} {Phys. Rev. Lett.}\ }\textbf {\bibinfo
  {volume} {116}},\ \bibinfo {pages} {147203} (\bibinfo {year}
  {2016})}\BibitemShut {NoStop}%
\bibitem [{\citenamefont {Gepr\"{a}gs}\ \emph {et~al.}(2016)\citenamefont
  {Gepr\"{a}gs}, \citenamefont {Kehlberger}, \citenamefont {Coletta},
  \citenamefont {Qiu}, \citenamefont {Guo}, \citenamefont {Schulz},
  \citenamefont {Mix}, \citenamefont {Meyer}, \citenamefont {Kamra},
  \citenamefont {Althammer}, \citenamefont {Huebl}, \citenamefont {Jakob},
  \citenamefont {Ohnuma}, \citenamefont {Adachi}, \citenamefont {Barker},
  \citenamefont {Maekawa}, \citenamefont {Bauer}, \citenamefont {Saitoh},
  \citenamefont {Gross}, \citenamefont {Goennenwein},\ and\ \citenamefont
  {Kl\"{a}ui}}]{Geprags_2016_NatCommun_7_1}%
  \BibitemOpen
  \bibfield  {author} {\bibinfo {author} {\bibfnamefont {S.}~\bibnamefont
  {Gepr\"{a}gs}}, \bibinfo {author} {\bibfnamefont {A.}~\bibnamefont
  {Kehlberger}}, \bibinfo {author} {\bibfnamefont {F.~D.}\ \bibnamefont
  {Coletta}}, \bibinfo {author} {\bibfnamefont {Z.}~\bibnamefont {Qiu}},
  \bibinfo {author} {\bibfnamefont {E.-J.}\ \bibnamefont {Guo}}, \bibinfo
  {author} {\bibfnamefont {T.}~\bibnamefont {Schulz}}, \bibinfo {author}
  {\bibfnamefont {C.}~\bibnamefont {Mix}}, \bibinfo {author} {\bibfnamefont
  {S.}~\bibnamefont {Meyer}}, \bibinfo {author} {\bibfnamefont
  {A.}~\bibnamefont {Kamra}}, \bibinfo {author} {\bibfnamefont
  {M.}~\bibnamefont {Althammer}}, \bibinfo {author} {\bibfnamefont
  {H.}~\bibnamefont {Huebl}}, \bibinfo {author} {\bibfnamefont
  {G.}~\bibnamefont {Jakob}}, \bibinfo {author} {\bibfnamefont
  {Y.}~\bibnamefont {Ohnuma}}, \bibinfo {author} {\bibfnamefont
  {H.}~\bibnamefont {Adachi}}, \bibinfo {author} {\bibfnamefont
  {J.}~\bibnamefont {Barker}}, \bibinfo {author} {\bibfnamefont
  {S.}~\bibnamefont {Maekawa}}, \bibinfo {author} {\bibfnamefont {G.~E.~W.}\
  \bibnamefont {Bauer}}, \bibinfo {author} {\bibfnamefont {E.}~\bibnamefont
  {Saitoh}}, \bibinfo {author} {\bibfnamefont {R.}~\bibnamefont {Gross}},
  \bibinfo {author} {\bibfnamefont {S.~T.~B.}\ \bibnamefont {Goennenwein}}, \
  and\ \bibinfo {author} {\bibfnamefont {M.}~\bibnamefont {Kl\"{a}ui}},\ }\href
  {\doibase 10.1038/ncomms10452} {\bibfield  {journal} {\bibinfo  {journal}
  {Nat. Commun.}\ }\textbf {\bibinfo {volume} {7}},\ \bibinfo {pages} {10452}
  (\bibinfo {year} {2016})}\BibitemShut {NoStop}%
\bibitem [{\citenamefont {Cramer}\ \emph {et~al.}(2017)\citenamefont {Cramer},
  \citenamefont {Guo}, \citenamefont {Gepr\"{a}gs}, \citenamefont {Kehlberger},
  \citenamefont {Ivanov}, \citenamefont {Ganzhorn}, \citenamefont
  {Della~Coletta}, \citenamefont {Althammer}, \citenamefont {Huebl},
  \citenamefont {Gross}, \citenamefont {Kosel}, \citenamefont {Kl\"{a}ui},\
  and\ \citenamefont {Goennenwein}}]{Cramer_2017_NanoLett_17_6}%
  \BibitemOpen
  \bibfield  {author} {\bibinfo {author} {\bibfnamefont {J.}~\bibnamefont
  {Cramer}}, \bibinfo {author} {\bibfnamefont {E.-J.}\ \bibnamefont {Guo}},
  \bibinfo {author} {\bibfnamefont {S.}~\bibnamefont {Gepr\"{a}gs}}, \bibinfo
  {author} {\bibfnamefont {A.}~\bibnamefont {Kehlberger}}, \bibinfo {author}
  {\bibfnamefont {Y.~P.}\ \bibnamefont {Ivanov}}, \bibinfo {author}
  {\bibfnamefont {K.}~\bibnamefont {Ganzhorn}}, \bibinfo {author}
  {\bibfnamefont {F.}~\bibnamefont {Della~Coletta}}, \bibinfo {author}
  {\bibfnamefont {M.}~\bibnamefont {Althammer}}, \bibinfo {author}
  {\bibfnamefont {H.}~\bibnamefont {Huebl}}, \bibinfo {author} {\bibfnamefont
  {R.}~\bibnamefont {Gross}}, \bibinfo {author} {\bibfnamefont
  {J.}~\bibnamefont {Kosel}}, \bibinfo {author} {\bibfnamefont
  {M.}~\bibnamefont {Kl\"{a}ui}}, \ and\ \bibinfo {author} {\bibfnamefont
  {S.~T.~B.}\ \bibnamefont {Goennenwein}},\ }\href {\doibase
  10.1021/acs.nanolett.6b04522} {\bibfield  {journal} {\bibinfo  {journal}
  {Nano Lett.}\ }\textbf {\bibinfo {volume} {17}},\ \bibinfo {pages} {3334}
  (\bibinfo {year} {2017})}\BibitemShut {NoStop}%
\bibitem [{\citenamefont {Kamra}\ \emph {et~al.}(2017)\citenamefont {Kamra},
  \citenamefont {Agrawal},\ and\ \citenamefont
  {Belzig}}]{Kamra_2017_PhysRevB_96_2}%
  \BibitemOpen
  \bibfield  {author} {\bibinfo {author} {\bibfnamefont {A.}~\bibnamefont
  {Kamra}}, \bibinfo {author} {\bibfnamefont {U.}~\bibnamefont {Agrawal}}, \
  and\ \bibinfo {author} {\bibfnamefont {W.}~\bibnamefont {Belzig}},\ }\href
  {\doibase 10.1103/physrevb.96.020411} {\bibfield  {journal} {\bibinfo
  {journal} {Phys. Rev. B}\ }\textbf {\bibinfo {volume} {96}},\ \bibinfo
  {pages} {020411} (\bibinfo {year} {2017})}\BibitemShut {NoStop}%
\bibitem [{\citenamefont {Maehrlein}\ \emph {et~al.}(2018)\citenamefont
  {Maehrlein}, \citenamefont {Radu}, \citenamefont {Maldonado}, \citenamefont
  {Paarmann}, \citenamefont {Gensch}, \citenamefont {Kalashnikova},
  \citenamefont {Pisarev}, \citenamefont {Wolf}, \citenamefont {Oppeneer},
  \citenamefont {Barker},\ and\ \citenamefont
  {Kampfrath}}]{Maehrlein_2018_SciAdv_4_7}%
  \BibitemOpen
  \bibfield  {author} {\bibinfo {author} {\bibfnamefont {S.~F.}\ \bibnamefont
  {Maehrlein}}, \bibinfo {author} {\bibfnamefont {I.}~\bibnamefont {Radu}},
  \bibinfo {author} {\bibfnamefont {P.}~\bibnamefont {Maldonado}}, \bibinfo
  {author} {\bibfnamefont {A.}~\bibnamefont {Paarmann}}, \bibinfo {author}
  {\bibfnamefont {M.}~\bibnamefont {Gensch}}, \bibinfo {author} {\bibfnamefont
  {A.~M.}\ \bibnamefont {Kalashnikova}}, \bibinfo {author} {\bibfnamefont
  {R.~V.}\ \bibnamefont {Pisarev}}, \bibinfo {author} {\bibfnamefont
  {M.}~\bibnamefont {Wolf}}, \bibinfo {author} {\bibfnamefont {P.~M.}\
  \bibnamefont {Oppeneer}}, \bibinfo {author} {\bibfnamefont {J.}~\bibnamefont
  {Barker}}, \ and\ \bibinfo {author} {\bibfnamefont {T.}~\bibnamefont
  {Kampfrath}},\ }\href {\doibase 10.1126/sciadv.aar5164} {\bibfield  {journal}
  {\bibinfo  {journal} {Sci. Adv.}\ }\textbf {\bibinfo {volume} {4}},\ \bibinfo
  {pages} {eaar5164} (\bibinfo {year} {2018})}\BibitemShut {NoStop}%
\bibitem [{\citenamefont {Seifert}\ \emph {et~al.}(2018)\citenamefont
  {Seifert}, \citenamefont {Jaiswal}, \citenamefont {Barker}, \citenamefont
  {Weber}, \citenamefont {Razdolski}, \citenamefont {Cramer}, \citenamefont
  {Gueckstock}, \citenamefont {Maehrlein}, \citenamefont {Nadvornik},
  \citenamefont {Watanabe}, \citenamefont {Ciccarelli}, \citenamefont
  {Melnikov}, \citenamefont {Jakob}, \citenamefont {M\"{u}nzenberg},
  \citenamefont {Goennenwein}, \citenamefont {Woltersdorf}, \citenamefont
  {Rethfeld}, \citenamefont {Brouwer}, \citenamefont {Wolf}, \citenamefont
  {Kl\"{a}ui},\ and\ \citenamefont {Kampfrath}}]{Seifert_2018_NatCommun_9_1}%
  \BibitemOpen
  \bibfield  {author} {\bibinfo {author} {\bibfnamefont {T.~S.}\ \bibnamefont
  {Seifert}}, \bibinfo {author} {\bibfnamefont {S.}~\bibnamefont {Jaiswal}},
  \bibinfo {author} {\bibfnamefont {J.}~\bibnamefont {Barker}}, \bibinfo
  {author} {\bibfnamefont {S.~T.}\ \bibnamefont {Weber}}, \bibinfo {author}
  {\bibfnamefont {I.}~\bibnamefont {Razdolski}}, \bibinfo {author}
  {\bibfnamefont {J.}~\bibnamefont {Cramer}}, \bibinfo {author} {\bibfnamefont
  {O.}~\bibnamefont {Gueckstock}}, \bibinfo {author} {\bibfnamefont {S.~F.}\
  \bibnamefont {Maehrlein}}, \bibinfo {author} {\bibfnamefont {L.}~\bibnamefont
  {Nadvornik}}, \bibinfo {author} {\bibfnamefont {S.}~\bibnamefont {Watanabe}},
  \bibinfo {author} {\bibfnamefont {C.}~\bibnamefont {Ciccarelli}}, \bibinfo
  {author} {\bibfnamefont {A.}~\bibnamefont {Melnikov}}, \bibinfo {author}
  {\bibfnamefont {G.}~\bibnamefont {Jakob}}, \bibinfo {author} {\bibfnamefont
  {M.}~\bibnamefont {M\"{u}nzenberg}}, \bibinfo {author} {\bibfnamefont
  {S.~T.~B.}\ \bibnamefont {Goennenwein}}, \bibinfo {author} {\bibfnamefont
  {G.}~\bibnamefont {Woltersdorf}}, \bibinfo {author} {\bibfnamefont
  {B.}~\bibnamefont {Rethfeld}}, \bibinfo {author} {\bibfnamefont {P.~W.}\
  \bibnamefont {Brouwer}}, \bibinfo {author} {\bibfnamefont {M.}~\bibnamefont
  {Wolf}}, \bibinfo {author} {\bibfnamefont {M.}~\bibnamefont {Kl\"{a}ui}}, \
  and\ \bibinfo {author} {\bibfnamefont {T.}~\bibnamefont {Kampfrath}},\ }\href
  {\doibase 10.1038/s41467-018-05135-2} {\bibfield  {journal} {\bibinfo
  {journal} {Nat. Commun.}\ }\textbf {\bibinfo {volume} {9}},\ \bibinfo {pages}
  {2899} (\bibinfo {year} {2018})}\BibitemShut {NoStop}%
\bibitem [{\citenamefont {Khan}\ \emph {et~al.}(2019)\citenamefont {Khan},
  \citenamefont {Kanamaru}, \citenamefont {Hsu}, \citenamefont {Kichise},
  \citenamefont {Fujii}, \citenamefont {Koreeda},\ and\ \citenamefont
  {Satoh}}]{Khan_2019_JPhysCondensMatter_31_27}%
  \BibitemOpen
  \bibfield  {author} {\bibinfo {author} {\bibfnamefont {P.}~\bibnamefont
  {Khan}}, \bibinfo {author} {\bibfnamefont {M.}~\bibnamefont {Kanamaru}},
  \bibinfo {author} {\bibfnamefont {W.-H.}\ \bibnamefont {Hsu}}, \bibinfo
  {author} {\bibfnamefont {M.}~\bibnamefont {Kichise}}, \bibinfo {author}
  {\bibfnamefont {Y.}~\bibnamefont {Fujii}}, \bibinfo {author} {\bibfnamefont
  {A.}~\bibnamefont {Koreeda}}, \ and\ \bibinfo {author} {\bibfnamefont
  {T.}~\bibnamefont {Satoh}},\ }\href {\doibase 10.1088/1361-648x/ab1665}
  {\bibfield  {journal} {\bibinfo  {journal} {J. Phys.: Condens. Matter}\
  }\textbf {\bibinfo {volume} {31}},\ \bibinfo {pages} {275402} (\bibinfo
  {year} {2019})}\BibitemShut {NoStop}%
\bibitem [{\citenamefont {Hsu}\ \emph {et~al.}(2020)\citenamefont {Hsu},
  \citenamefont {Shen}, \citenamefont {Fujii}, \citenamefont {Koreeda},\ and\
  \citenamefont {Satoh}}]{Hsu_2020_PhysRevB_102_17}%
  \BibitemOpen
  \bibfield  {author} {\bibinfo {author} {\bibfnamefont {W.-H.}\ \bibnamefont
  {Hsu}}, \bibinfo {author} {\bibfnamefont {K.}~\bibnamefont {Shen}}, \bibinfo
  {author} {\bibfnamefont {Y.}~\bibnamefont {Fujii}}, \bibinfo {author}
  {\bibfnamefont {A.}~\bibnamefont {Koreeda}}, \ and\ \bibinfo {author}
  {\bibfnamefont {T.}~\bibnamefont {Satoh}},\ }\href {\doibase
  10.1103/physrevb.102.174432} {\bibfield  {journal} {\bibinfo  {journal}
  {Phys. Rev. B}\ }\textbf {\bibinfo {volume} {102}},\ \bibinfo {pages}
  {174432} (\bibinfo {year} {2020})}\BibitemShut {NoStop}%
\bibitem [{\citenamefont {Kim}\ \emph {et~al.}(2017)\citenamefont {Kim},
  \citenamefont {Kim}, \citenamefont {Hirata}, \citenamefont {Oh},
  \citenamefont {Tono}, \citenamefont {Kim}, \citenamefont {Okuno},
  \citenamefont {Ham}, \citenamefont {Kim}, \citenamefont {Go}, \citenamefont
  {Tserkovnyak}, \citenamefont {Tsukamoto}, \citenamefont {Moriyama},
  \citenamefont {Lee},\ and\ \citenamefont {Ono}}]{Kim_2017_NatMater_16_12}%
  \BibitemOpen
  \bibfield  {author} {\bibinfo {author} {\bibfnamefont {K.-J.}\ \bibnamefont
  {Kim}}, \bibinfo {author} {\bibfnamefont {S.~K.}\ \bibnamefont {Kim}},
  \bibinfo {author} {\bibfnamefont {Y.}~\bibnamefont {Hirata}}, \bibinfo
  {author} {\bibfnamefont {S.-H.}\ \bibnamefont {Oh}}, \bibinfo {author}
  {\bibfnamefont {T.}~\bibnamefont {Tono}}, \bibinfo {author} {\bibfnamefont
  {D.-H.}\ \bibnamefont {Kim}}, \bibinfo {author} {\bibfnamefont
  {T.}~\bibnamefont {Okuno}}, \bibinfo {author} {\bibfnamefont {W.~S.}\
  \bibnamefont {Ham}}, \bibinfo {author} {\bibfnamefont {S.}~\bibnamefont
  {Kim}}, \bibinfo {author} {\bibfnamefont {G.}~\bibnamefont {Go}}, \bibinfo
  {author} {\bibfnamefont {Y.}~\bibnamefont {Tserkovnyak}}, \bibinfo {author}
  {\bibfnamefont {A.}~\bibnamefont {Tsukamoto}}, \bibinfo {author}
  {\bibfnamefont {T.}~\bibnamefont {Moriyama}}, \bibinfo {author}
  {\bibfnamefont {K.-J.}\ \bibnamefont {Lee}}, \ and\ \bibinfo {author}
  {\bibfnamefont {T.}~\bibnamefont {Ono}},\ }\href {\doibase 10.1038/nmat4990}
  {\bibfield  {journal} {\bibinfo  {journal} {Nat. Mater.}\ }\textbf {\bibinfo
  {volume} {16}},\ \bibinfo {pages} {1187} (\bibinfo {year}
  {2017})}\BibitemShut {NoStop}%
\bibitem [{\citenamefont {Donges}\ \emph {et~al.}(2020)\citenamefont {Donges},
  \citenamefont {Grimm}, \citenamefont {Jakobs}, \citenamefont {Selzer},
  \citenamefont {Ritzmann}, \citenamefont {Atxitia},\ and\ \citenamefont
  {Nowak}}]{Donges_2020_PhysRevRes_2_1}%
  \BibitemOpen
  \bibfield  {author} {\bibinfo {author} {\bibfnamefont {A.}~\bibnamefont
  {Donges}}, \bibinfo {author} {\bibfnamefont {N.}~\bibnamefont {Grimm}},
  \bibinfo {author} {\bibfnamefont {F.}~\bibnamefont {Jakobs}}, \bibinfo
  {author} {\bibfnamefont {S.}~\bibnamefont {Selzer}}, \bibinfo {author}
  {\bibfnamefont {U.}~\bibnamefont {Ritzmann}}, \bibinfo {author}
  {\bibfnamefont {U.}~\bibnamefont {Atxitia}}, \ and\ \bibinfo {author}
  {\bibfnamefont {U.}~\bibnamefont {Nowak}},\ }\href {\doibase
  10.1103/physrevresearch.2.013293} {\bibfield  {journal} {\bibinfo  {journal}
  {Phys. Rev. Res.}\ }\textbf {\bibinfo {volume} {2}},\ \bibinfo {pages}
  {013293} (\bibinfo {year} {2020})}\BibitemShut {NoStop}%
\bibitem [{\citenamefont {Hinzke}\ and\ \citenamefont
  {Nowak}(2011)}]{Hinzke_2011_PhysRevLett_107_2}%
  \BibitemOpen
  \bibfield  {author} {\bibinfo {author} {\bibfnamefont {D.}~\bibnamefont
  {Hinzke}}\ and\ \bibinfo {author} {\bibfnamefont {U.}~\bibnamefont {Nowak}},\
  }\href {\doibase 10.1103/physrevlett.107.027205} {\bibfield  {journal}
  {\bibinfo  {journal} {Phys. Rev. Lett.}\ }\textbf {\bibinfo {volume} {107}},\
  \bibinfo {pages} {027205} (\bibinfo {year} {2011})}\BibitemShut {NoStop}%
\bibitem [{\citenamefont {Schlickeiser}\ \emph {et~al.}(2014)\citenamefont
  {Schlickeiser}, \citenamefont {Ritzmann}, \citenamefont {Hinzke},\ and\
  \citenamefont {Nowak}}]{Schlickeiser_2014_PhysRevLett_113_9}%
  \BibitemOpen
  \bibfield  {author} {\bibinfo {author} {\bibfnamefont {F.}~\bibnamefont
  {Schlickeiser}}, \bibinfo {author} {\bibfnamefont {U.}~\bibnamefont
  {Ritzmann}}, \bibinfo {author} {\bibfnamefont {D.}~\bibnamefont {Hinzke}}, \
  and\ \bibinfo {author} {\bibfnamefont {U.}~\bibnamefont {Nowak}},\ }\href
  {\doibase 10.1103/physrevlett.113.097201} {\bibfield  {journal} {\bibinfo
  {journal} {Phys. Rev. Lett.}\ }\textbf {\bibinfo {volume} {113}},\ \bibinfo
  {pages} {097201} (\bibinfo {year} {2014})}\BibitemShut {NoStop}%
\bibitem [{\citenamefont {Selzer}\ \emph {et~al.}(2016)\citenamefont {Selzer},
  \citenamefont {Atxitia}, \citenamefont {Ritzmann}, \citenamefont {Hinzke},\
  and\ \citenamefont {Nowak}}]{Selzer_2016_PhysRevLett_117_10}%
  \BibitemOpen
  \bibfield  {author} {\bibinfo {author} {\bibfnamefont {S.}~\bibnamefont
  {Selzer}}, \bibinfo {author} {\bibfnamefont {U.}~\bibnamefont {Atxitia}},
  \bibinfo {author} {\bibfnamefont {U.}~\bibnamefont {Ritzmann}}, \bibinfo
  {author} {\bibfnamefont {D.}~\bibnamefont {Hinzke}}, \ and\ \bibinfo {author}
  {\bibfnamefont {U.}~\bibnamefont {Nowak}},\ }\href {\doibase
  10.1103/physrevlett.117.107201} {\bibfield  {journal} {\bibinfo  {journal}
  {Phys. Rev. Lett.}\ }\textbf {\bibinfo {volume} {117}},\ \bibinfo {pages}
  {107201} (\bibinfo {year} {2016})}\BibitemShut {NoStop}%
\bibitem [{\citenamefont {Shokr}\ \emph {et~al.}(2019)\citenamefont {Shokr},
  \citenamefont {Sandig}, \citenamefont {Erkovan}, \citenamefont {Zhang},
  \citenamefont {Bernien}, \citenamefont {\"{U}nal}, \citenamefont {Kronast},
  \citenamefont {Parlak}, \citenamefont {Vogel},\ and\ \citenamefont
  {Kuch}}]{Shokr_2019_PhysRevB_99_21}%
  \BibitemOpen
  \bibfield  {author} {\bibinfo {author} {\bibfnamefont {Y.~A.}\ \bibnamefont
  {Shokr}}, \bibinfo {author} {\bibfnamefont {O.}~\bibnamefont {Sandig}},
  \bibinfo {author} {\bibfnamefont {M.}~\bibnamefont {Erkovan}}, \bibinfo
  {author} {\bibfnamefont {B.}~\bibnamefont {Zhang}}, \bibinfo {author}
  {\bibfnamefont {M.}~\bibnamefont {Bernien}}, \bibinfo {author} {\bibfnamefont
  {A.~A.}\ \bibnamefont {\"{U}nal}}, \bibinfo {author} {\bibfnamefont
  {F.}~\bibnamefont {Kronast}}, \bibinfo {author} {\bibfnamefont
  {U.}~\bibnamefont {Parlak}}, \bibinfo {author} {\bibfnamefont
  {J.}~\bibnamefont {Vogel}}, \ and\ \bibinfo {author} {\bibfnamefont
  {W.}~\bibnamefont {Kuch}},\ }\href {\doibase 10.1103/physrevb.99.214404}
  {\bibfield  {journal} {\bibinfo  {journal} {Phys. Rev. B}\ }\textbf {\bibinfo
  {volume} {99}},\ \bibinfo {pages} {214404} (\bibinfo {year}
  {2019})}\BibitemShut {NoStop}%
\bibitem [{\citenamefont {Zhang}\ \emph {et~al.}(2016)\citenamefont {Zhang},
  \citenamefont {Zhou},\ and\ \citenamefont
  {Ezawa}}]{Zhang_2016_NatCommun_7_1}%
  \BibitemOpen
  \bibfield  {author} {\bibinfo {author} {\bibfnamefont {X.}~\bibnamefont
  {Zhang}}, \bibinfo {author} {\bibfnamefont {Y.}~\bibnamefont {Zhou}}, \ and\
  \bibinfo {author} {\bibfnamefont {M.}~\bibnamefont {Ezawa}},\ }\href
  {\doibase 10.1038/ncomms10293} {\bibfield  {journal} {\bibinfo  {journal}
  {Nat. Commun.}\ }\textbf {\bibinfo {volume} {7}},\ \bibinfo {pages} {10293}
  (\bibinfo {year} {2016})}\BibitemShut {NoStop}%
\bibitem [{\citenamefont {Hirata}\ \emph {et~al.}(2019)\citenamefont {Hirata},
  \citenamefont {Kim}, \citenamefont {Kim}, \citenamefont {Lee}, \citenamefont
  {Oh}, \citenamefont {Kim}, \citenamefont {Nishimura}, \citenamefont {Okuno},
  \citenamefont {Futakawa}, \citenamefont {Yoshikawa}, \citenamefont
  {Tsukamoto}, \citenamefont {Tserkovnyak}, \citenamefont {Shiota},
  \citenamefont {Moriyama}, \citenamefont {Choe}, \citenamefont {Lee},\ and\
  \citenamefont {Ono}}]{Hirata_2019_NatNanotechnol_14_3}%
  \BibitemOpen
  \bibfield  {author} {\bibinfo {author} {\bibfnamefont {Y.}~\bibnamefont
  {Hirata}}, \bibinfo {author} {\bibfnamefont {D.-H.}\ \bibnamefont {Kim}},
  \bibinfo {author} {\bibfnamefont {S.~K.}\ \bibnamefont {Kim}}, \bibinfo
  {author} {\bibfnamefont {D.-K.}\ \bibnamefont {Lee}}, \bibinfo {author}
  {\bibfnamefont {S.-H.}\ \bibnamefont {Oh}}, \bibinfo {author} {\bibfnamefont
  {D.-Y.}\ \bibnamefont {Kim}}, \bibinfo {author} {\bibfnamefont
  {T.}~\bibnamefont {Nishimura}}, \bibinfo {author} {\bibfnamefont
  {T.}~\bibnamefont {Okuno}}, \bibinfo {author} {\bibfnamefont
  {Y.}~\bibnamefont {Futakawa}}, \bibinfo {author} {\bibfnamefont
  {H.}~\bibnamefont {Yoshikawa}}, \bibinfo {author} {\bibfnamefont
  {A.}~\bibnamefont {Tsukamoto}}, \bibinfo {author} {\bibfnamefont
  {Y.}~\bibnamefont {Tserkovnyak}}, \bibinfo {author} {\bibfnamefont
  {Y.}~\bibnamefont {Shiota}}, \bibinfo {author} {\bibfnamefont
  {T.}~\bibnamefont {Moriyama}}, \bibinfo {author} {\bibfnamefont {S.-B.}\
  \bibnamefont {Choe}}, \bibinfo {author} {\bibfnamefont {K.-J.}\ \bibnamefont
  {Lee}}, \ and\ \bibinfo {author} {\bibfnamefont {T.}~\bibnamefont {Ono}},\
  }\href {\doibase 10.1038/s41565-018-0345-2} {\bibfield  {journal} {\bibinfo
  {journal} {Nat. Nanotechnol.}\ }\textbf {\bibinfo {volume} {14}},\ \bibinfo
  {pages} {232} (\bibinfo {year} {2019})}\BibitemShut {NoStop}%
\bibitem [{\citenamefont {Jungwirth}\ \emph {et~al.}(2016)\citenamefont
  {Jungwirth}, \citenamefont {Marti}, \citenamefont {Wadley},\ and\
  \citenamefont {Wunderlich}}]{Jungwirth_2016_NatNanotechnol_11_3}%
  \BibitemOpen
  \bibfield  {author} {\bibinfo {author} {\bibfnamefont {T.}~\bibnamefont
  {Jungwirth}}, \bibinfo {author} {\bibfnamefont {X.}~\bibnamefont {Marti}},
  \bibinfo {author} {\bibfnamefont {P.}~\bibnamefont {Wadley}}, \ and\ \bibinfo
  {author} {\bibfnamefont {J.}~\bibnamefont {Wunderlich}},\ }\href {\doibase
  10.1038/nnano.2016.18} {\bibfield  {journal} {\bibinfo  {journal} {Nat.
  Nanotechnol.}\ }\textbf {\bibinfo {volume} {11}},\ \bibinfo {pages} {231}
  (\bibinfo {year} {2016})}\BibitemShut {NoStop}%
\end{thebibliography}%
\end{document}